\documentclass[a4paper,article,10pt]{memoir}
\usepackage[english]{babel}
\usepackage{CCLAuthor}
\usepackage[reqno]{amsmath}% AMS-LaTeX-first 
\usepackage{amssymb}
\usepackage{amsthm}  
\theoremstyle{plain}

\theoremstyle{definition}

\numberwithin{theorem}{chapter}% AMS-LaTeX-last
\usepackage[round,authoryear]{natbib} % bibliography
\usepackage[]{graphicx}               % graphics
\usepackage{color}
\usepackage{mathabx}

\definecolor{ThemeColor}{rgb}{0.3294,0.5373,0.7216}

\usepackage[font={small,rm},nooneline,margin=0.cm]{caption}  %[2008/04/01]   %margin=0.5cm,
\DeclareCaptionFont{blue}{\color{ThemeColor}}
\DeclareCaptionFormat{myformat}{#1#2\vspace*{0.1cm}\hrule\vspace*{0.1cm} #3}
\graphicspath{{}}	
\divide\pdfpxdimen by 96

\begin{document}

%
%----------------------------------------------
%
%    declare the TITLE of your contribution
%    (see Contribs.tex)
%
\title{\textbf{Fluid dynamics of Earth's core:\\ geodynamo, inner core dynamics,\\ core formation.}}
%    
%    AUTHOR(s) and AFFILIATION(s)
%    (see Contribs.tex for a nontrivial example with
%    two authors and three affiliations)
%
\author{%
    Renaud Deguen \textsuperscript{*}
    and 
    Marine Lasbleis \textsuperscript{\dag}
    % end authors
    \\ \smallskip\small% some space
    % begin affiliations
    \textsuperscript{*} Universit\'e de Lyon, UCBL, ENSL, CNRS,\\ \small Laboratoire de G\'eologie de Lyon: Terre, Plan\`etes, Environnement,\\ \small  Villeurbanne, France  
    \\ \smallskip\small
    \textsuperscript{\dag} 
    Laboratoire de Plan\'etologie et G\'eodynamique,\\ \small UMR-CNRS 6112, Universit\'e de Nantes, Nantes, France
    \\
    }% end affiliations
    \maketitle
%
%----------------------------------------------
%
%   ABSTRACT: comment the lines between ABS-first and ABS-last 
%   if your contribution has no abstract
%    
% 	abstract is essential for database indexing % ap140610
%
    \begin{abstract}
    This chapter is build from three 1.5 hours lectures given in Udine in april 2018 on various aspects of Earth's core dynamics. 
The chapter starts with a short historical note on the discovery of Earth's magnetic field and core (section \ref{Section:Introduction}).
We then turn to an introduction of magnetohydrodynamics (section \ref{Section:MHD}),  introducing and discussing the \textit{induction equation} and
the form and effects of the Lorentz force.
Section \ref{Section:GeomagneticFieldGeometry} is devoted to the description of Earth's magnetic field, introducing its spherical harmonics description and showing how it can be used to demonstrate the internal origin of the geomagnetic origin.
We then move to an introduction of the convection-driven model of the geodynamo (section \ref{Section:CoreDynamics}), discussing our current understanding of the dynamics of Earth's core, obtaining heuristically the Ekman dependency of the critical Rayleigh number for natural rotating convection, and introducing the equations and non-dimensional parameters used to model a convectively driven dynamo. 
The following section deals with the energetics of the geodynamo (section \ref{Section:Energetics}).
The final two section deal with the dynamics of the inner core, focusing on the effect of the magnetic field (section \ref{Section:InnerCore}), and with the formation of the core (section \ref{Section:CoreFormation}).

Given the wide scope of this chapter and the limited time available,  this introduction to Earth's core dynamics is by no means intended to be comprehensive. 
For more informations, the interested reader may refer to \cite{Jones2011}, \cite{Olson2013}, or \cite{Christensen2015} on the geomagnetic field and the geodynamo,  to \cite{Sumita2015}, \cite{Deguen2012} and \cite{lasbleis2015b} on the dynamics of the inner core, and to \cite{Rubie2015} on core formation.
    \end{abstract}
% 
%
%----------------------------------------------
%   SECTIONING commands
%   numbered section:          \CCLsection{...}
%   unnumbered section:        \CCLsection*{...}
%   
%   numbered subsection:       \CCLsubsection{...}
%   unnumbered subsection:     \CCLsubsection*{...} 
%   
%   there are only unnumbered 
%   subsubsections             \CCLsubsubsection{...}
%   
%---

\chapter{Introduction}

\label{Section:Introduction}

\subsection{The birth of geomagnetism}

\textit{Magnetism}, the study of magnetic fields, may be argued to start with the discovery and description of natural magnets, or lodestones. 
Magnets have been known since at least $\sim 600$ BC in Greece, and $\sim 300$ BC in ancient China.
Aristote (384-322 BC) attributes to Thales ($\sim 600$ BC) the observation that ``magnets exert a force on iron''. 
The birth of \textit{geomagnetism} requires the realisation that there is an ambient magnetic field at Earth's scale. 
The primary observation is that a magnet allowed to rotate freely in a horizontal plane points toward the north (as defined by the direction of the polar star).
This seems to have been known since at least the I$^\text{st}$ century in China;  the compass was used in navigation since at least the X$^\text{th}$ century in China, and since the XII$^\text{th}$ century in Europe.

In the western world, the first scientific treatise describing the properties of magnets is a letter (\textit{Epistola}) written by Pierre de Maricourt (also known as Petrus Peregrinus) in 1269 to a friend of him, Sygerus de Foucaucourt.
Pierre de Maricourt wrote his \textit{epistola} during a  military campaign led by Charles d'Anjou, while the french army was besieging the town of Lucera, in southern Italia.
In his letter, Pierre de Maricourt describes some of the most important properties of natural magnets: (i) a magnet has two poles (North and South); (ii) the South pole of a magnet attracts the North pole of another magnet, two identical poles repel each others; (iii) the two poles of a magnet cannot be isolated: breaking a magnet into two pieces gives two magnets, each with two poles (in modern terms, this means that there is no magnetic \textit{monopole}); (iv) a magnet free to rotate points toward the North pole.
In a second part of his letter, Pierre de Maricourt describes the design of a perpetual motion machine using the properties of magnets. 

The exact geometry of Earth's magnetic field turns out to be more complex than what early observations suggested:
\begin{enumerate}
\item
Compasses do not point exactly toward the geographic north. 
The \textit{magnetic declination} -- the angle between the direction of the true geographic north and the direction given by a compass -- has been known since at least the end of XI$^\mathrm{th}$'s century in China.
Christopher Colombus was the first to observe that the magnetic declination varies spatially: while sailing westward from the old world, he observed that the direction given by his compasses changed with longitude along his path. 
The declination was positive in Europe (\textit{i.e.} a compass points slightly to the west of the true north), decreased gradually along Columbus' path, reached zero somewhere in the middle of the Atlantic ocean, and became negative on the western side of the Atlantic.
\item
Furthermore, the actual direction of the magnetic field is tilted from the horizontal plane: when allowed to rotate around an horizontal axis, a magnetised needle points toward a direction which makes a well defined angle with the horizontal, called the  \textit{magnetic inclination}.
This has been noticed by Georg Hartmann in 1544 and precisely measured by Robert Norman, who found a declination of 71$^\mathrm{o}$50' in London in 1581.
\end{enumerate}

\begin{figure}[t]
\centering
\includegraphics[width=0.5\linewidth]{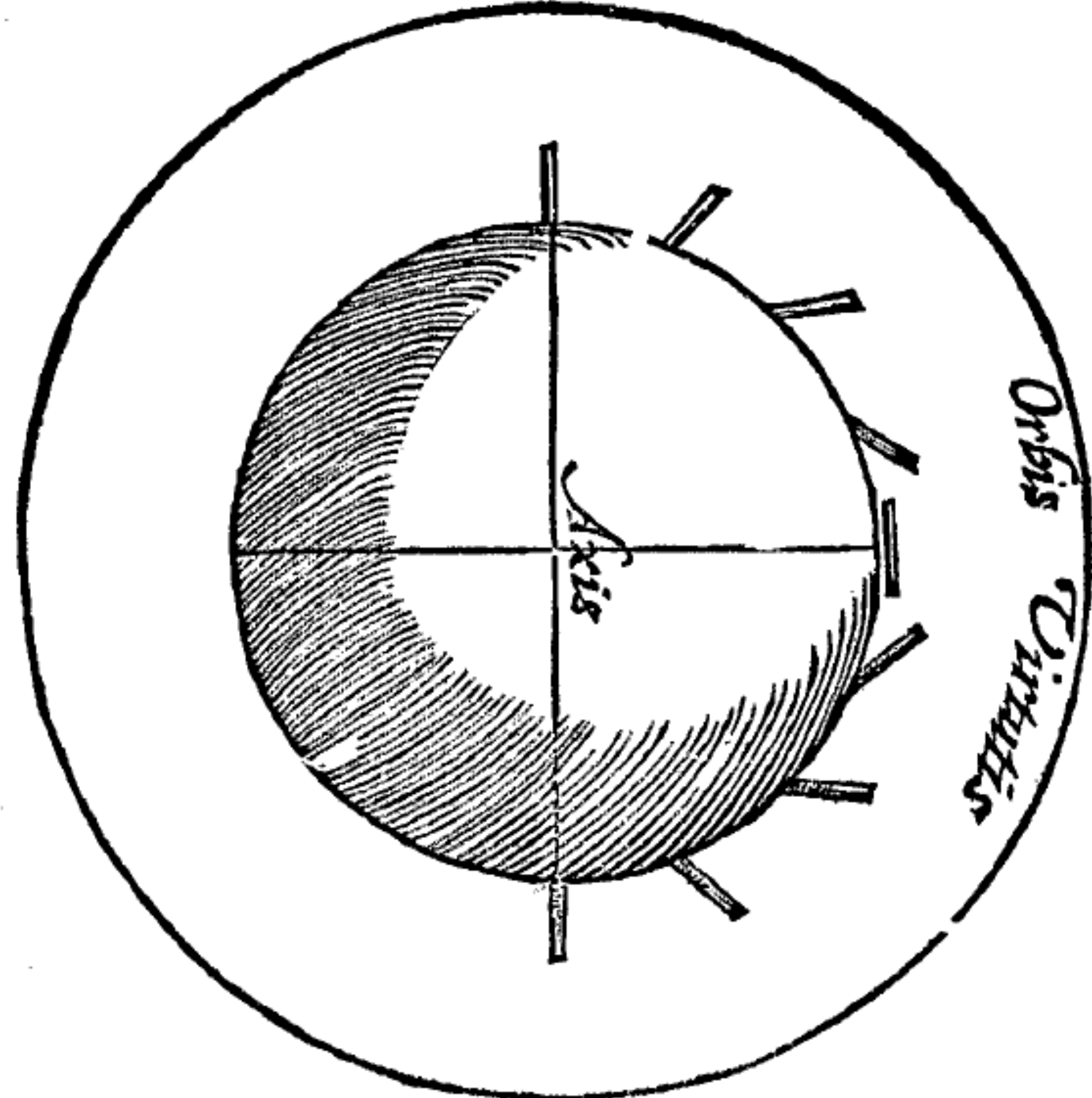}
\caption{Gilbert's \textit{terella}, a natural magnet carved into a spherical shape (\textit{De Magnete}, 1600). The direction of the magnetic field at the surface of the \textit{terella} is indicated on the right-hand side of the figure: the magnetic field is perpendicular to the surface at the poles, and parallel to the surface at the equator. 
}
\label{Fig:Terella}
\end{figure}

\begin{figure}
\centering
\includegraphics[width=0.85\linewidth]{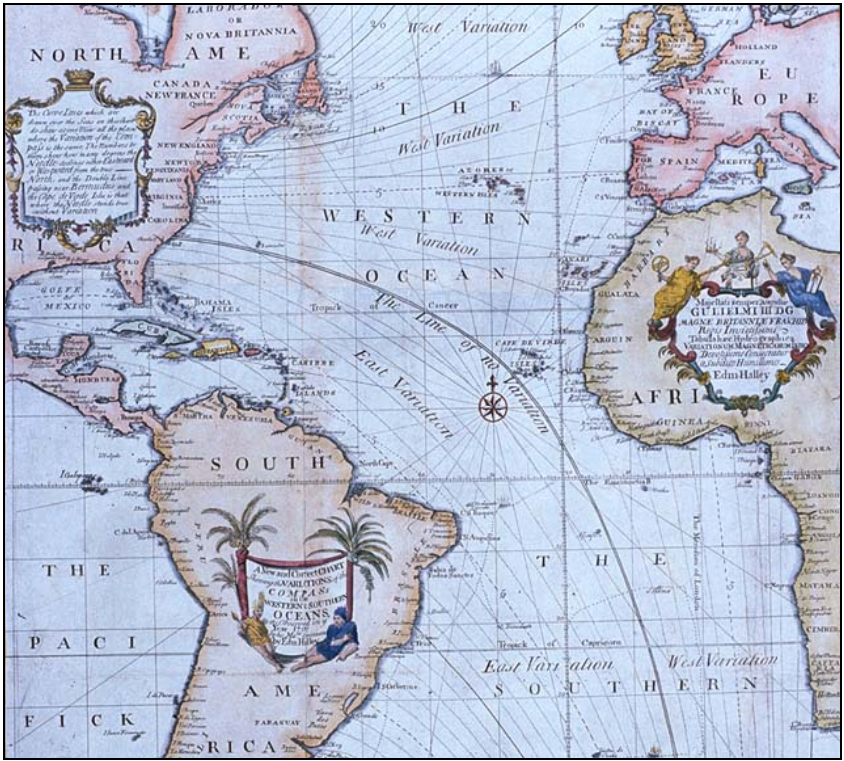}

\caption{Halley's map of magnetic declination, published in 1701.
The thick black line with the annotation ``The line of no Variation'' is the line of zero declination.
}
\label{Fig:Halley}
\end{figure}

Perhaps the first scientifically grounded theory of the origin of the geomagnetic field is Gilbert's theory published in 1600 in his \textit{De Magnete}, which states that \textit{Earth is a magnet}.
Gilbert's claim was based on the similarity between the latitudinal variations of the inclination of Earth's magnetic field  and the variations of the magnetic field orientation he measured on a model of Earth, his \textit{terella}, obtained by carving a lodestone into the shape of a sphere (figure \ref{Fig:Terella}).
The geometry of Earth's magnetic field indeed strongly resembles that of a magnet, but it has soon been clear that this theory is at best incomplete.
Perhaps the strongest argument against Gilbert's theory comes from the observation that the geomagnetic field varies with time.
Measurements of the declination in London across the XVII$^\mathrm{th}$ century have shown variations of several degrees in a few decades; 
variations of the declination and inclination on a decadal timescale have been well documented by the mid XVIII$^\mathrm{th}$ century. 
Other evidences of fast magnetic field variations include
 the westward drift of a line of zero declination inferred to be in the middle of the Atlantic ocean in 1701 by Halley (figure \ref{Fig:Halley}), which had reached the South American continent about 60 years after the publication of Halley's map. 
Modern measurements and inferences have confirmed that the geomagnetic field varies on timescales spanning a very wide range, including monthly to decadal variations.
Such fast time variations are hardly compatible with Gilbert's theory.

\subsection{The discovery of the core}

The basic structure of the Earth -- a solid, rocky mantle surrounding an iron-alloy core which is molten except for a solid inner core at its center -- has not been known before the early XX$^\mathrm{th}$ century. 
In the XVIII$^\mathrm{th}$ and early XIX$^\mathrm{th}$ centuries, it was widely believed that most of the Earth is molten, except for a thin crust on which we live. 
A qualitative argument in favour of this view is the observation of molten lava erupting from volcanoes, which is straightforwardly explained if the interior of the Earth is indeed molten.
A more quantitative argument is given by  the observation that the temperature increases with depth in mines, at a rate such that the melting temperature of most crustal rocks would be reached at a depth lower than 50 km.\footnote{The temperature actually does not reach the melting temperature because: (i) the melting temperature increases with pressure; and (ii) the high temperature gradient observed in mines is limited to depth of $\sim 30$ km or less, the temperature gradient becoming much less pronounced at deeper depth because of convective motions in the mantle.} 

This view was challenged by scientists including Amp\`ere, Poisson, or Kelvin, Amp\`ere arguing that such a thin crust cannot withstand the deviatoric stresses induced by the lunar tides. 
The fact that the interior of the Earth is solid to a large extent has been demonstrated in the middle of the XIX$^\mathrm{th}$ century by  Lord Kelvin and George Darwin (the son of Charles), by considering the deformation of the interior of the Earth in response to tidal forcing. 
The oceanic tides that we see at Earth's surface are the relative motion of the oceans with respect to the crust.
We would not see them if the Earth was fully molten since it would deform in the same way as the oceans.
The fact that we do see oceanic tides means that the interior of the Earth resists deformation, \textit{i.e.} has a finite rigidity. 
Calculations based on available data led Kelvin to state that ``the Earth is as rigid as steel''.

The first quantitative model of the interior of the Earth including a rocky mantle and an iron-rich core was proposed in 1896 by Emil Wiechert based on the following constraints and assumptions.  
The mean density of the Earth (5.6, obtained from the value $g$ at Earth's surface) is significantly higher than the density of rocks near the surface ($\sim 3$). 
Wiechert believed that compression alone cannot increase the density of rocks to such extent, and argued that a change of chemical composition at some depth is required. 
Since only metals were known to have densities higher than 5.6, it was natural for Wiechert to assume the deepest layer (the core) of his model to be metallic. 
The estimated moment of inertia of the Earth provides further constraints on the repartition of mass within the Earth.
By considering a two-layer model with constant density in each layer, the constraints coming from the mean density and moment of inertia of the Earth enable to determine the size of the core and densities of the core and mantle.
Wiechert obtained a core radius of about 4970 km, and mantle and core densities of $3.2$ and $8.2$, respectively.
His estimate of the core density being close to that of iron at low pressure ($7.8$), Wiechert proposed that the core is composed of iron, and attributes the difference of density to compression.

The end of XIX$^\mathrm{th}$ century and the beginning of the XX$^\mathrm{th}$ century witnessed the emergence of seismology as a new and growing branch of geophysics. 
Milne, Wiechert, Oldham and others realised that the study of seismic waves\footnote{Compression waves, called P-waves, and shear waves, called S-waves.} propagating within the Earth offers a way to probe the interior of the Earth. 
By studying the time of arrival of P-waves as a function of the distance from Earthquakes' epicenters, Oldham (1906) discovered that no  P-wave were observed at epicentral distances between $\sim 105^\mathrm{o}$ and $\sim 140^\mathrm{o}$.
He explained this \textit{shadow zone} by the presence of a discontinuity (which he interpreted as the boundary between the core and mantle of Wiechert's model), at which the wave propagation velocity decreases, thus refracting the waves downward. 
Gutenberg (who was a student of Wiechert) estimated the radius of the core to be about $3470\pm20$ km, which is consistent with the modern estimate of 3480 km.
At that time the core was believed to be solid, mostly because of Kelvin and Darwin's estimates of the rigidity of the Earth. 
The molten state of the core has been established by Jeffrey in 1926 by comparing the mean rigidity of the Earth to the rigidity of the mantle obtained from estimates of the speed of propagation of seismic waves in the mantle.
The mean rigidity of the Earth is significantly smaller than the rigidity of the mantle, which implies that the rigidity of the core must be low. 

The final major piece of the basic structure of the Earth has been discovered by Inge Lehmann, a danish seismologist.
In spite of what has been said in the previous paragraph, P-waves are sometimes observed in the core shadow zone.
These anomalous arrivals, called P', have been attributed by \cite{Lehmann1936} to P-waves reflected at a discontinuity inside the core at which the seisimic waves velocity increases. 
She thus discovered the \textit{inner core}, a sphere 1221 km in radius (Inge Lehmann's estimate was 1400 km), which was later shown to be solid \citep{Dziewonski1971}.

By comparing seismological estimates of the density and elastic moduli in the core to the results of high pressure experiments, later work established that the liquid and solid parts of the core are both made predominantly of iron, alloyed with lighter elements (Si, O, S, C, H, ...) \citep[\textit{e.g.}][]{Birch1952,Birch1964,Jephcoat1987,Alfe2000,Badro2007}.
 These light elements are more abundant in the liquid outer core (about 10 wt.\%) than in the solid inner core (less than 5 wt.\%), which is consistent with the idea that the inner core is crystallising from the outer core \citep{Birch1940,Jacobs1953} since the impurities in an alloy are usually partitioned into the melt upon crystallisation.

%--------------------------------------------------------------------------------------------------------------------------------------------%
%--------------------------------------------------------------------------------------------------------------------------------------------%

%--------------------------------------------------------------------------------------------------------------------------------------------%
%--------------------------------------------------------------------------------------------------------------------------------------------%
\chapter{A short introduction to magnetohydrodynamics} % (MHD)}
%--------------------------------------------------------------------------------------------------------------------------------------------%
%--------------------------------------------------------------------------------------------------------------------------------------------%
\label{Section:MHD}

Magnetohydrodynamics concerns the interactions between the flow of an electrically conducting fluid and a magnetic field.
It thus combines hydrodynamics, which is modelled with the Navier-Stokes and continuity equations, and electromagnetism, which is modelled with Maxwell's equations, supplemented with Ohm's law and the conservation of electric charges. 
Interactions between the flow and magnetic field are twofold: (i) the motion of an electrically conducting material can act as a source term for the evolution of the magnetic field, which can be quantified with the \textit{induction equation}, a prognostic equation for the evolution of the magnetic field; and (ii) the magnetic field generates a force on the electrically conducting material, the \textit{Lorentz force}.
The goal of this section is to introduce the induction equation, and discuss the form and effects of the Lorentz force. 

\section{Classical electromagnetism}

Maxwell's equations are
\begin{align}
\boldsymbol{\nabla} \cdot \mathbf{E} &= \frac{\rho_{c}}{\epsilon_{0}}	&	\text{\quad Gauss law}, \label{Eq:GaussLaw} \\
\boldsymbol{\nabla} \times  \mathbf{E} &= - \frac{\partial  \mathbf{B}}{\partial t} & \text{\quad Faraday's law of induction}, \label{Eq:FaradayLaw}\\
\boldsymbol{\nabla} \cdot \mathbf{B} &= 0		& \text{\quad No magnetic monopole}, \\
\boldsymbol{\nabla} \times \mathbf{B} &= \mu_{0}\,  \mathbf{j} + \mu_{0} \epsilon_{0} \frac{\partial  \mathbf{E}}{\partial t} & \text{\quad Ampere's law + displacement current},  \label{Eq:Ampere}
\end{align}
where $\mathbf{E}$ is the electric field, $\mathbf{B}$ is the magnetic field, $\rho_{c}$ is the electric charge density, $\mathbf{j}$ is the electric current density, $\mu_{0}$ is the permeability of free space, and $\epsilon_{0}$ is the permittivity of free space.
Note that $\mu_{0} \epsilon_{0} =1/c^{2}$, where $c$ is the speed of light.
Maxwell's equations are supplemented by an equation expressing that electric charges are conserved, which we write 
\begin{equation}
\frac{\partial \rho_{c}}{\partial t} + \boldsymbol{\nabla} \cdot \mathbf{j} = 0,
\label{Eq:ChargeConservation}
\end{equation}
and by Ohm's law, which can be written in a moving electrically conducting material as
\begin{equation}
\mathbf{j} = \rho_{c} \mathbf{u} + \sigma \left( \mathbf{E} + \mathbf{u} \times \mathbf{B} \right),
\label{Eq:OhmLaw0}
\end{equation}
where $\sigma$ is the electrical conductivity, and $\mathbf{u}$ the velocity of the conducting material.

\subsection{The charge density in an electrically conducting material}
Replacing $\mathbf{j}$ in equation \eqref{Eq:ChargeConservation} by its expression from Ohm's law and then using Gauss law (equation \eqref{Eq:GaussLaw}), we obtain the following equation for $\rho_{c}$, 
\begin{equation}
\frac{\epsilon_{0}}{\sigma} \frac{D \rho_{c}}{D t} + \rho_{c} = - \epsilon_{0} \boldsymbol{\nabla}\cdot \left(\mathbf{u}\times\mathbf{B}\right),
\end{equation}
which means that the charge density follows time variations of $\epsilon _{0}\boldsymbol{\nabla}\cdot \left(\mathbf{u}\times\mathbf{B}\right)$  with a response time, or time lag, $\epsilon_{0}/\sigma$.  
In most liquid metals, $\epsilon_{0}/\sigma$ is between $10^{-16}$ and $10^{-18}$ s and the charge density responds almost instantaneously to changes in $\mathbf{u}$ and $\mathbf{B}$. 
The charge density must therefore be very close to
\begin{equation}
\rho_{c} = - \epsilon_{0} \boldsymbol{\nabla}\cdot \left(\mathbf{u}\times\mathbf{B}\right) .
\label{Eq:ChargeDensityQuasiStatic}
\end{equation}

This estimate can be used to discuss the importance of the term $\rho_{c} \mathbf{u}$ in Ohm's law.
Denoting by $L$ and $U$ typical length and velocity scales of a given problem, comparing $\rho_{c} \mathbf{u}$  and $\sigma \mathbf{u} \times \mathbf{B}$ shows that
\begin{equation}
\frac{|\rho_{c} \mathbf{u}|}{|\sigma \mathbf{u} \times \mathbf{B}|} \sim \frac{|\epsilon_{0} \boldsymbol{\nabla}\cdot \left(\mathbf{u}\times\mathbf{B}\right) \mathbf{u}|}{|\sigma \mathbf{u} \times \mathbf{B}|} \sim \frac{\epsilon_{0}/\sigma}{L/U}.
\end{equation}
The term $\rho_{c} \mathbf{u}$ can therefore be neglected if the macroscopic timescale $L/U$ is large compared to the charge response time $\epsilon_{0}/\sigma$.
Since, again, $\epsilon_{0}/\sigma \sim 10^{-18}-10^{-16}$ s, $\rho_{c} \mathbf{u}$ can be neglected in any situation typically encountered in MHD problems, including planetary core dynamics. 
Ohm's law can therefore be written as
\begin{equation}
\mathbf{j} =  \sigma \left( \mathbf{E} + \mathbf{u} \times \mathbf{B} \right).
\label{Eq:OhmLaw}
\end{equation}

\subsection{The non-relativistic limit}

When applied to the geodynamo problem (as well as most MHD problems), Amp\`ere's law (equation \eqref{Eq:Ampere}) can be simplified as follows in the non-relativistic limit.
Consider a problem with $\mathbf{B}$ and $\mathbf{E}$ having magnitude $B$ and $E$ varying on typical lengthscale $L$ and timescale $T$.
The ratio of the displacement current $\mu_{0} \epsilon_{0} \partial_{t}  \mathbf{E}$ to the curl of $\mathbf{B}$ is on the order of
\begin{equation}
\frac{\left| \dfrac{1}{c^{2}} \dfrac{\partial  \mathbf{E}}{\partial t} \right|}{\left| \boldsymbol{\nabla} \times \mathbf{B} \right|} \sim \frac{L}{c^{2} T} \frac{E}{B}.
\end{equation}
Using Faraday's law of induction (equation \eqref{Eq:FaradayLaw}), which implies that $E/L\sim B/T$, we obtain
\begin{equation}
\frac{\left| \dfrac{1}{c^{2}} \dfrac{\partial  \mathbf{E}}{\partial t} \right|}{\left| \boldsymbol{\nabla} \times \mathbf{B} \right|} \sim \left(\frac{L/T}{c} \right)^{2}.
\end{equation}
The ratio $L/T$ is an estimate of typical velocities, which in the non-relativistic limit is $\ll c$. This implies that
\begin{equation}
\left|\frac{1}{c^{2}}  \frac{\partial  \mathbf{E}}{\partial t} \right| \ll \left| \boldsymbol{\nabla} \times \mathbf{B} \right|.
\end{equation}
This will always be assumed here, and we will use the non-relativistic Maxwell's equations:
\begin{align}
\boldsymbol{\nabla} \cdot \mathbf{E} &= \frac{\rho_{c}}{\epsilon_{0}}	&	\text{\quad Gauss law}, \\
\boldsymbol{\nabla} \times  \mathbf{E} &= - \frac{\partial  \mathbf{B}}{\partial t} & \text{\quad Faraday's law of induction} ,\\
\boldsymbol{\nabla} \cdot \mathbf{B} &= 0		& \text{\quad No magnetic monopole}, \\
\boldsymbol{\nabla} \times \mathbf{B} &= \mu_{0}\,  \mathbf{j}  & \text{\quad Ampere's law}. \label{Eq:AmpereLaw}
\end{align}

\section{The induction equation} 
\label{Section:InductionEquation}

A prognostic equation for $\mathbf{B}$ can be obtained from Ohm's law and Maxwell's equations. 
Taking the curl of Ohm's law and using Faraday and Amp\`ere's laws, one can obtain the so-called \textit{induction equation}, which writes
\begin{equation}
\frac{\partial \mathbf{B}}{\partial t} = \boldsymbol{\nabla} \times (\mathbf{u} \times \mathbf{B}) + \eta \nabla^{2} \mathbf{B}, \label{Eq:Induction1}
\end{equation}
where $\eta$, the \textit{magnetic diffusivity}, is defined as
\begin{equation}
\eta = \frac{1}{\mu_{0}\sigma}.
\end{equation}
An alternative -- and sometimes more useful -- form of the induction equation can be obtained by noting that $\boldsymbol{\nabla} \times (\mathbf{u} \times \mathbf{B}) = \left(\mathbf{B} \cdot \boldsymbol{\nabla} \right) \mathbf{u} - \left(\mathbf{u} \cdot \boldsymbol{\nabla} \right) \mathbf{B}$, which allows to transform equation \eqref{Eq:Induction1} into
\begin{equation}
 \frac{\partial \mathbf{B}}{\partial t} +\underbrace{ \left(\mathbf{u}\cdot\boldsymbol{\nabla}\right) \mathbf{B}}_{\text{Advection}} = \underbrace{\left(\mathbf{B}\cdot\boldsymbol{\nabla}\right)\mathbf{u}}_{\text{Stretching}} + \eta \nabla^{2} \mathbf{B},
\end{equation}
or
\begin{equation}
\frac{D \mathbf{B}}{D t} =  \left(\mathbf{B}\cdot\boldsymbol{\nabla}\right)\mathbf{u}+ \eta \nabla^{2} \mathbf{B}.
\end{equation}
This shows that the Lagrangian derivative of $\mathbf{B}$ depends on a competition between two terms: a \textit{stretching} term $\left(\mathbf{B} \cdot \boldsymbol{\nabla} \right) \mathbf{u}$, and a diffusion term $ \eta \nabla^{2} \mathbf{B}$.
Diffusion will always tend to smooth out spatial variations of $\mathbf{B}$.
In contrast, the stretching term can increase the magnitude of $\mathbf{B}$. 

To estimate the relative importance of these two terms, we denote by $B$ and $U$ the magnitudes of the magnetic and velocity fields, which are both assumed to vary over the same  length scale $L$.
Forming the ratio of the stretching and diffusion terms, we obtain
\begin{equation}
\frac{|\boldsymbol{\nabla} \times (\mathbf{u} \times \mathbf{B})|}{|\eta \nabla^{2} \mathbf{B}|} \sim \frac{U B / L}{\eta B/L^{2}} = \frac{U L}{\eta}=R_{m},
\label{Eq:Rm}
\end{equation} 
which defines the magnetic Reynolds number $R_{m}$.
$R_{m}$ is a measure of the relative importance of stretching and advection of the magnetic field to its diffusion  (it could also have been called a ``magnetic P\'eclet number'').

To understand the induction equation, we will first consider  the limits of small and large $R_{m}$, before considering a more general case.  

\subsection{The $R_{m}\ll 1$ limit}

If $R_{m}\ll 1$, the advection and stretching terms are both negligible compared to the diffusion term, and the induction equation becomes a simple diffusion equation:
\begin{equation}
\frac{\partial \mathbf{B}}{\partial t} = \eta \nabla^{2} \mathbf{B}.
\end{equation}
In this limit, a magnetic field varying over a length scale $L$ would smooth out by diffusion in a timescale %$\sim \tau_{\eta} = L^{2}/\eta$.
\begin{equation}
{\tau_{\eta} = \frac{L^{2}}{\eta}} = \mu_{0} \sigma L^{2}.
\end{equation}

In the Earth's core ($L\sim 3000$ km, $\eta\sim 1$ m$^{2}$.s$^{-1}$), the diffusion timescale is $\tau_{\eta} \sim 300\,000$ years.
Properly taking into account the spherical geometry of the core yields a smaller timescale, $\tau_{\eta}=L^{2}/(\pi^{2}\eta) \sim 30\,000$ years,  which implies that a spatially varying magnetic field can perdure only during a few tenths of thousand years. Since this is small compared to the age of the Earth (4.56 Gy), this implies that the geomagnetic field  cannot be of primordial origin. 

In contrast, the magnetic field would diffuse very rapidly in any reasonable size laboratory experiment: in a one meter size experiment using liquid sodium (which has a very low magnetic diffusivity, $\eta \sim 0.1$ m$^{2}$), the diffusion time is $\sim 10$ s.

\subsection{The $R_{m}\gg 1$ limit}

At $R_{m}\gg 1$, the induction equation becomes:
\begin{equation}
\frac{D \mathbf{B}}{D t} =  \frac{\partial \mathbf{B}}{\partial t} +\underbrace{ \left(\mathbf{u}\cdot\boldsymbol{\nabla}\right) \mathbf{B}}_{\text{Advection}} = \underbrace{\left(\mathbf{B}\cdot\boldsymbol{\nabla}\right)\mathbf{u}}_{\text{Stretching}}.
\label{Eq:InductionInfiniteRm}
\end{equation}
Let us consider the evolution of an initially uniform magnetic field $\mathbf{B} = B_{0} \mathbf{e}_{z}$ in response to a flow with velocity $\mathbf{u}=(u_{x},u_{z})$.
At time $t=0$, at which $\mathbf{B} = B_{0} \mathbf{e}_{z}$, 
the term $\left(B\cdot\boldsymbol{\nabla}\right)\mathbf{u}$  writes
\begin{equation}
B_{0}
\left[
\begin{array}{c}
\partial_{z} u_{x} \\
\partial_{z} u_{z} 
\end{array}
\right]
\end{equation}
From this one can see that:
\begin{enumerate}
\item
shearing the magnetic field ($\mathbf{u} = u_{x}(z) \mathbf{e}_{x}$) produces a magnetic field component perpendicular to the initial direction of the magnetic field.
Solving the induction equation with a simple shear flow ($u_{x}=\dot\gamma z$, where $\dot\gamma$ is the shear rate) gives
\begin{align}
B_{x} &= \dot\gamma B_{0} t , \\
B_{z} &= B_{0}.
\end{align}
The magnetic lines are tilted by the shear, and tends to be aligned with the shear direction. 
Note also that $|\mathbf{B}|$ increases as  $|\mathbf{B}|=B_{0}\sqrt{1+(\dot\gamma t)^{2}}$. 
There is therefore a net production -- and not just a redistribution -- of magnetic energy.
\item
stretching (resp. compressing) the magnetic field increases (resp. decreases) its magnitude.
Solving the induction equation with $\mathbf{u} = u_{z}(z)\mathbf{e}_{z}$ gives
\begin{align}
B_{x} &= 0, \\
B_{z} &= B_{0} \exp\left( \int_{0}^{t} \frac{\partial u_{z}}{\partial z} dt \right).
\end{align}
A constant $\partial u_{z}/\partial z$ results in exponential growth (if it is positive) or decrease (if negative) of $B_{z}$.
\end{enumerate}

In the limit of infinite R$_{m}$, one can obtain two useful theorems:
\begin{description}
\item[Helmholtz theorem:] \textit{magnetic lines are material lines.}\footnote{
%\item[Proof:]  
\textbf{Proof:} consider a small vector $\boldsymbol{\delta} $ having material end-points, advected by the flow. 
At time $t+dt$ this small vector will be equal to
\begin{align}
\boldsymbol{\delta} (t+dt) &= - \mathbf{u}(\mathbf{x},t)dt + \boldsymbol{\delta} (t) + \mathbf{u}(\mathbf{x}+\boldsymbol{\delta} ,t)dt, \\
	&= \boldsymbol{\delta} (t) + \boldsymbol{\delta} \cdot \boldsymbol{\nabla} \mathbf{u}\,dt + \mathcal{O}(\boldsymbol{\delta} ^{2}).
\end{align}
Taking the $dt \rightarrow 0$, $\boldsymbol{\delta} \rightarrow 0$ limit gives
\begin{equation}
\frac{D \boldsymbol{\delta} }{Dt} = \left(\boldsymbol{\delta} \cdot \boldsymbol{\nabla} \right) \mathbf{u}.  %\quad \leftarrow \text{same eq. as $\mathbf{B}$}
\end{equation}
$\boldsymbol{\delta} $ therefore evolves according to the same equation as $\mathbf{B}$ (equation \eqref{Eq:InductionInfiniteRm}).
}
\item[Alfven's frozen flux theorem:] \textit{a magnetic tube is material and its magnetic flux is conserved: denoting by $S$ any cross-section of a magnetic tube, then the magnetic flux $\int_{S} \mathbf{B}\cdot d\mathbf{S}$ (which is constant along the tube because $\boldsymbol{\nabla}\cdot \mathbf{B}=0$) does not vary with time}.\footnote{\textbf{Proof:} 
The fact that a magnetic tube is material follows directly from Helmholtz theorem.
To show that its magnetic flux does not vary with time, write its time derivative as
\begin{equation}
\frac{d}{dt} \int_{S} \mathbf{B}\cdot d\mathbf{S} = \int_{S} \frac{\partial \mathbf{B}}{\partial t}\cdot d\mathbf{S} + \oint_{\mathcal{C}} \mathbf{B} \cdot \mathbf{u}\times  d\mathbf{l},
\label{Eq:TimeDerivativeMagneticFlux}
\end{equation}
use the diffusion-free induction equation to write the first term on the RHS as
\begin{equation}
\int_{S} \frac{\partial \mathbf{B}}{\partial t}\cdot d\mathbf{S} = \int_{S} \boldsymbol{\nabla} \times (\mathbf{u}\times \mathbf{B})\cdot d\mathbf{S},
\end{equation}
and the identity $(\mathbf{A}\times\mathbf{B})\cdot\mathbf{C}=-\mathbf{B}\cdot(\mathbf{A}\times \mathbf{C})$ plus Stokes' theorem to write the second term as
\begin{align}
\oint_{\mathcal{C}} \mathbf{B} \cdot \mathbf{u}\times  d\mathbf{l} &= -  \oint_{\mathcal{C}} (\mathbf{u}  \times \mathbf{B})\cdot  d\mathbf{l}, \\
	&= - \int_{S} \boldsymbol{\nabla} \times (\mathbf{u}\times \mathbf{B})\cdot d\mathbf{S},
\end{align}
which is equal to the opposite of the first term on the RHS of equation \eqref{Eq:TimeDerivativeMagneticFlux}.}
\end{description}

\subsection{An example of a kinematic solution of the full induction equation}

Let us now consider one simple example of solution of the full induction equation.
We consider a velocity field of the form $\mathbf{u} = U \sin(2\pi z/L) \mathbf{e}_{x}$ (\textit{i.e.} a pure shear flow with a shear rate varying periodically with $z$).
With a velocity field of this form, the induction equation writes
\begin{align}
\frac{\partial B_{x}}{\partial t} &=  2 \pi \frac{U}{L}  \cos(2\pi z/L) B_{z}  + \eta\frac{\partial^{2} B_{x}}{\partial z^{2}}, \\
\frac{\partial B_{z}}{\partial t} &= \eta\frac{\partial^{2} B_{z}}{\partial z^{2}}.
\end{align}
The magnetic field is assumed to be initially uniform, $\mathbf{B}(t=0)=B_{0}\mathbf{e}_{z}$.
Solving these equations (by looking for a solution of the form $B_{x}=f(t) \cos\left(2\pi z/L\right)$) gives
\begin{align}
B_{x} &= B_{0} R_{m} \left( 1 - \mathrm{e}^{-t/\tau}\right) \cos\left(2\pi z/L\right), \\
B_{z} &= B_{0},
\end{align}
where
\begin{equation}
R_{m} = \frac{U L}{\eta}, \quad 
\tau = \frac{1}{4\pi^{2}} \frac{L^{2}}{\eta}.
\end{equation}
The magnitude of $B_{x}$ increases linearly with $t$ when $t\ll \tau$, and saturates at a value $R_{m} B_{0}$  when vertical diffusion balances magnetic field production, which happens at $t\sim \tau$. 
Compared to the infinite $R_{m}$ limit, there is still a tendency to align $\mathbf{B}$ with the shear, but this is now mitigated by diffusion.

\section{The Lorentz force}
\label{Section:LorentzForce}

Experiments show that a charged particle ($q$) moving with velocity $\mathbf{u}$ in an electric field $\mathbf{E}$ and magnetic field $\mathbf{B}$ experiences a force 
\begin{equation}
\mathbf{F}_{L} = q \left( \mathbf{E} + \mathbf{u} \times \mathbf{B} \right)
\end{equation}
called the Lorentz force.
This can be generalised to the case of a charged continuous medium: electric and magnetic fields produce a volumetric force given by
\begin{equation}
\mathbf{f}_{L} = \rho_{c} \mathbf{E} + \mathbf{j} \times \mathbf{B}.
\end{equation}

\subsection{The non-relativistic limit}

In the non-relativistic limit, one can first use Faraday's law (without the current displacement) to write
\begin{equation}
\mathbf{f}_{L} = \rho_{c} \mathbf{E} +\frac{1}{\mu_{0}} (\boldsymbol{\nabla} \times \mathbf{B} ) \times \mathbf{B}.
\label{Eq:LorentzForce0}
\end{equation}
Using the estimate of $\rho_{c}$ for electrical conductors (equation \eqref{Eq:ChargeDensityQuasiStatic}), the relative importance of the two terms of the Lorentz force can be estimated as
\begin{equation}
\frac{|\rho_{c} \mathbf{E} |}{\left| \dfrac{1}{\mu_{0}} (\boldsymbol{\nabla} \times \mathbf{B} ) \times \mathbf{B} \right|} 
\sim \frac{|\epsilon_{0} \boldsymbol{\nabla}\cdot \left(\mathbf{u}\times\mathbf{B}\right)  \mathbf{E} |}{\left| \dfrac{1}{\mu_{0}} (\boldsymbol{\nabla} \times \mathbf{B} ) \times \mathbf{B} \right|} 
\sim \frac{U E}{B/(\mu_{0}\epsilon_{0})}
\sim \left( \frac{U}{c} \right)^{2},
\end{equation}
where we have used the scaling relation $E/B \sim L/T \sim U$ obtained from Faraday's law.
In a non-relativistic problem, we can therefore safely neglect the $\rho_{c} \mathbf{E}$ term and write the Lorentz force as
\begin{equation}
\mathbf{f}_{L} = \frac{1}{\mu_{0}} (\boldsymbol{\nabla} \times \mathbf{B} ) \times \mathbf{B}.
\label{Eq:LorentzForce}
\end{equation}

\subsection{Magnetic pressure and tension}

A bit of algebra on $(\boldsymbol{\nabla} \times \mathbf{B} ) \times \mathbf{B}$ allows to decompose the Lorentz force into two terms:
\begin{equation}
\mathbf{f}_{L} = -  \boldsymbol{\nabla} \underbrace{ \left( \frac{B^{2}}{2 \mu_{0}} \right)}_{\substack{\text{magnetic}\\ \text{pressure}}}+ \underbrace{ \frac{1}{\mu_{0}} \left( \mathbf{B}\cdot \boldsymbol{\nabla} \right)\mathbf{B} }_{\substack{\text{magnetic}\\ \text{tension}}}.
\end{equation}
The first term acts as a pressure term : a force equal to the gradient of the \textit{magnetic pressure} $B^{2}/(2 \mu_{0})$.
The second term is called \textit{magnetic tension}, for reasons which  will be explained below.
This decomposition is often useful because in many cases the gradient of the magnetic pressure can be balanced to a large extent by pressure gradients, so that only the magnetic tension has a strong dynamic effect (see section \ref{Section:ICLorentz} for example). 
Only the magnetic tension can produce vorticity.

\begin{figure}[t]
\centering
\includegraphics[width=0.45\linewidth]{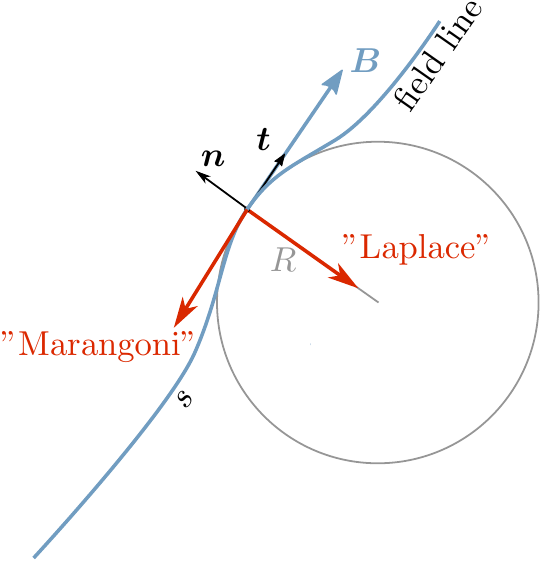}
\caption{The magnetic tension induced by bending a field line can be decomposed as the sum of a ``Laplace'' term oriented perpendicularly to the field line, with a magnitude proportional to the field line curvature, and  of a ``Marangoni'' term, which acts parallel to the field line.}
\label{Fig:MagneticTension}
\end{figure}

We restrict here our analysis to 2D for the sake of simplicity, but the following reasoning can be generalised to 3D.
Let us consider a curvilinear Frenet coordinate system $(\mathbf{t},\mathbf{n})$ attached 
to a given magnetic field line (figure \ref{Fig:MagneticTension}).
We denote by $s$ the coordinate along this field line.
In this coordinate system, we can write
\begin{align}
\frac{1}{\mu_{0}} \left( \mathbf{B}\cdot \boldsymbol{\nabla} \right)\mathbf{B} &= \frac{B}{\mu_{0}} \frac{\partial}{\partial s}\left( B \mathbf{t} \right),   \\
&=\frac{B^{2}}{\mu_{0}}  \frac{\partial \mathbf{t}}{\partial s} + \frac{\partial}{\partial s} \left( \frac{B^{2}}{2 \mu_{0}} \right) \mathbf{t} , \\ 
&= -  2 \frac{B^{2}}{2 \mu_{0}} \,  \mathcal{K}  \mathbf{n} + \boldsymbol{\nabla}_{\mathbf{t}} \left( \frac{B^{2}}{2 \mu_{0}} \right),   \label{Eq:MagneticTension} 
\end{align}
where  $\mathcal{K} =\boldsymbol{\nabla} \cdot \mathbf{n}$  	
is the (signed) curvature of the field line,  the derivative of $\mathbf{t}$  according to $s$ being equal to $-\mathcal{K}  \mathbf{n}$ . 
The operator $\boldsymbol{\nabla}_{\mathbf{t}}(...)=\boldsymbol{\nabla}(...) - \mathbf{n} (\mathbf{n} \cdot \boldsymbol{\nabla})(...)$ is the component of the gradient parallel to the field line.

This equation bears a strong resemblance with the expression of the stress jump induced by interfacial tension across the interface between two immiscible fluids, which is equal to
\begin{equation}
 -  \gamma  \mathcal{K}  \mathbf{n} + \boldsymbol{\nabla}_{\mathbf{t}} \gamma,
\end{equation}
where $\gamma$ is the interfacial tension.
The first term is the \textit{Laplace pressure} term, which expresses the fact that interfacial tension induces a pressure jump across a curved interface. This pressure jump is proportional to the interfacial tension  and to the curvature of the interface. 
The second term is a stress tangential to the interface associated with gradients of interfacial tension, which is responsible of \textit{Marangoni} effects.

Equation \eqref{Eq:MagneticTension} has a similar mathematical form with $B^{2}/(2 \mu_{0})$ taking the role of interfacial tension.
The analog of the \textit{Laplace pressure} implies that deforming a magnetic line induces a volumetric force proportional to $B^{2}$ normal to the line, and directed toward the center of curvature of the field lines.
This force thus acts against deformation of the magnetic lines and tends to straighten them.
The analog of the \textit{Marangoni} tension is a force acting parallel to the magnetic lines toward regions of higher magnetic field intensity.
It produces a tension parallel to the fields lines if the magnetic field intensity varies along field lines.

Putting back the magnetic pression and magnetic tension terms together, the Lorentz force writes
\begin{align}
\mathbf{j} \times \mathbf{B} &= -  \boldsymbol{\nabla} \left( \frac{B^{2}}{2 \mu_{0}} \right) +  \frac{1}{\mu_{0}} \left( \mathbf{B}\cdot \boldsymbol{\nabla} \right)\mathbf{B} , \\
		&= -  \boldsymbol{\nabla} \left( \frac{B^{2}}{2 \mu_{0}} \right) + \boldsymbol{\nabla}_{\mathbf{t}} \left( \frac{B^{2}}{2 \mu_{0}} \right) - \frac{B^{2}}{\mu_{0}} \ \mathcal{K}  \mathbf{n} ,  \\
		&= -  \boldsymbol{\nabla}_{\mathbf{n}} \left( \frac{B^{2}}{2 \mu_{0}} \right) - \frac{B^{2}}{\mu_{0}}  \mathcal{K}  \mathbf{n} ,
\end{align}
where $\boldsymbol{\nabla}_{\mathbf{n}}$ is the component of the gradient normal to the field line.
Both terms acts perpendicularly to the field lines (consistently with the $\frac{1}{\mu_{0}}(\boldsymbol{\nabla} \times \mathbf{B} ) \times \mathbf{B}$ expression of the Lorentz force).

\subsection{Alfv\`en waves}

\begin{figure}[t]
\centering
\includegraphics[width=0.6\linewidth]{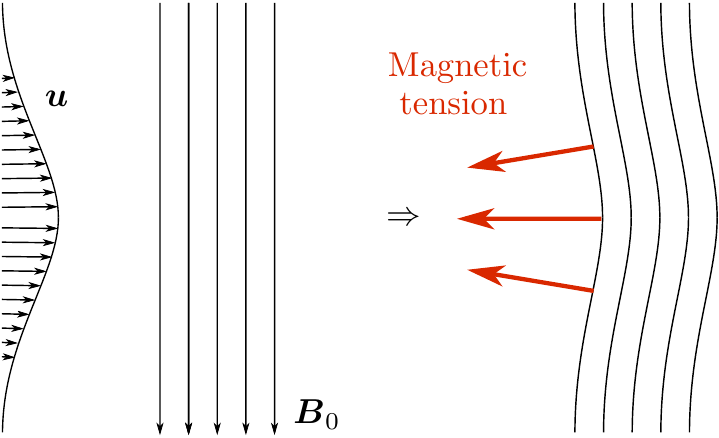}
\caption{The bending of magnetic field lines produces a Lorentz force acting against the flow that deformed the field lines.
}
\label{Fig:AlfvenWaves}
\end{figure}

The relationship between the Lorentz force and the magnetic field lines curvature suggests that a magnetised fluid can carry waves: if magnetic lines are deformed from an initial equilibrium state by a flow with velocity $\mathbf{u}$, deformation of the lines will produce a restoring  force (the Lorentz force) normal to the line and of magnitude proportional to the curvature of the lines (figure \ref{Fig:AlfvenWaves}). 
Only a velocity field perpendicular to the magnetic lines would curve the lines and thus produce a restoring force.
Taking into account that \textit{a velocity perturbation propagating as a plane wave in an \textit{incompressible} fluid must be a transverse wave}\footnote{
%Let us start by noting that a velocity perturbation propagating as a plane wave in an \textit{incompressible} fluid must be a transverse wave.
To show this, consider a plane wave propagating in the $z-$direction, with fluid velocity 
\begin{equation}
\mathbf{u}=\mathcal{R}\{ \mathbf{u}_{0}  \exp[ i (k_{z} z - \omega t) ] \},
\end{equation}
where $\mathcal{R}$ indicates the real part.
If the fluid is incompressible, $\boldsymbol{\nabla}\cdot \mathbf{u}=0$, and hence
\begin{equation}
\frac{\partial u_{z}}{\partial z} = - \frac{\partial u_{x}}{\partial x} - \frac{\partial u_{y}}{\partial y},
\end{equation}
which is equal to 0 since the plane wave assumption implies that  the velocity field is a function of $z$ and $t$ only.
This implies that the component $u_{z}$ of the velocity field must be spatially uniform. In other words, only the component of the velocity field perpendicular to the propagation direction can oscillate spatially: the wave must be transverse. 
}, 
we can thus expect velocity perturbations perpendicular to the field lines to propagate in the form of a transverse wave travelling in the direction of magnetic lines. 
These are called \textit{Alfv\`en waves}.
The celerity of these waves can be obtained from dimensional analysis: the propagation velocity $v_{A}$ can be a function of the magnetic field intensity $B_{0}$ [T=kg s$^{-2}$ A$^{-1}$], of the magnetic permeability $\mu_{0}$ [m kg s$^{-2}$ A$^{-2}$], of the density $\rho$ [kg m$^{-3}$] of the fluid, and of the wave number $k$ [m$^{-1}$] of the wave, which are the only parameters of the problem.
There is only one way to build a group of parameters with the dimension of velocity  from this list of parameters (this is proved with the Vaschy-Buckingham theorem), which is
\begin{equation}
v_{A} \sim \frac{B_{0}}{\sqrt{\rho \mu_{0}}}.
\end{equation}
An important point is that it is not possible to build a velocity involving the wave number $k$, which implies that Alfv\`en waves must be non-dispersive. 

A classical way of deriving the dispersion equation of these waves is to work directly from the Navier-Stokes (including the Lorentz force) and induction equations.\footnote{Consider small perturbations of the velocity and magnetic fields, linearise the Navier-Stokes and induction equations, take the curl of these two equations, and combine them after taking the time derivative of the curled Navier-Stokes equation (vorticity equation).
%The demonstration can for example be found in \cite{}.
}
As an alternative, we will here obtain the wave equation from an analysis based on magnetic tension. 
We consider a conducting, incompressible fluid of uniform density permeated by a magnetic field,
 and neglect magnetic diffusion in the induction equation (thus taking the infinite R$_{m}$ limit).
Noting that  $\mathbf{u}\cdot \boldsymbol{\nabla} \mathbf{u} =0$ (since we are considering plane transverse waves), 
the Navier-Stokes and induction equations then write
\begin{align}
\rho \frac{\partial \mathbf{u}}{\partial t} &= -\boldsymbol{\nabla} p -  \boldsymbol{\nabla}_{\mathbf{n}} \left( \frac{B^{2}}{2 \mu_{0}} \right) - \frac{B^{2}}{\mu_{0}}  \mathcal{K}  \mathbf{n} , \label{Eq:NSAlven1} \\
\frac{\partial \mathbf{B}}{\partial t} &=   \left(\mathbf{B} \cdot \boldsymbol{\nabla} \right) \mathbf{u} - \left(\mathbf{u} \cdot \boldsymbol{\nabla} \right) \mathbf{B} . \label{Eq:InductionAlven1}		
\end{align}
We consider a uniform background magnetic field $\mathbf{B}_{0}=B_{0}\mathbf{e}_{z}$, which is slightly perturbed by a small velocity field perturbation. By ``slightly pertubed'', we mean here that the curvature $\mathcal{K}$ of the magnetic lines is assumed to remain small. 
The restoring Lorentz force being perpendicular to $\mathbf{B}$, we expect oscillations of the fluid perpendicular to $\mathbf{B}$, in the $x$ direction, and, since only transverse waves can be carried by an incompressible fluid, propagation parallel to the magnetic fields lines, in the $z$ direction.
A transverse wave propagating along a field line will shear the magnetic line and produce by induction a magnetic field in the $x$ direction.
We thus consider small  velocity and magnetic fields perturbations of the form $\mathbf{u}=u_{x}(z,t)\mathbf{e}_{x}$ and $\bold{b}=b_{x}(z,t)\mathbf{e}_{x}$, small curvature $\mathcal{K}(z)$, 
and linearize equations \eqref{Eq:NSAlven1} and \eqref{Eq:InductionAlven1} to obtain
\begin{align}
\rho \frac{\partial u_{x}}{\partial t} &= - \frac{B_{0}^{2}}{\mu_{0}} \mathcal{K} , \\
\frac{\partial b_{x}}{\partial t} &= B_{0} \frac{\partial u_{x}}{\partial z}.
\end{align}
We denote by $\delta_{x}$ the $x$-displacement of a magnetic field line. 
Since in the infinite R$_{m}$ limit magnetic field lines are material lines, the horizontal displacement of the field lines is linked to the velocity field by
\begin{equation}
u_{x} = \frac{\partial \delta_{x}}{\partial t}.
\end{equation}
With $\mathbf{n} \simeq \mathbf{e}_{x} - \frac{\partial \delta_{x}}{\partial z} \mathbf{e}_{z}$, the curvature is 
\begin{equation}
\mathcal{K}  = \boldsymbol{\nabla} \cdot \mathbf{n} \simeq - \frac{\partial^{2} \delta_{x}}{\partial z^{2}}. 
\end{equation}
The $x$-component of Navier-Stokes then writes 
\begin{equation}
\frac{\partial^{2} \delta_{x}}{\partial t^{2}} - \frac{B_{0}^{2}}{\rho \mu_{0}} \frac{\partial^{2} \delta_{x}}{\partial z^{2}} = 0,
\end{equation}
which is a non-dispersive wave equation, with celerity  
\begin{equation}
v_{A} = \frac{B_{0}}{\sqrt{\rho \mu_{0}}}.
\end{equation}
The period $T$ of theses waves is related to their wavelength $\lambda$ by
\begin{equation}
T = \frac{\sqrt{\rho \mu_{0}} \lambda}{B_{0}}.
\end{equation}

%--------------------------------------------------------------------------------------------------------------------------------------------%
%--------------------------------------------------------------------------------------------------------------------------------------------%
\chapter{The geometry of Earth's magnetic field}
\label{Section:GeomagneticFieldGeometry}
%--------------------------------------------------------------------------------------------------------------------------------------------%
%--------------------------------------------------------------------------------------------------------------------------------------------%

\begin{figure}
\includegraphics[width=\linewidth]{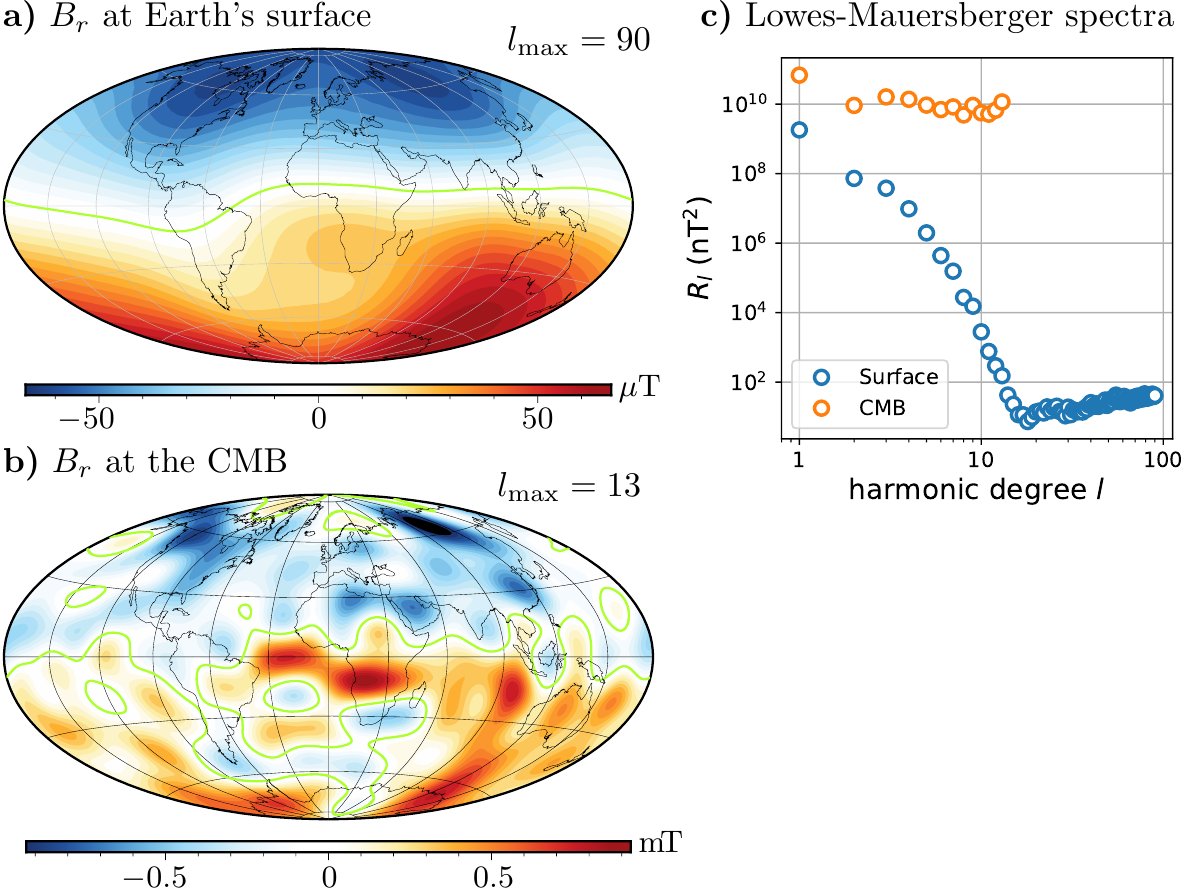}
\caption{\textbf{a)} radial component $B_{r}$ of the geomagnetic field at the surface of the Earth in 2003 up to spherical harmonic degree $l=90$, according to the Postdam Magnetic Model of the Earth (POMME) \citep{maus2006}; \textbf{b)} radial component $B_{r}$ of the geomagnetic field at the core-mantle boundary, obtained from the model POMME; \textbf{c)} Lowes-Mauersberger spectra for the  magnetic field at the surface (blue circles)  and at the core-mantle boundary (orange circles).}
\label{Fig:GeomagneticField}
\end{figure}

\begin{figure}
\centering
\includegraphics[width=0.65\linewidth]{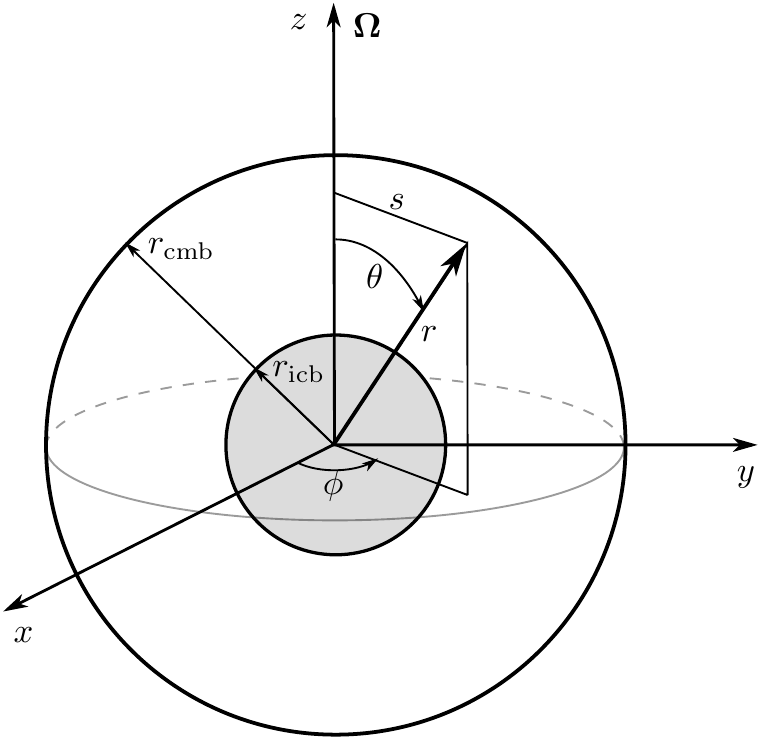}
\caption{Spherical $(r,\theta,\phi)$ and cylindrical $(s,\phi,z)$ coordinate systems.}
\label{Fig:CoordinateSystems}
\end{figure}

\section{Spherical harmonics decomposition} 

Figure \ref{Fig:GeomagneticField}a shows a map of the intensity of the radial component $B_{r}$ of the geomagnetic field at the surface of the Earth.
As already discussed in the introduction, and first noted by Gilbert in 1600, Earth's magnetic field strongly resembles the magnetic field which would be produced by a magnet inside the Earth and aligned with the rotation axis: the field is strongly dipolar, close to axisymmetric, with negative $B_{r}$ in the northern hemisphere and positive $B_{r}$ in the southern hemisphere.
However, closer inspection of figure \ref{Fig:GeomagneticField}a reveals deviations from a NS-oriented dipolar field and smaller scales details. 
For example, the position of the magnetic equator (the line at the surface where $\mathbf{B}$ is horizontal) wanders quite far from the geographic equator. 
Figure \ref{Fig:GeomagneticField}a also shows patches of stronger field intensity under North America, Siberia, and the south Indian ocean.

The fact that the mantle, crust, and lower atmosphere of the Earth can be considered as current-free regions (they are good electric insulators)  greatly simplifies the mathematical description of the geomagnetic field: 
Amp\`ere's law (equation \eqref{Eq:AmpereLaw}) with $\mathbf{j}=0$ writes $\boldsymbol{\nabla} \times \mathbf{B} = 0$, which implies that $\mathbf{B}$ can be written as the gradient of a scalar field, the geomagnetic potential $V$:
\begin{equation}
\mathbf{B} = - \boldsymbol{\nabla} V.
\end{equation}
Since  $\mathbf{B}$ is a divergence free vector field and $\boldsymbol{\nabla} \cdot \boldsymbol{\nabla} (...)=\Delta (...)$, the geomagnetic potential obeys Laplace equation,
\begin{equation}
\Delta V = 0.
\end{equation}
The general solution of this equation can be written as
\begin{equation}
%V(r,\theta,\phi) =  \sum_{l=1}^{\infty} \sum_{m=-l}^{l} \left( A_{lm} r^{l} + \frac{B_{lm}}{r^{l+1}} \right) Y_{lm}(\theta,\phi),
V(r,\theta,\phi) =  \sum_{l=1}^{\infty} \sum_{m=-l}^{l} \left( \frac{A_{lm}}{r^{l+1}} + B_{lm} r^{l}  \right) Y_{lm}(\theta,\phi),
\label{Eq:VSH1}
\end{equation}
where $(r,\theta,\phi)$ are the usual spherical coordinate systems (radius, colatitude, longitude) as defined on figure \ref{Fig:CoordinateSystems}, and $Y_{lm}(\theta,\phi)$ are the spherical harmonics, $l$ and $m$ being the degree and order, which form a complete set of orthogonal functions on the sphere.

\subsection{Demonstrating the internal origin of the geomagnetic field}

Written this way, the geomagnetic potential can be seen as the sum of terms of internal origins,  which decrease with increasing $r$ (the $A_{lm}$ terms), and terms of external origins, which increase with increasing $r$ (the $B_{lm}$ terms).
By determining the $A_{lm}$ and $B_{lm}$ coefficients, it is therefore possible to determine whether the magnetic field observed at the surface of the Earth is predominantly of internal or external origin. 
In his \textit{General Theory of Terrestrial Magnetism} published in 1839, Gauss introduced the geomagnetic potential and its spherical harmonics expansion.
Considering only terms of internal origin, Gauss determined the coefficients of the expansion up to degree $l=4$ using least-square inversion from intensity, inclination, and declination maps available at that time. 
From the excellent agreement between the observations and his spherical harmonics expansion, he concluded that  the geomagnetic field must be of internal origin.

\begin{figure}
\centering
\includegraphics[width=0.5\linewidth]{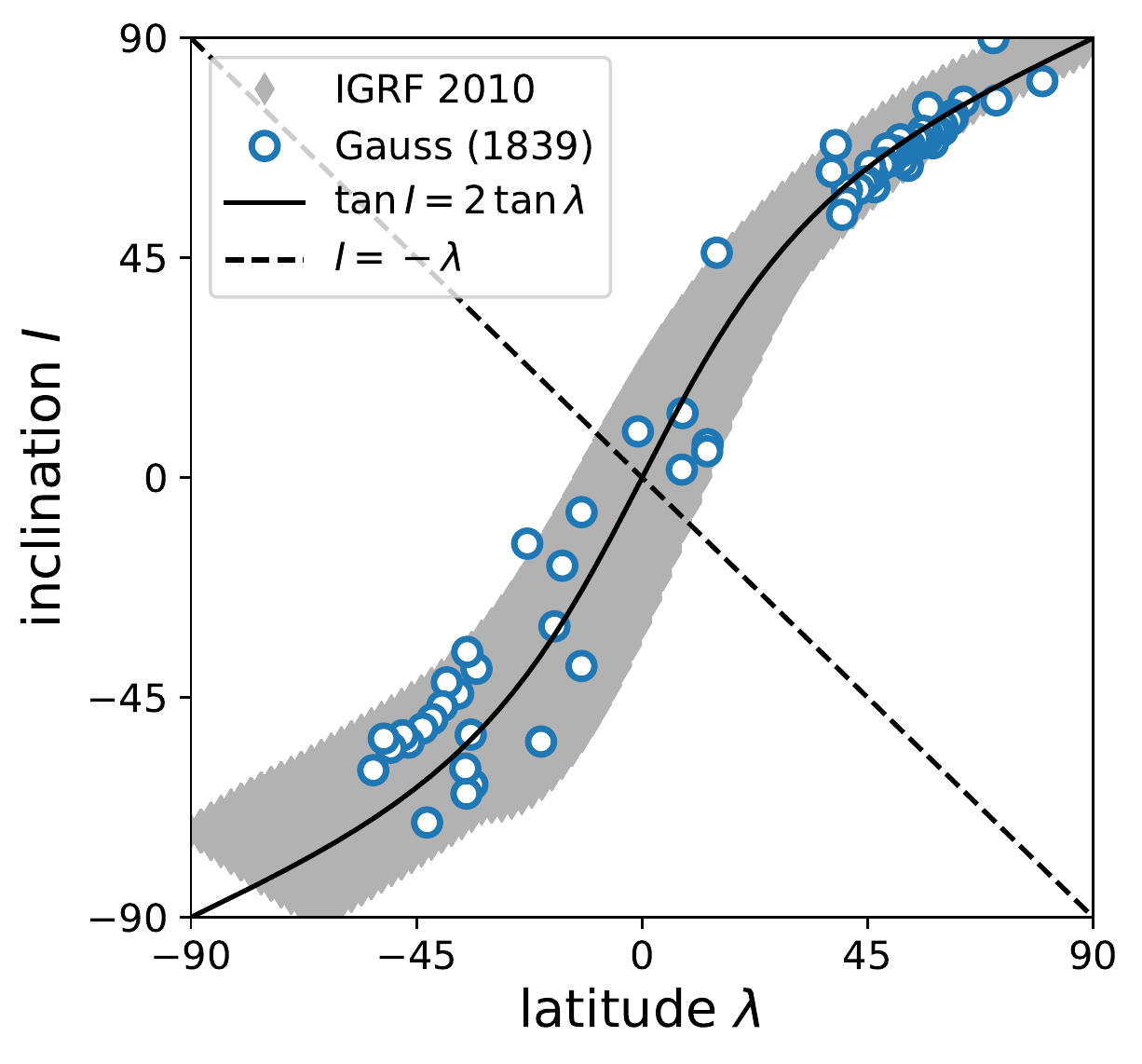}

\caption{Inclination $I$ as a function of latitude $\lambda$ for IGRF 2010 geomagnetic model (gray) and for magnetic field measurements used by Gauss (blue circles).
Also shown are the predictions for a $l=1$, $m=0$ magnetic field of internal (solid black line) and external (dashed black line) origins.
}
\label{Fig:ILat}
\end{figure}

The  internal and external terms only differ by the way they vary with $r$, and it may not be obvious at first sight how it is possible to differentiate between them from measurements all made at the same radius. 
But the actual data are measurements of the magnetic field $\mathbf{B} = - \boldsymbol{\nabla} V$, and not $V$. The latitudinal variations of the direction of $\mathbf{B}$ at a given $r$ happens to be  sensitive to the origin of the magnetic field. 
To get a feeling of how this works, let us consider the $l=1$ terms only, and write the potential as
\begin{equation}
V =  \left[  \frac{A_{10}}{r^{2}} + B_{10} {r}  \right] Y_{10}(\theta) = \left[  \frac{A_{10}}{r^{2}} + B_{10} {r}  \right]  \cos \theta.
\end{equation}
From this we obtain the corresponding magnetic field components:
\begin{align}
B_{r} &=  -\frac{\partial V}{\partial r}=-  \left[ -2 A_{10}/r^{3}  + B_{10}  \right] Y_{10} =\left[ 2 A_{10}/r^{3}  - B_{10}   \right]\cos \theta, \\ 
B_{\theta} & = - \frac{1}{r}\frac{\partial V}{\partial \theta}= -  \left[ A_{10}/r^{3}  + B_{10} \right] \frac{\partial Y_{10}}{\partial \theta}=  \left[ A_{10}/r^{3}  + B_{10} \right] \sin \theta ,  \\
B_{\phi} &= -\frac{1}{r \sin\theta} \frac{\partial V}{\partial \phi}= 0,
\end{align}
and calculate the \textit{inclination} $I$ (the angle between $\mathbf{B}$ and the horizontal plane) as
\begin{equation}
\tan I =\frac{B_{r}}{B_{\theta}} =  \frac{2 A_{10}/r^{3}  - B_{10}  }{A_{10}/r^{3}  + B_{10} }\tan \lambda,
\end{equation}
where $\lambda=\pi/2-\theta$ is the latitude. 
If the geomagnetic field is either of internal origin only ($B_{10}=0$), or of external origin only ($A_{10}=0$), then 
\begin{align}
\tan I &= 2 \tan \lambda \quad &\text{if $\mathbf{B}$ is of internal origin},\\ 
 I &= - \lambda \quad &\text{if $\mathbf{B}$ is of external origin}.
\end{align}
Figure \ref{Fig:ILat} shows the inclination $I$ as a function of $\lambda$ for the IGRF 2010 geomagnetic model (grey diamonds), and for the data used by Gauss (blue diamonds).
%One can see that the relation predicting $I$ for a magnetic field of internal origin better explains the data. 
The data points are quite dispersed around the $\tan I = 2 \tan \lambda$ curve, but this is mostly due to the fact that the geomagnetic dipole is actually tilted from the rotation axis, and also because Earth's magnetic field includes non-negligible higher degree terms.

From now on we will thus keep only the contributions of internal origin.
The spherical harmonics can be written as $Y_{lm}=\mathrm{e}^{i m \phi} P_{lm}(\cos \theta)$, where $P_{lm}$ are the associated Legendre polynomials. 
Keeping only the terms in $r^{-l-1}$ in equation \eqref{Eq:VSH1} and using real coefficients, one can write the magnetic potential as
\begin{equation}
V(r,\theta,\phi) = r_{\Earth} \sum_{l=1}^{\infty} \sum_{m=-l}^{l}  \left(\frac{r_{\Earth}}{r}\right)^{l+1} \left[ g_{l}^{m} \cos m \phi + h_{l}^{m} \sin m \phi \right]  P_{lm}(\cos\theta),
\label{Eq:VGaussCoeffs}
\end{equation}
where $g_{l}^{m}$ and $h_{l}^{m}$ are the \textit{Gauss coefficients}, and $r_{\Earth}$ is the radius of the Earth.

It is useful to look at the spectrum of magnetic energy as a function of the spherical harmonic degree $l$.
Using the fact that spherical harmonics form an orthogonal basis, one can show that the magnetic energy at a radius $r$ corresponding to all components of degree $l$ is given by
\begin{equation}
R_{l}(r) = \left( \frac{r_{\Earth}}{r} \right)^{2l+4} (l+1)\sum_{m=0}^{l} \left( g_{lm}^{2}  + h_{lm}^{2} \right).
\end{equation}
The resulting spectrum is called the \textit{Lowes-Mauersberger} spatial power spectrum.
Figure \ref{Fig:GeomagneticField}c shows this spectrum at Earth's surface in blue circles. 
The energy of the dipolar components ($l=1$) is more than an order of magnitude larger than the $l=2$ component.
$R_{l}$ then decreases with $l$ up to $l$ equal 13 or 14 where it reaches a plateau.
At $l> 13$, the magnetic energy is dominated by contributions of the crustal magnetisation, which obscures the field from deeper origin.
Only the $l\leq 13$ spherical harmonics components can be attributed to the magnetic field originating from the core. 

\subsection{The magnetic field at the core-mantle boundary}

One very interesting property of a curl-free magnetic field is that it is possible to extrapolate to other radii the magnetic potential once we know the coefficients of the spherical harmonics decomposition.
Since the mantle of the Earth can be considered to be current-free, it is in particular possible to extrapolate the surface magnetic field down to the base of the mantle, though this can be done only for harmonic degrees $l\leq 13$ since the higher degrees components of the core field are obscured by the crustal field. 
The potential for $l\leq 13$ at the core-mantle boundary can be obtained by writing  $V$ at $r=r_\mathrm{cmb}$ from equation \eqref{Eq:VGaussCoeffs}.
From this one can obtain the associated magnetic field map and spectrum. 

Figure \ref{Fig:GeomagneticField}b shows the radial component of the field at the  core-mantle boundary (CMB). 
Though the field is still quite dipolar, it exhibits much more smaller scale variations than the surface field (figure \ref{Fig:GeomagneticField}a). 
This is confirmed by inspection of the energy spectrum (figure \ref{Fig:GeomagneticField}c,  orange circles): the spectrum is about flat for $l\geq 2$, with the energy of the dipolar components still about an order of magnitude larger than the higher $l$ components.
The fact the dipolar components still stand out at the CMB is an important result.
Since the spherical components decrease with $r$ as $r^{-l-1}$, any locally produced magnetic field would appear dipolar when seen from far enough.
The fact that the field at the CMB is still dominated by the $l=1$ terms suggests that  the dipolar nature of the geomagnetic field is a robust feature of the geodynamo. %, and not just due 

\subsection{The field within the core: poloidal-toroidal decomposition}

In planetary cores, $\mathbf{B}$ cannot be considered to be curl-free anymore ($\mathbf{j} \neq 0$), and thus cannot be written as the gradient of a potential. 
However, the fact that $\mathbf{B}$ is a divergence-free vector field allows to write it as
\begin{equation}
\mathbf{B} = \boldsymbol{\nabla} \times (T r \mathbf{e}_{r}) + \boldsymbol{\nabla} \times \boldsymbol{\nabla} (P r \mathbf{e}_{r})
\end{equation}
where $T$ and $P$ are the \textit{poloidal} and \textit{toroidal} potentials. 
Note that the toroidal part has no radial component, which means that the magnetic field reconstructed at the CMB only corresponds to the poloidal part.
We have no direct constraints on the toroidal part of the magnetic field in the core. 

\section{Core flow inversion}

The magnetic field at the CMB evolves with time in a measurable way.
These variations can be interpreted with the help of the induction equation, remembering that at the CMB only the radial component of the magnetic field is known.
Taking the dot product of the induction equation with $\mathbf{e}_{r}$ gives
\begin{equation}
\frac{\partial B_{r}}{\partial t} = - \boldsymbol{\nabla}_{h} \cdot (B_{r} \mathbf{u}_{h}) + \eta \frac{1}{r} \nabla^{2} \left( r B_{r} \right),
\label{Eq:InductionEquation_r}
\end{equation}
where $\boldsymbol{\nabla}_{h} \cdot (...)$ is the horizontal part of the divergence operator, and $\mathbf{u}_{h}=(0,u_{\theta},u_{\phi})$ is the horizontal part of the velocity field.  
The diffusion term is usually neglected, which can be justified \textit{a posteriori} on the basis that the magnetic Reynolds number based on the smallest spatial scale considered ($\sim 1000$ km) and estimated velocity is $\sim 500$.
Dropping the diffusion term, equation \eqref{Eq:InductionEquation_r} writes
\begin{equation}
\frac{\partial B_{r}}{\partial t} = - \boldsymbol{\nabla}_{h} \cdot (B_{r} \mathbf{u}_{h}) .
\label{Eq:InductionEquation_r2}
\end{equation}
Knowing $B_{r}$ and its time derivative, one can in principle invert equation \eqref{Eq:InductionEquation_r2} to obtain the horizontal velocity field just below the CMB. 
This happens to be a severely ill-posed inverse problem, the most obvious reason being that we are trying to estimate a two-components vector field ($\mathbf{u}_{h}$) from a scalar field ($B_{r}$).
In other words, we have only one equation for two unknowns.
Inverting equation \eqref{Eq:InductionEquation_r2} thus requires a second equation for $\mathbf{u}_{h}$. 
Various assumptions have been made (\textit{e.g.} steady flow, toroidal flow, tangentially  geostrophic flow, columnar flow, quasi-geostrophic flow, ...), and the choice does impact the resulting flow pattern (see \cite{holme2015} for a review).
Robust features of the inverted flows include a strong westward flow under the Atlantic, a much weaker flow under the Pacific, and some degree of symmetry about the equatorial plane.
The RMS flow velocity is around  $12$ to $14$ km per year, or about $4\times 10^{-4}$ m.s$^{-1}$.

%--------------------------------------------------------------------------------------------------------------------------------------------%
%--------------------------------------------------------------------------------------------------------------------------------------------%
\chapter{Basics of planetary core dynamics}
\label{Section:CoreDynamics}
%--------------------------------------------------------------------------------------------------------------------------------------------%
%--------------------------------------------------------------------------------------------------------------------------------------------%

\section{The geodynamo hypothesis}
\label{Section:GeodynamoHypothesis}

As discussed in section \ref{Section:Introduction}, Gilbert's claim that \textit{Earth is a magnet} has been dismissed by the observation of fast temporal changes of the geomagnetic field.
In addition, we now know that permanent magnets (\textit{i.e.} ferromagnetic or ferrimagnetic materials) lose their permanent magnetic properties (by becoming paramagnetic) above a critical temperature called the \textit{Curie temperature}.
Magnetite and iron have Curie temperatures of 858 K and 1043 K respectively.
The temperature in the Earth exceeds these temperatures at depths larger than around $100-150$ km \citep{jaupart2010}, which confines ferromagnetism to rather shallow depths.
Magnetisation of crustal material can in fact be quite strong, but cannot explain the large scale part of Earth' s magnetic field. 

Leaving aside \textit{ad-hoc} theories, we are thus left with the MHD equations  introduced in section \ref{Section:MHD} to understand the origin of Earth's magnetic field.  
We have already shown (section \ref{Section:InductionEquation}) that the geomagnetic field cannot be of primordial origin, and must be sustained in some way against ohmic dissipation: solving the induction equation with no velocity field indeed shows that spatial variations of the magnetic field would be smoothed out by diffusion on a timescale of $\sim 30\,000$ years, while we know from paleomagnetism that the geomagnetic field has been sustained for (at least) the last $\sim 3.5$ Gy. 

\subsection{Self-exciting dynamos} 

The current theory for the origin of the Sun and Earth's magnetic fields has been proposed in 1919 by Sir Joseph Larmor in a meeting of the \textit{British Association for the Advancement of Science}. 
In the short report of his presentation \citep{Larmor1919}, he wrote:
\begin{quotation}
``In the case of the Sun, surface phenomena point to the existence of a residual internal circulation mainly in meridian planes. 
Such internal motion induces an electric field acting on the moving matter; and if any conducting path around the solar axis happens to be open, an electric current will flow round it, which may in turn increase the inducing magnetic field.
In this way it is possible for the internal cyclic motion to act after the manner of the cycle of a self-exciting dynamo, and maintain a significant magnetic field from insignificant beginnings, at the expense of some of the energy of the internal circulation.

[...]

The very extraordinary feature of the Earth magnetic field is its great and rapid changes, comparable with its whole amount.
[...] [\textit{a self-exciting dynamo}] would account for magnetic change, sudden or gradual, on the Earth merely by change of internal conducting channels: though, on the other hand, it would require fluidity and residual circulation in deep-seated regions.''
\end{quotation}
Larmor's proposition was inspired by the \textit{self-exciting dynamos} developed in the second half of 19$^\mathrm{th}$ century, which were the first efficient electric generators.
The underlying mechanism 
can be understood as a positive feedback loop involving Faraday's law of induction and Amp\`ere's law.
Consider an electrically conducting material moving into an arbitrarily small seed magnetic field.
Faraday's law of induction tells that time variations of a magnetic field in an electrically conducting material produces electric currents.
Equivalently, the motion of the conducting material into the seed magnetic field produces electric currents. 
According to Amp\`ere's law, these electric currents would themselves produce a magnetic field.
If the orientation of the electric current is such that the induced magnetic field reinforces the initial seed magnetic field, then the intensity of the magnetic field %and the magnetic energy 
can grow. 

\begin{figure}
\centering
\includegraphics[width=\linewidth]{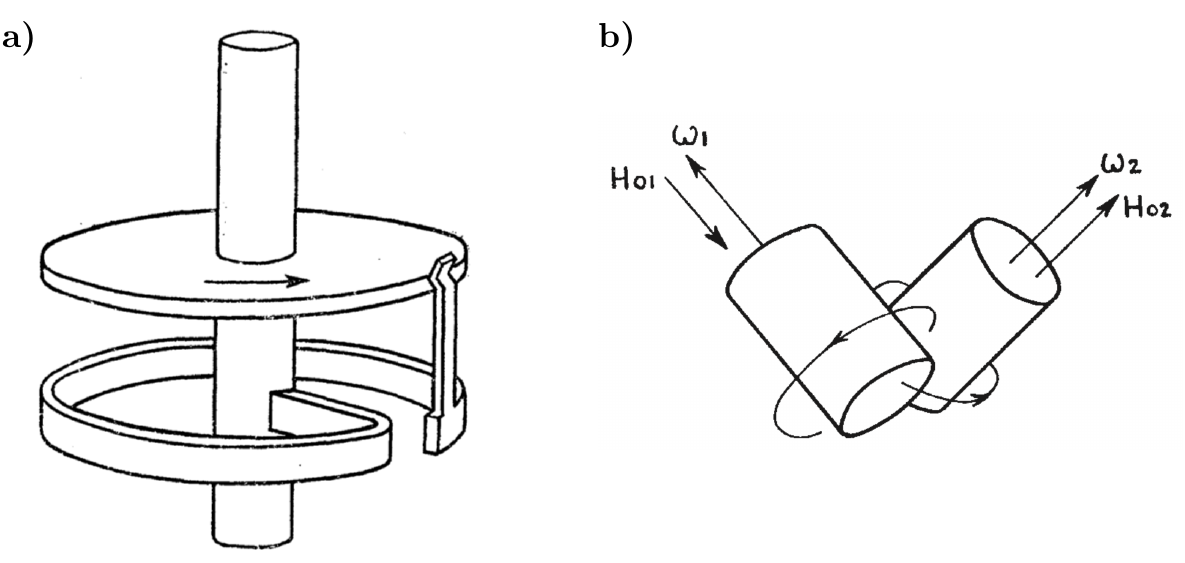}
\caption{\textbf{a)} Bullard's disk dynamo \citep{Bullard1954}. It consists in a Faraday disk whose center and periphery are connected by a conducting wire forming a loop around the axle of the disk.  
In the presence of a 'seed' magnetic field oriented parallel to the axis, rotating the copper disk produces an electric current, which intensity depends on the rate of rotation of the disk and of the electric resistance $R$ of the wire. 
The arrangement of the electric wire loop is such that the induced magnetic field reinforces the seed magnetic field. 
\textbf{b)}
A schematic of Lowes \& Wilkinson's experimental dynamo \citep{Lowes1963}, based on \cite{Herzenberg1958}'s kinematic dynamos.
}
\label{Fig:BullardLowesDynamos}
\end{figure}

The word dynamo was coined by Faraday to name the electric generator he invented, now called the \textit{Faraday disk}. 
Faraday's disk consists in a copper disk rotating within a magnetic field produced by a horseshoe magnet. 
The rotating motion of the copper disk produces a difference of electric potential between its centre and periphery by virtue of Faraday's law of induction, and hence an electrical current if the center and periphery of the disk are linked trough an electric circuit. 
Self-exciting dynamos are based on a similar principle 
except that the permanent magnet is replaced by electromagnets fed by the induced electric current. 
This creates a positive feedback loop which increases the intensity of the magnetic field, thus increasing the currents produced by induction. 
The concept has been  formulated by Anyos Jedlik  around 1856; practical designs of working self-exciting dynamos have been patented by Varley in 1866, and presented by Wheatstone and Siemens in 1867.
These self-exciting dynamos were able to produce a much higher power output than permanent magnet dynamos, which opened the way to the industrial use of electricity. 
A simple conceptual model of a self-exciting dynamo was proposed by \cite{Bullard1954} as a toy model of the geodynamo (figure \ref{Fig:BullardLowesDynamos}).

Compared to self-exciting dynamos such as Bullard's, the concept of \textit{geodynamo}  faces several additional difficulties. 
To quote \cite{Bullard1954}: ``A central problem [...] is to determine whether there exist motions of a simply connected, symmetrical fluid body which is homogeneous and isotropic that will cause it to act as a self-exciting dynamo [...]. We call such dynamos 'homogeneous' to distinguish them from the dynamos of the electrical engineer, which are multiply connected and of low symmetry.''
In industrial dynamos, the electric currents produced by induction are fed into electric wires which are arranged in such a way that
the induced magnetic field indeed reinforces the seed field.
In Earth's core, no such wiring exists and the electric currents path is determined from Ohm's law by the direction of the electromotive force (and thus ultimately by the geometry of the velocity and magnetic fields). 
The motion of molten iron in the core has to be such that, on average, the induced electric currents can maintain the geomagnetic field.

As formulated by \cite{Bullard1954}, the question is not (yet) whether a free-flowing fluid can maintain a magnetic field in spite of the expected feedback of the Lorentz force onto the flow.
At first no such feedback was considered: early proponents of the geodynamo hypothesis first tried to find \textit{kinematic dynamos}, where only the induction equation is solved, the velocity field being chosen in the hope that it could produce dynamo action. 
This is a difficult problem, even if the velocity field is prescribed, but working kinematic dynamos have indeed been found. 
Classical exemples include the dynamos found by \cite{Herzenberg1958}, G.O. \cite{Roberts1972}, and \cite{Ponomarenko1973}.

The first laboratory demonstration  of homogeneous dynamo action was given by \cite{Lowes1963} with an apparatus inspired by Herzenberg's dynamo.
Lowes \& Wilkinson's apparatus consists in two rotating solid cylinders made of a high magnetic permeability iron alloy (``Perminvar'') with their axes at right angles, embedded in a solid block of the same material (figure \ref{Fig:BullardLowesDynamos}). The electrical contact between the cylinders and the block is ensured by a thin layer of liquid mercury.  
Denoting by $\omega$ the rotation rate of the cylinders, $a$ their radii, and $\lambda = 1/(\mu \sigma)$ the magnetic diffusivity of the  ``Perminvar'' alloy, 
\cite{Lowes1963} found that their apparatus operates as a self-exciting dynamo when the non dimensional number $\omega a^{2}/\lambda$ exceeds a critical value. 
This number is a magnetic Reynolds number (with velocity scale $\omega a$), so the fact that it must exceed a critical value for dynamo action is consistent with our analysis of the induction equation (section \ref{Section:InductionEquation}).
In spite of being made of two materials (Perminvar plus a thin layer of mercury), 
this is effectively a homogeneous dynamo in the sense that the electric currents are not forced into specific paths. 

A next critical step for establishing the viability of the geodynamo hypothesis has been to obtain dynamos where the velocity field is not imposed \textit{a priori}, and where the feedback of the magnetic field on the flow is taken into account  (\textit{via} the Lorentz force). 
Fluid dynamos obtained numerically by self-consistently solving  the coupled Navier-Stokes, induction, and heat transfer equations have started to appear in the eighties in the contexts of the solar dynamo \citep{Gilman1981,Glatzmaier1984,Glatzmaier1985,Glatzmaier1985b} and geodynamo \citep{Zhang1988,Glatzmaier1995a,Kageyama1995}.
In spite of a number of shortcomings, these numerical dynamos  driven by convective motions have been successful in reproducing some of the more salient features of Earth's magnetic field (more on this in section \ref{Section:ConvectiveDynamos}).

Experimental liquid homogeneous self-exciting dynamos have been obtained  more recently, in two apparatus developed in Riga (Latvia) \citep{Gailitis2001} and Karlsruhe (Germany) \citep{Stieglitz2001}.
In both experiments liquid sodium was forced by propellers into a system of pipes, which effectively imposed the velocity field. 
Liquid sodium is the best electrically conducting liquid available in large quantities (it has a magnetic diffusivity $\eta=0.1$ m$^{2}$.s$^{-1}$, about ten times lower than molten iron), and has  been systematically used as working fluid in dynamo experiments. 
In both apparatus the imposed flow is inspired by known kinematic dynamos: the Riga experiment is based upon the Ponomarenko dynamo, and the Karlsruhe experiment is based upon the G.O. Roberts dynamo. 
The Riga experiment used 2 m$^{3}$ of sodium and required a power input of 200 kW; the Karlsruhe experiment used $1.6$ m$^{3}$ of sodium and required a power input of 630 kW.

Obtaining experimentally a dynamo with a much less constrained flow proved to be even more arduous.
To date  the only successful  free-flowing experiment is the VKS experiment developed in Cadarache (France) \citep{Monchaux2007,Berhanu2007}.
Though numerical dynamo calculations are now well established, liquid metal experiments are still very valuable tools for understanding planetary core dynamics \citep[\textit{e.g.}][]{Nataf2008,Cabanes2014}. 
On certain aspects liquid metal experiments are dynamically closer to Earth's core conditions than numerical simulations. 
In particular, liquid sodium has a magnetic Prandtl number (the ratio of kinematic viscosity to the magnetic diffusivity) of about $10^{-6}$, similar to that of Earth's core, while working numerical dynamos have been so far limited to magnetic Prandtl number values above $\sim 0.1$.

\subsection{What drives the geodynamo?}

In parallel to the question of the feasibility of a liquid homogeneous dynamo, a central question has been the source of power of the geodynamo. 
What drives the motion and provides the energy that is being lost by ohmic dissipation? 

In industrial devices or laboratory experiences, the power is provided by an external mean and therefore is not, theoretically speaking, an issue (though it can be a practical issue, since driving a liquid dynamo in the laboratory does require a quite large power input).

In Earth's core, the possible sources of motion and power fall into two broad categories: (i) natural convection, either of thermal origin \citep{bullard1949,bullard1950,Verhoogen1961}  or compositional origin \citep{Braginsky1963,Gubbins1977,Loper1978,ORourkeStevenson2016}; and (ii) stirring produced by astronomical forcing, \textit{i.e.} motions forced in the core by either tidal forcing or changes of the mantle rotation vector (in direction -- precession or nutation -- or intensity -- libration) \citep{Bondi1948,bullard1949,Malkus1963,Malkus1968}.
The second class of models is the subject of the chapter by Le Reun and Le Bars; we will focus here on natural convection. 

Core convection, whether it be thermal or compositional, is controlled by the rate at which the core loses heat to the mantle.
From a thermal point of view, the core is the slave of the mantle: The heat flux from the core to the mantle is dictated by the efficiency of heat transport by mantle convection, and core convection is thus tied to mantle convection.
Cooling of the core can drive convective motions is several ways:
\begin{enumerate}
\item
Cooling the core from above can potentially produce thermal convection, if the imposed heat flux is larger than the flux which can be conducted along an adiabat in the core (see section \ref{Section:RotatingConvection}).
\item
The secular cooling of the core is responsible for its slow solidification and the formation of the solid inner core. 
In spite of the core being colder at the CMB than at deeper depth, the core started to solidify at its center and the inner core is now growing outward.
The reason for this is that the solidification temperature of the core material increases with pressure faster  than the actual temperature \citep{Jacobs1953}.
If the core started hot and fully molten and then gradually cooled down, the temperature first reached the solidification temperature at the center of the core, which allowed the inner core to nucleate. 
Further cooling results in the slow growth of the  inner core.

One key point here is that the core is made of an iron-rich alloy rather than pure iron. 
Since the density of the core is lower than that of pure iron, we know that the main impurities are ``light elements'' (\textit{i.e.} lighter than iron), likely a mixture of mainly oxygen, sulfur, and silicium.
Upon solidification, these elements are partitioned preferentially into the liquid outer core: solidification of the inner core thus results in an outward flux of light elements at the inner core boundary (ICB), which can drive \textit{compositional} convection.
In addition, the latent heat of solidification helps thermal convection by slowing down the cooling of the inner core boundary. 
\item
Finally, cooling of the core may also be at the origin of a compositional flux across the core-mantle boundary.

Light elements like oxygen, silicium, or magnesium, are present both in the core and mantle (composed predominently of silicates and oxydes of iron and magnesium).
Since the core and mantle materials in contact at the core-mantle boundary should be very close to thermodynamic equilibrium, the relative abundance of these elements in the core and silicates at the CMB is set by the partitioning coefficient of these elements between silicates and the core alloy.
This in general depends on temperature: cooling of the core thus modifies this chemical equilibrium and  results in a flux of elements between mantle and core. 
If light elements are transferred from the mantle to the core, then a stably stratified layer may form below the CMB. 
If in the other way, the flux of light elements would drive compositional convection in the core.
The actual flux of element is limited by the rate at which convection in the mantle provides ``fresh'' material which can react with the molten iron of the core; again, mantle convection controls to a large part the buoyancy flux available to drive core convection.  

Another possible mechanism is exsolution of light elements from the core.
It has been proposed that MgO \citep{ORourkeStevenson2016,Badro2016} and SiO$_{2}$ \citep{Hirose2017} can precipitate (or exsolve) from the core alloy if initially abundant enough.
The saturation concentration of these species happens to decrease with decreasing temperature, so a gradual cooling of the core results in the progressive removal of magnesium, oxygen, and silicon from the core.
If exsolution happens predominantly in the vicinity of the core-mantle boundary, then removing MgO and SiO$_2$ from the core induces a buoyancy flux (by leaving behind an iron-rich, dense melt) which can drive convection in the core. 
\end{enumerate}

\section{Rotating convection}
\label{Section:RotatingConvection}

\subsection{Governing equations and non-dimensional parameters}

The liquid outer core is modelled as a spherical shell of outer and inner radii $r_\mathrm{cmb}$ and $r_\mathrm{icb}$. The shell thickness is denoted by $H=r_\mathrm{cmb}-r_\mathrm{icb}$.
In what follows, we will use either a spherical coordinate system $(r,\theta,\phi)$ or a cylindrical coordinate system $(s,\phi,z)$ (figure \ref{Fig:CoordinateSystems}).
The core is rotating at a rate $\Omega$, and the gravity field is $\mathbf{g}=-g(r) \mathbf{e}_{r}$. 
We assume that there is a positive temperature difference $\Delta T$ between the inner and outer boundaries.
We denote by $\nu$ the kinematic viscosity of the liquid core, by $\rho$ its density, by $\kappa$ its thermal diffusivity, by $\alpha$ its thermal expansion coefficient, and by $c_{p}$ is specific heat capacity.
Though this is a somewhat dubious assumption, the outer core is often modelled as a Boussineq fluid 
(which in particular includes the assumption that the fluid is incompressible). 
Molten iron in the outer core is also typically assumed to behave as a newtonian fluid with temperature-independent kinematic  viscosity. 

Rotating thermal convection is governed by the Navier-Stokes equation in a rotating frame of reference, mass conservation, and a transport equation for temperature $T$.
Denoting by $\mathbf{u}$ the velocity field, by $P$ the pressure, and by $T$ the temperature, 
this set of equations can be written (under the Boussinesq approximation), as
\begin{align}
\rho \frac{D \mathbf{u}}{Dt} + \underbrace{2 \rho \mathbf{\Omega} \times \mathbf{u}}_{\text{\normalsize Coriolis}} &= - \boldsymbol{\nabla} P + \underbrace{\alpha \rho T g \mathbf{e}_{r}}_{\text{\normalsize buoyancy}} + \rho \nu \nabla^{2} \mathbf{u}, \label{Eq:NS1} \\
\boldsymbol{\nabla} \cdot \mathbf{u} &= 0,	\\
\frac{D T}{Dt} &= \kappa \nabla^{2} T.
\label{Eq:HeatEquation}
\end{align}
These equations can be made dimensionless with a variety of different choices of scales.
For exemple, scaling lengths by the outer core thickness $H$, time by $H^{2}/\nu$, velocity by $\nu/H$, temperature by the difference of temperature $\Delta T$ across the shell, and pressure by $\rho \Omega \nu$, gives
\begin{align}
 E \frac{D \mathbf{u}}{Dt} + 2  \mathbf{e}_{z} \times \mathbf{u} &= - \boldsymbol{\nabla} P + \frac{E}{Pr} Ra\, T \mathbf{e}_{r} + E \nabla^{2} \mathbf{u} , \label{Eq:NS_adim}\\
\boldsymbol{\nabla} \cdot \mathbf{u} &= 0,	\\
\frac{D T}{Dt} &= \frac{1}{Pr}  \nabla^{2} T, 
\end{align}
where the Ekman, Rayleigh, and Prandtl numbers are defined as
\begin{align}
\text{Ekman number } E &= \dfrac{\nu}{\Omega H^{2}}, \\
\text{Rayleigh number } Ra &=\dfrac{\alpha g H^{3} \Delta T}{\kappa \nu}, \label{Eq:Rayleigh1} \\
\text{Prandtl number } Pr &= \dfrac{\nu}{\kappa}.
\end{align}

The quantity $Ra E /Pr$ is sometimes taken as a ``modified Rayleigh number'' defined as
\begin{equation}
Ra^{*} = \frac{E}{Pr} Ra = \frac{\alpha g \Delta T H}{\Omega \nu}.
\end{equation}
If a heat flux $q$  is imposed at the boundaries rather than a temperature difference, $Ra$ and $Ra^{*}$ can be modified to give flux-based Rayleigh numbers by replacing $\Delta T$ by $q H/k$, where $k$ is thermal conductivity. 

The Ekman number is a measure of the ratio of the viscous forces and Coriolis acceleration, if the viscous forces are estimated assuming a flow varying spatially on a length scale $\sim H$:
\begin{equation}
\frac{\text{viscous forces}}{\text{Coriolis}} \sim \frac{|\rho \nu \nabla^{2} \mathbf{u}|}{|2 \rho \mathbf{\Omega} \times \mathbf{u}|} \sim \frac{\nu V/H^{2}}{\Omega V} \sim \frac{\nu}{\Omega H^{2}} .
\end{equation}
It is about $10^{-15}$ for the outer core, which means that viscous forces would become of importance only at relatively small length scales (more about this later).

It can also be useful to consider the vorticity equation, obtained by taking the curl of equation \eqref{Eq:NS_adim},
\begin{equation}
 \frac{D \boldsymbol{\zeta}}{D t} = (2 \mathbf{\Omega} + \boldsymbol{\zeta}) \cdot \boldsymbol{\nabla} \mathbf{u}+ \alpha g \boldsymbol{\nabla} T \times \mathbf{e}_{r} + \nu \boldsymbol{\nabla}^{2}\boldsymbol{\zeta},  
\end{equation}
which can usually be simplified to
\begin{equation}
 \frac{D \boldsymbol{\zeta}}{D t} = 2 \Omega \frac{\partial \mathbf{u}}{\partial z} + \alpha g \boldsymbol{\nabla} T \times \mathbf{e}_{r} + \nu \boldsymbol{\nabla}^{2}\boldsymbol{\zeta},  
 \label{Eq:VorticityEquation}
\end{equation}
by assuming that the vorticity $\boldsymbol{\zeta}$ is small compared to the rotation rate $\mathbf{\Omega}$.
Here the term $2 \Omega \frac{\partial \mathbf{u}}{\partial z}$ is the production of vorticity through stretching of planetary vorticity, 
$\alpha g \boldsymbol{\nabla} T \times \mathbf{e}_{r}$ is the \textit{baroclinic} production of vorticity, and $\nu \boldsymbol{\nabla}^{2}\boldsymbol{\zeta}$ is the viscous diffusion of vorticity.

\subsection{Geostrophy}

The very small value of the Ekman number of the core suggests that viscous forces may be neglected, at least as long as only large scale motions are considered. 
Keeping only the Coriolis acceleration and the pressure gradient (thus neglecting also inertia and buoyancy), the Navier-Stokes equation reduces to
\begin{equation}
2 \rho \mathbf{\Omega} \times \mathbf{u} = - \boldsymbol{\nabla} p,
\label{Eq:GeostrophicBalance}
\end{equation}
which is called the geostrophic balance. 
Taking the curl of equation \eqref{Eq:GeostrophicBalance}, which gives
\begin{equation*}
\mathbf{\Omega} \cdot \boldsymbol{\nabla} \mathbf{u} = \Omega \frac{\partial \mathbf{u}}{\partial z}=0,  
\end{equation*}
shows that the velocity field is invariant along the rotation axis if this balance holds. 
This is known as the \textit{Taylor-Proudman theorem}.

In a rotating container with sloped boundaries, such as planetary cores, the motion is further restricted by the condition that no fluid can cross the boundary.
In a spherical container, this condition writes $\mathbf{u} \cdot \mathbf{e}_{r} = 0$.
In other words, the velocity field at the boundary can only have $\theta$ and $\phi$ components.
Of these, only the $\phi$ component is allowed by the Taylor-Proudman constraint. 
This means that the only possible motions are longitudinal (zonal) motions of the form
\begin{equation*}
\mathbf{u} = u_{\phi}(s,\phi,t) \mathbf{e}_{\phi}.
\end{equation*}
Mass conservation ($\boldsymbol{\nabla}\cdot\mathbf{u}=(r \sin \theta)^{-1}\partial u_{\phi}/\partial \phi=0$) further implies that $\mathbf{u}$ cannot depend on $\phi$.
The only possible geostrophic flows in a spherical container thus consist in rigid cylinders rotating about Earth's axis of rotation, called \textit{geostrophic cylinders}:
\begin{equation*}
\mathbf{u} = u_{\phi}(s,t) \mathbf{e}_{\phi}.
\end{equation*}

\subsubsection{Torsional oscillations}

\begin{figure}[t]
\includegraphics[width=\linewidth]{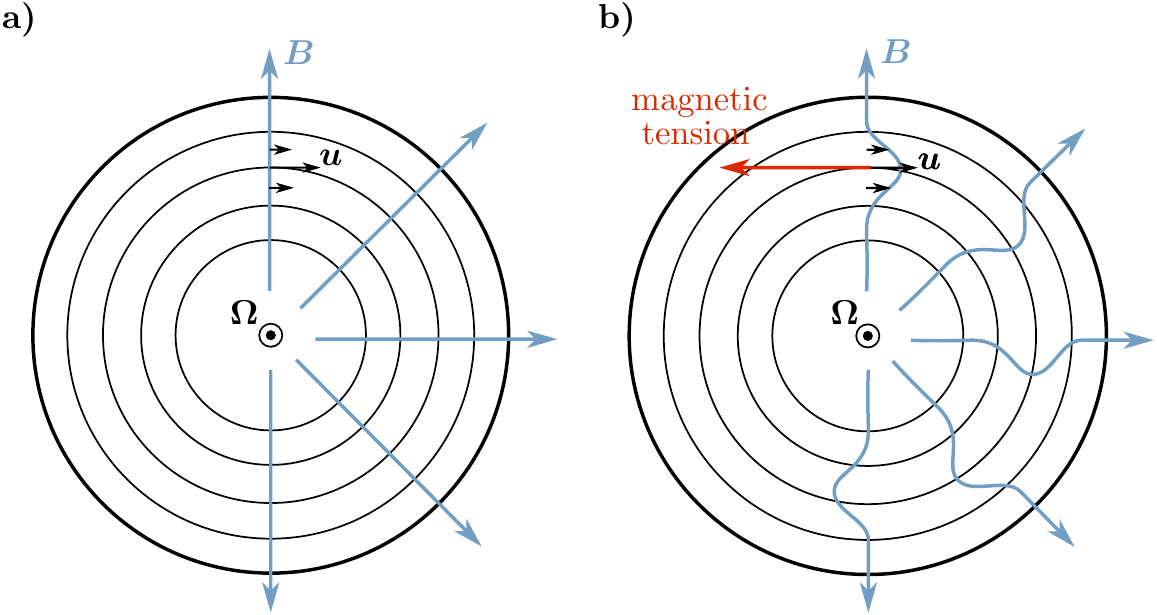}
\caption{Torsional waves mechanism.}
\label{Fig:TorsionalWaves}
\end{figure}

Though it is somewhat of a side note in the context of a thermal convection-oriented section, it is worth discussing quickly the effect of a magnetic field on geostrophic motions. 
The addition of a magnetic field permeating the geostrophic cylinders allows the propagation of a subclass of Alfv\`en waves called \textit{torsional waves}, or \textit{torsional oscillations}  \citep{Braginsky1970}.

The basic mechanism behind these waves can be easily understood using the concept of magnetic tension introduced in section \ref{Section:LorentzForce}.
Assume one geostrophic cylinder rotates at a different rate (figure \ref{Fig:TorsionalWaves}a).
The differential rotation of this cylinder with respect to its neighbours will bend the magnetic lines as shown on figure \ref{Fig:TorsionalWaves}b.
The resulting magnetic tension will  produce a restoring force of magnitude inversely proportional to the azimuthal displacement of the cylinder.
The net effect on the geostrophic cylinder is a restoring torque.
As shown in section \ref{Section:LorentzForce} Alfv\`en waves are longitudinal waves; torsional waves therefore must propagate in the $s$-direction. 

A dispersion equation for these waves can be obtained by applying conservation of angular momentum to a geostrophic cylinder of inner radius $s$ and thickness $ds$. 
By restricting the analysis to small relative motions of the cylinders, one find that the angular velocity $\omega(s)$ obeys the following equation:
\begin{equation}
\frac{\partial^{2}\omega}{\partial t^{2}} = \frac{1}{\mu_{0} \rho s^{2} H(s)} \frac{\partial}{\partial s} \left( s^{3} H(s) \bar{B}_{s}\frac{\partial \omega}{\partial s} \right),
\end{equation}
where
\begin{equation}
 \bar{B}_{s} = \sqrt{
 \frac{1}{4\pi H}
 \int_{-H}^{H} \int_{0}^{2\pi} B_{s}^{2}  d\phi dz
}.
\end{equation}
is the r.m.s. value of $B_{s}$ on the geostrophic cylinder of radius $s$ \citep{Braginsky1970}.

The propagation velocity of these oscillations being  proportional to $ \bar{B}_{s}$, identifying torsional oscillations in core flow inversions can provide an estimate of the magnetic fields intensity \textit{within} the  core.
\cite{Gillet2010} found a 6-year period signal in their core flow inversions which they interpreted as a torsional wave. It takes about 4 years for this wave to propagate through the outer core; the resulting propagation velocity gives $ \bar{B}_{s}\sim 2$ mT.
For comparison, the radial component of the magnetic field at the CMB peaks at about 1 mT (figure \ref{Fig:GeomagneticField}).

\subsection{Onset of thermal convection}  

We now turn to the question of the initiation of thermal convection in a rotating planetary core.  
Linear stability analysis shows that at low Ekman number the critical Rayleigh number $\mathrm{Ra}_\mathrm{cr}$ and the horizontal length scale $\ell_{c}$ of the first unstable mode scale as
\begin{align}
\mathrm{Ra}_\mathrm{cr} &\sim E^{-4/3} , \\
\ell_{c} &\sim H E^{1/3}. \label{Eq:l_c}
\end{align}
The linear stability analysis is challenging \citep{Roberts1968,Busse1970,Jones2000,Dormy2004}; to try to understand this result at a lower mathematical cost, we will take a heuristic approach and derive the Ekman number dependency of $\mathrm{Ra}_\mathrm{cr}$ from simple physical arguments.

We have seen above that geostrophic motions in a spherical container are restricted to zonal flows of the form $\mathbf{u} = v_{\phi}(s,t) \mathbf{e}_{\phi}$.
This obviously cannot carry heat radially; thermal convection therefore necessarily involves deviations from geostrophy.
This can result from buoyancy forces, inertia, viscous forces, or Lorentz force. 
We will not consider here the effect of the Lorentz force since we are interested in the initiation of convection in a planetary core, 
which here is assumed to predate the generation of the magnetic field.  
Furthermore, we can safely neglect inertia at the onset of convection since it is quadratic in $\mathbf{u}$ while the Coriolis and viscous forces are linear in $\mathbf{u}$. 
We therefore cannot rely on inertia to break the geostrophic balance, and we are left with viscous and buoyancy forces. 

Let us first consider the vorticity equation \eqref{Eq:VorticityEquation}, which, neglecting the inertia-derived terms, is 
\begin{equation}
0 = 2 \Omega \frac{\partial \mathbf{u}}{\partial z} + \alpha g \boldsymbol{\nabla} T \times \mathbf{e}_{r} + \nu \boldsymbol{\nabla}^{2}\boldsymbol{\zeta}.  
\label{Eq:VorticityNoInertia}
\end{equation}
Denoting by $U$ the characteristic scale of convective velocities, $\ell$ the characteristic horizontal length scale of convective motions, and $\delta T$ the characteristic scale of horizontal temperature variations, the first term is on the order of $\Omega U/H$, the baroclinic term on the order of $\alpha g \delta T/\ell$, and the viscous term on the order of $\nu U/\ell^{3}$  (the vorticity $\boldsymbol{\zeta}=\boldsymbol{\nabla} \times \mathbf{u}$ is on the order of $U/\ell$).
Since buoyancy is the driver of the flow, we expect the baroclinic vorticity production to be always of importance, and balanced by either the viscous term or the Coriolis term, or both.

A balance between baroclinic production of vorticity and diffusion of vorticity gives 
\begin{equation}
U \sim \frac{\alpha g}{\nu} \ell^{2} \delta T,
\label{Eq:VelocityScalingStokes}
\end{equation}
which is basically a Stokes velocity.
Balancing baroclinic production of vorticity and stretching of the planetary vorticity gives
\begin{equation}
U \sim \frac{\alpha g}{\Omega} \frac{H}{\ell} \delta T,
\label{Eq:VelocityScalingThermalWind}
\end{equation}
(a \textit{thermal wind} balance).

The two above velocity scales are equal when the three terms in equation \eqref{Eq:VorticityNoInertia} are of equal importance, which happens when the flow length scale is on the order of $\ell_{c} \sim (\nu H/\Omega)^{1/3}= H E^{1/3}$. 
At $\ell \gg \ell_{c}$, the viscosity term is small compared to the planetary vorticity stretching term and the relevant velocity scaling is equation \eqref{Eq:VelocityScalingThermalWind};
at $\ell \ll \ell_{c}$, the viscosity term is large compared to the planetary vorticity stretching term and the relevant velocity scaling is equation \eqref{Eq:VelocityScalingStokes}.
Since the velocity scale increases with $\ell$ while $\ell < \ell_{c}$ (equation \eqref{Eq:VelocityScalingStokes}), and then decreases with increasing $\ell$ (equation \eqref{Eq:VelocityScalingThermalWind}), $\ell_{c}$ also happen to be the convection length scale at which the velocity would be maximal.

One key point here is that the $2 \Omega \frac{\partial \mathbf{u}}{\partial z} \sim \alpha g \boldsymbol{\nabla} T \times \mathbf{e}_{r}$ balance would yield a velocity field  $\mathbf{u}$ with no radial component. 
This will not carry heat radially. In the absence of inertia, radial convection would therefore requires viscous effects to be important.
This may seem counter-intuitive, but viscous effects are actually necessary to initiate a Rayleigh-B\'enard-type convection in a rotating sphere !
This would suggest that an horizontal scale smaller than $\ell_{c}$ is necessary for the initiation of convection.
Since in addition the velocity scale at $\ell < \ell_{c}$ decreases with decreasing $\ell$, we may expect that the optimal scale for the initiation of convection is on the order of $\ell_{c}$: radial motions are damped by rotation effects at larger scale, while smaller scale motions are slower due to viscous friction. 

Let us now consider a liquid parcel of size $\ell$ and temperature $T$ larger than the surrounding temperature $\bar T(r)$ (temperature excess $\delta T = T - \bar T$). %\bar T(r)+ \delta T$. 
$\bar T(r)$ is the background conductive temperature profile. 
The  radial velocity of the parcel is given by
\begin{equation}
u_{r} \sim \frac{\alpha g \ell^{2}}{\nu} ( T - \bar T)
\label{Eq:Scaling_ur_stability}
\end{equation}
since, as argued above, viscous effects are necessary to the initiation of convection in a rotating sphere. 
To see how the velocity of the parcel evolves, take its Lagrangian derivative:
\begin{equation}
\frac{D u_{r}}{Dt} \sim  \frac{\alpha  g \ell^{2}}{\nu} \left( \frac{DT}{Dt} - \frac{D\bar T}{Dt} \right).
\end{equation}
The first term in the parenthesis can be estimated from the heat equation,
\begin{equation}
\frac{DT}{Dt} = \kappa \nabla^{2} T \sim - \kappa \frac{T-\bar T}{\ell^{2}},
\label{Eq:ScalingHeatTransport}
\end{equation}
while the second term is
\begin{equation}
 \frac{D\bar T}{Dt} =  \frac{\partial \bar T}{\partial t} + \mathbf{u}\cdot \boldsymbol{\nabla} T = u_{r} \frac{d\bar T}{dr} \sim   \frac{\alpha  g \ell^{2}}{\nu} \left( T - \bar T \right) \frac{d\bar{T}}{dr}.
\end{equation}
Putting everything together, this gives
\begin{align}
\frac{D u_{r}}{Dt} &\sim  \frac{\alpha  g \ell^{2}}{\nu} \left( - \kappa \frac{T-\bar T}{\ell^{2}} - \frac{\alpha  g \ell^{2}}{\nu} \left( T - \bar T \right) \frac{d\bar{T}}{dr} \right), \\
&\sim  \frac{\alpha  g \ell^{2} \left( T - \bar T \right)}{\nu \ell^{2}/\kappa} \left(  - \frac{\alpha  g \ell^{4}}{\kappa \nu} \frac{d\bar{T}}{dr} - 1 \right) .
\end{align}
This can be rearranged to give an estimate of the growth rate of the vertical velocity:
\begin{equation}
\frac{1}{u_{r}} \frac{D u_{r}}{Dt} \sim \frac{\kappa}{\ell^{2}} \left( - \frac{\alpha  g \ell^{4} }{\kappa \nu} \frac{d\bar{T}}{dr}  - 1 \right). 
\label{Eq:HeuristicGrowthRate0}
\end{equation}
With $\frac{d\bar{T}}{dr} \sim - \Delta T/H$, this gives
\begin{equation}
\frac{1}{u_{r}} \frac{D u_{r}}{Dt} \sim \frac{\kappa}{\ell^{2}} \left( \frac{Ra}{(H/\ell)^{4}} -1  \right). \label{Eq:HeuristicGrowthRate}
\end{equation}  
A parcel displaced upward (resp. downward) will keep rising (resp. sinking) 
if $\frac{1}{u_{r}} \frac{D u_{r}}{Dt} >0$, \textit{i.e.} if the term within the parenthesis in equation \eqref{Eq:HeuristicGrowthRate} is positive.
This condition can be recast as a condition for the Rayleigh number, which must exceed a critical value $\mathrm{Ra}_\mathrm{cr}$ which is a function of the length scale $\ell$ of the perturbation:
\begin{equation}
  Ra > \mathrm{Ra}_\mathrm{cr}(\ell) \sim \left( \frac{H}{\ell} \right)^{4}.
  \label{Eq:CriticalRa_Ek21}
\end{equation}
Perturbations with the largest length scale $\ell$ will have the lowest critical Rayleigh number and will then be favoured. 
In non-rotating convection, the only limit to $\ell$ is the size $H$ of the convecting layer, so the lowest $\mathrm{Ra}_\mathrm{cr}$ will correspond to perturbations with $\ell \sim H$.
The initiation of convection then simply requires Ra to exceed some critical value, which depends only on the geometry and boundary conditions (it is for example $27 \pi^{4}/4\simeq 657.5$ for Rayleigh-B\'enard convection in a plane layer with free-slip boundaries). 
In a rotating sphere with radial gravity, driving radial motion requires initiating the convection with a horizontal length scale $\lesssim \ell_{c} = H E^{1/3}$.
Since the critical Rayleigh number is a decreasing function of $\ell$, we expect the length scale of the fastest growing mode to be $\sim \ell_{c}$, which gives
\begin{equation}
\mathrm{Ra}_\mathrm{cr} \sim E^{-4/3},
\label{Eq:CriticalRa_Ek2}
\end{equation}
as predicted by linear stability analysis.
 
 \subsection{Compressibility effect}
 
In the above analysis, we have left aside compressibility effects on the initiation of convection.  
 Rather than the Boussinesq version of the heat transfer equation \eqref{Eq:HeatEquation}, we now consider its more general form
\begin{equation}
 \frac{D T}{Dt} = \kappa \nabla^{2} T + \frac{\alpha T}{\rho c_{p}} \frac{D P}{Dt} + \dot \epsilon : \tau,
 \label{Eq:EntropyBalance}
\end{equation}
where $\dot \epsilon : \tau$ is the viscous dissipation ($\tau$ is the stress tensor and $\dot \epsilon$ the deformation rate tensor). 
Equation \eqref{Eq:EntropyBalance} can be obtained from the entropy balance \citep[for its derivation, see \textit{e.g.}][]{Ricard2007}. 

\subsubsection{Schwarzschild's stability criterion for thermal convection}

\begin{figure}[t]
\centering
\includegraphics[width=\linewidth]{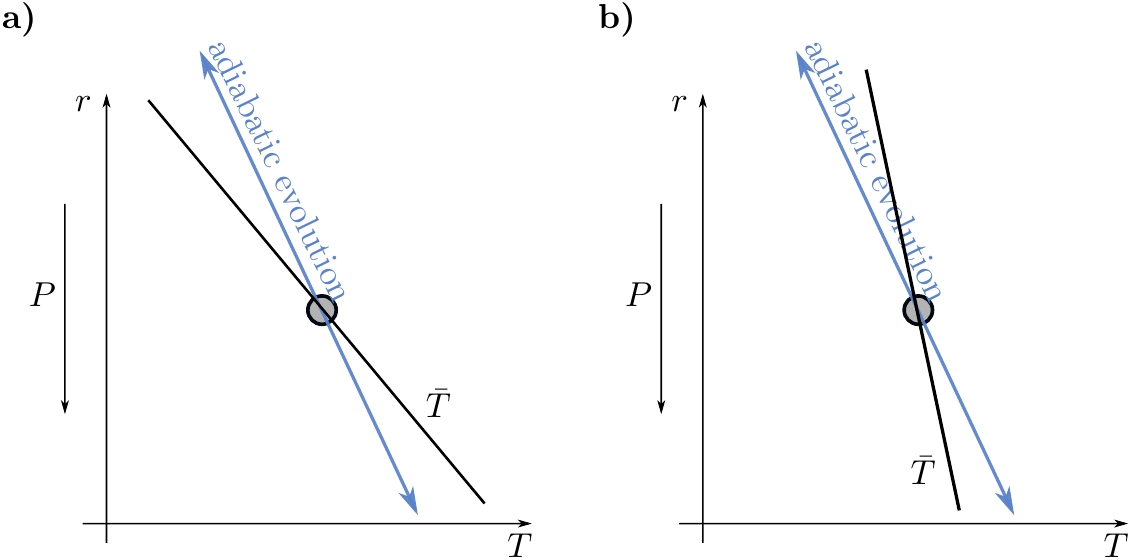}
\caption{Schwarzschild's criterion for thermal convection in a compressible fluid: a parcel of fluid displaced vertically and evolving adiabatically would find itself at a temperature higher than its surrounding if $\partial \bar T/\partial P > \partial T_\mathrm{ad}/\partial P$ (figure \textbf{a}), and lower than its surrounding if $\partial \bar T/\partial P < \partial T_\mathrm{ad}/\partial P$ (figure \textbf{b}).}
\label{Fig:Schwarzschild}
\end{figure}

To see the effect of the pressure term, let us consider a parcel of fluid displaced upward with negligible viscous dissipation and heat diffusion (thus following an isentropic path). 
It follows directly from  equation \eqref{Eq:EntropyBalance} that the parcel's temperature will vary with its pressure as
\begin{equation}
 \frac{D T}{DP} = \frac{\alpha T}{\rho c_{p}} \equiv \frac{\partial T_\mathrm{ad}}{\partial P},
 \label{Eq:AdiabaticGradientP}
\end{equation}
which corresponds to adiabatic heating (cooling) due to compression (decompression).\footnote{The evolution of the parcel is actually even an isentropic process, since the parcel's evolution is both adiabatic (no transfer of heat and mass with its surrounding) and reversible (no friction).}
This is equivalent to a radial gradient, the \textit{adiabatic gradient}, given by
\begin{equation}
\frac{\partial T_\mathrm{ad}}{\partial r} = - \frac{\alpha T g}{c_{p}}.
\label{Eq:AdiabaticGradient_r}
\end{equation}
The adiabatic temperature difference $\Delta T_\mathrm{ad}$ across the outer core is obtained by integrating one of equations \eqref{Eq:AdiabaticGradientP} or \eqref{Eq:AdiabaticGradient_r} (here assuming that $\alpha/c_{p}$ is constant and that $g(r)=g_\mathrm{cmb} r/r_\mathrm{cmb}$):
\begin{align}
\Delta T_\mathrm{ad}  &= T_\mathrm{icb}  \left[ \exp \left( - \int_{r_\mathrm{icb}}^{r_\mathrm{cmb}}  \frac{\alpha g}{c_{p}} dr \right) - 1 \right] , \\
&= T_\mathrm{icb}  \left[ \exp \left( - \frac{1}{2} \frac{\alpha g_\mathrm{cmb}}{c_{p}} \frac{r_\mathrm{cmb}^{2}- r_\mathrm{icb} ^{2}}{r_\mathrm{cmb}} \right) - 1 \right] ,
\end{align}
where $T_\mathrm{icb} $ is the temperature at the Inner Core Boundary.
In Earth's core, the adiabatic temperature difference is $\Delta T_\mathrm{ad} \sim 1000$ K, so this is a significant effect.

If displaced fast enough for negligible heat diffusion, a parcel of fluid displaced vertically from its initial state will follow an adiabatic path.
Its temperature $T$ will then vary with pressure $P$ according to equation \eqref{Eq:AdiabaticGradientP}.
If the background temperature gradient $\bar T$ is steeper than the adiabatic gradient, \textit{i.e.} if
\begin{equation}
\frac{\partial \bar T}{\partial P}>\frac{\partial T_\mathrm{ad}}{\partial P}, 
\label{Eq:Schwarzschild_1}
\end{equation}
then the temperature of a parcel of fluid displaced upward will becomes larger than the background temperature, with the temperature difference increasing with the upward displacement of the parcel (figure \ref{Fig:Schwarzschild}\textbf{a}). The parcel will thus become less and less dense than the surrounding, and will keep rising. 
In contrast, if 
\begin{equation}
\frac{\partial \bar T}{\partial P}<\frac{\partial T_\mathrm{ad}}{\partial P},
\label{Eq:Schwarzschild_2}
\end{equation}
then a parcel of fluid displaced upward will have a temperature lower than the background temperature (figure \ref{Fig:Schwarzschild}\textbf{b}). It will thus be denser than the surrounding fluid, and its buoyancy will eventually send it back toward its initial position. 
A temperature profile satisfying the condition \eqref{Eq:Schwarzschild_2} is thus stable against thermal convection.

Equation \eqref{Eq:Schwarzschild_1} is thus a \textit{necessary} condition for thermal convection, known as \textit{Schwarzschild's stability criterion}\footnote{Alternatively, equation \eqref{Eq:Schwarzschild_2} is a sufficient condition for stability.}. 
It is not a \textit{sufficient} condition since it assumes that fluid parcels follow adiabatic and reversible paths when displaced upward or downward.
In a real fluid, the parcels would exchange heat with the surrounding by diffusion, and viscous dissipation would be non-zero.

\subsubsection{Critical Rayleigh number with compressibility effects}
 
 Going back to our heuristic derivation of the critical Rayleigh number, we consider again the evolution of a fluid parcel displaced from its initial position, but now assume that the evolution of its temperature is governed by 
 \begin{equation}
 \frac{D T}{Dt} = \kappa \nabla^{2} T + \frac{\alpha T}{\rho c_{p}} \frac{D P}{Dt},
 \label{Eq:EntropyBalanceNoDissipation}
\end{equation}
ignoring the viscous dissipation term $\dot \epsilon : \tau$ included in equation \eqref{Eq:EntropyBalance}.
Viscous dissipation is quadratic in $\mathbf{u}$, and can therefore be neglected when looking at the initiation of convection.

In equation \eqref{Eq:EntropyBalanceNoDissipation}, the pressure derivative can be estimated as
 \begin{equation}
 \frac{D P}{Dt} \simeq u_{r} \frac{\partial P}{\partial r} \simeq - \rho g u_{r}
 \end{equation}
 neglecting terms in pressure fluctuations. 
 Equation \eqref{Eq:EntropyBalanceNoDissipation} thus gives
 \begin{equation}
 \frac{D T}{Dt} \simeq - \kappa \frac{T-\bar T}{\ell^{2}} -  u_{r} \frac{dT_\mathrm{ad} }{dr}.
 \end{equation}
Using this relation instead of equation \eqref{Eq:ScalingHeatTransport} and going through the same steps (equations \eqref{Eq:Scaling_ur_stability} to \eqref{Eq:CriticalRa_Ek2}),
the rate of change of the parcel's velocity is now given by 
\begin{equation}
\frac{1}{u_{r}} \frac{D u_{r}}{Dt} \sim \frac{\kappa}{\ell^{2}} \left( - \frac{\alpha  g \ell^{4} }{\kappa \nu} \frac{d\left(\bar{T}-T_\mathrm{ad}\right)}{dr}  - 1 \right)
\end{equation}
instead of equation \eqref{Eq:HeuristicGrowthRate0}.
Denoting by $\Delta T_\mathrm{>ad} = \Delta T - \Delta T_\mathrm{ad}$ the excess temperature
difference above the adiabatic temperature across the outer core (or {super-adiabatic temperature difference}), we have $\frac{d(\bar{T}-T_\mathrm{ad})}{dr} \sim - \Delta T_\mathrm{>ad}/H$, and 
\begin{equation}
\frac{1}{u_{r}} \frac{D u_{r}}{Dt} \sim \frac{\kappa}{\ell^{2}} \left( \frac{Ra_\mathrm{>ad}}{(H/\ell)^{4}} -1  \right), \label{Eq:HeuristicGrowthRateCompressible}
\end{equation}  
where
\begin{equation}
Ra_\mathrm{>ad} = \dfrac{\alpha g H^{3} \Delta T_\mathrm{>ad}}{\kappa \nu} 
\end{equation}
is the super-adiabatic Rayleigh number.
The instability criterion is thus generalised by simply replacing the classical Rayleigh number by the super-adiabatic Rayleigh number.
In a compressible fluid at low Ekman, convection will thus start if 
\begin{equation}
Ra_\mathrm{>ad} > \mathrm{Ra}_\mathrm{cr} \sim E^{-4/3}.
\label{Eq:InstabilityCondiction_wCompressibility}
\end{equation}

 \subsection{Application to Earth's core}
 
\subsubsection{Thermal convection}

\begin{table}[t]
\caption{Representative values of some properties of the outer core.}
\centering
\begin{tabular}{@{}lcc@{}}
\toprule
radius of the core-mantle boundary & $r_\mathrm{cmb}$	& 3480 km \\
radius of the inner core boundary & $r_\mathrm{icb}$	& 1221 km \\
acceleration of gravity at CMB	& $g_\mathrm{cmb}$	& $10.68$ m.s$^{-2}$ \\
kinematic viscosity			& $\nu$		& $10^{-6}$ m$^{2}$.s$^{-1}$ \\
thermal diffusivity			& $\kappa$ 	& $5\times10^{-6}$  m$^{2}$.s$^{-1}$ \\
compositional diffusivity 		& $\kappa_{\chi}$	& $10^{-9\pm 2}$ m$^{2}$.s$^{-1}$ 	\\
magnetic diffusivity			& $\eta$ 		& $1$  m$^{2}$.s$^{-1}$ \\
thermal expansion coefficient 	& $\alpha$ 	& $10^{-5}$ K$^{-1}$ \\
compositional expansion coefficient 	& $\beta$ 	& $1$ wt\%$^{-1}$ \\
\bottomrule
\end{tabular}
\label{Table:properties}
\end{table}

How much superadiabatic the core needs to be to meet this requirement?
Using the definitions of $Ra_\mathrm{>ad}$ and $E$, the condition \eqref{Eq:InstabilityCondiction_wCompressibility} can be recast as a condition for the superadiabatic temperature difference:
\begin{equation}
\Delta T_\mathrm{>ad} \gtrsim \frac{\Omega^{4/3} \kappa}{\alpha g H^{1/3} \nu^{1/3}},
\end{equation}
With parameter values from table \ref{Table:properties}, the required $\Delta T_\mathrm{>ad}$ is only on the order of $10^{-7} - 10^{-6}$ K: this is tiny, compared for example to the adiabatic temperature variation across the core, which is $\sim 1000$ K.

In practice this means that Schwarzschild's criterion is a very good indicator of the likelihood of core convection. 
In a planetary core, the temperature difference across the core is not imposed.
What is imposed is the heat flux $Q_\mathrm{cmb}$ at the Core mantle Boundary, which is controlled by mantle convection.
A requirement for thermal convection is therefore that the heat flux imposed by mantle convection is larger than the heat flux $Q_\mathrm{cmb}^\mathrm{ad} $ which can be carried by diffusion along an adiabatic temperature profile:
\begin{align}
Q_\mathrm{cmb} > Q_\mathrm{cmb}^\mathrm{ad}  &= - 4\pi r_\mathrm{cmb}^{2} k \frac{\partial T_\mathrm{ad}}{\partial r}, \\
&= 4\pi r_\mathrm{cmb}^{2} \frac{\alpha\, g_\mathrm{cmb}\, T_\mathrm{cmb}}{c_{p}}	k , \label{Eq:AdiabaticFlux}
\end{align}
where $k$ is the thermal conductivity of the core.
The value of the thermal conductivity of the core is currently quite debated. 
Until quite recently, available estimates were in the range $30-60$ W.m$^{-1}$.K$^{-1}$ \citep{Stacey2001,Stacey2007}, but much higher values ($> 150$  W.m$^{-1}$.K$^{-1}$) have since been proposed by several independent groups \citep{de-Koker2012,pozzo2012,Gomi2013,gomi2016}.
Other groups \citep{seagle2013,konopkova2016} still favour a relatively low value of $k$.

Finally, it is also instructive to estimate the horizontal scale $\ell_{c}$ at which convection is initiated.
From equation \eqref{Eq:l_c}, this is $\ell_{c} \sim H E^{1/3} \sim 2000\ \mathrm{km} \times (10^{-15})^{1/3} \sim 20$ m, a factor $10^{5}$ smaller than the outer core shell thickness!

\subsubsection{Compositional convection}

Though equations \eqref{Eq:NS1}-\eqref{Eq:HeatEquation} have been written for thermal convection, the same set of equations can be used to describe compositional convection if $T$ is replaced by a concentration (in light elements) $\chi$, $\alpha$ by a coefficient of compositional expansion $\beta$, and $\kappa$ by a compositional diffusivity $\kappa_{\chi}$.
There is no direct compositional analog of the adiabatic gradient, so the condition for compositional convection is simply 
\begin{equation}
Ra_{\chi} = \dfrac{\beta g H^{3} \Delta \chi}{\kappa_{\chi} \nu} > \mathrm{Ra}_\mathrm{cr} \sim E^{-4/3},
\label{Eq:Ra_cr_compo}
\end{equation}
where $\Delta \chi$ is the light element concentration difference across the outer core.
Written in terms of $\Delta \chi$, condition \eqref{Eq:Ra_cr_compo} is equivalent to the following condition:
\begin{equation}
\Delta \chi \gtrsim \frac{\Omega^{4/3} \kappa_{\chi}}{\beta g H^{1/3} \nu^{1/3}} \simeq 10^{-12} \ \text{wt.\%}.
\end{equation}
A tiny difference in composition can drive core convection.

\section{Convective dynamos}
\label{Section:ConvectiveDynamos}

\subsection{Governing equations}

In its simplest form, the set of governing equations for a convectively driven dynamo consists of the three equations governing rotating convection with the addition of the Lorentz force in the Navier-Stokes equation, plus the induction equation:
\begin{align}
\rho \frac{D \mathbf{u}}{Dt} + 2 \rho \mathbf{\Omega} \times \mathbf{u} &= - \boldsymbol{\nabla} p + \alpha \rho T g \mathbf{e}_{r} + \frac{1}{\mu_{0}} (\boldsymbol{\nabla} \times \mathbf{B} ) \times \mathbf{B} +  \rho \nu \nabla^{2} \mathbf{u}, \\
\boldsymbol{\nabla} \cdot \mathbf{u} &= 0,	\\
\frac{D T}{Dt} &= \kappa \nabla^{2} T, \\
\frac{D \mathbf{B}}{D t} &=  \left(\mathbf{B}\cdot\boldsymbol{\nabla}\right)\mathbf{u}+ \eta \nabla^{2} \mathbf{B}.
\end{align}
This set of equations can be made dimensionless by using the characteristic scales already used in section \ref{Section:RotatingConvection} in the case of thermal convection for lengths ($H$), time ($H^{2}/\nu$), temperature ($\Delta T$ or $q H/k$) and pressure ($\rho \Omega \nu$).
A magnetic field scale can for example be obtained by assuming a balance between the Coriolis acceleration and the Lorentz force.
Writing the Lorentz force as $\mathbf{j} \times \mathbf{B}$ and estimating $\mathbf{j}$ from Ohm's law as $\mathbf{j} \sim \sigma \mathbf{u} \times \mathbf{B}$ gives a Lorentz force $\sim \sigma U B^{2}$. 
Balancing this with the Coriolis force which is $\sim \rho \Omega U$ gives a magnetic field scale $B_{0} = \sqrt{\rho \Omega/\sigma}$.
Using this set of scales, we obtain the following dimensionless set of equations:
\begin{align}
 E \frac{D \mathbf{u}}{Dt} + 2  \mathbf{e}_{z} \times \mathbf{u} &= - \boldsymbol{\nabla} P + \frac{E}{Pr} Ra\, T \mathbf{e}_{r} + (\boldsymbol{\nabla} \times \mathbf{B} ) \times \mathbf{B} + E \nabla^{2} \mathbf{u} , \label{Eq:NS_adim2}\\
\boldsymbol{\nabla} \cdot \mathbf{u} &= 0,	\\
\frac{D T}{Dt} &= \frac{1}{Pr}  \nabla^{2} T, \\
\frac{D \mathbf{B}}{D t} &=  \left(\mathbf{B}\cdot\boldsymbol{\nabla}\right)\mathbf{u}+ \frac{1}{P_{m}} \nabla^{2} \mathbf{B}, \label{Eq:Induction_adim}
\end{align}
where the magnetic Prandtl number $P_{m}$ is defined as the ratio of the kinematic viscosity $\nu$ to the magnetic diffusivity:
\begin{equation}
P_{m} = \frac{\nu}{\eta} .
\end{equation}
$P_{m}$ is about $10^{-6}$ in Earth's core: the magnetic field diffuses much faster than momentum, and we therefore expect the magnetic field to vary over larger length scales that the velocity field.

In addition to the \textit{input} non-dimensional numbers ($E$, $Ra$, $Pr$, $P_{m}$), it is also often useful to consider \textit{output} non-dimensional numbers based on measured dynamical quantities such as some  averages of the velocity  and magnetic field, $\langle u\rangle$ and $\langle B \rangle$.
Examples of useful output non-dimensional numbers include the followings:
\begin{align}
\text{Reynolds number } Re &= \frac{\text{inertia}}{\text{viscous forces}}= \frac{H \langle u\rangle}{\nu} , \\
\text{magnetic Reynolds number }Rm &= \frac{\text{stretching of $\mathbf{B}$}}{\text{diffusion $\mathbf{B}$}}= \frac{H \langle u\rangle}{\eta} = P_{m} Re,\\
\text{Rossby number } Ro &= \frac{\text{inertia}}{\text{Coriolis force}}= \frac{\langle u \rangle}{\Omega H} = E Re,  \\
\text{Elsasser number }\Lambda &= \frac{\text{Lorentz force}}{\text{Coriolis force}}= \frac{\sigma \langle B \rangle^{2}}{\rho \Omega}.
\end{align}
In this expression of the Elsasser number, the Lorentz force has been estimated from the expression $\mathbf{j}\times\mathbf{B}$ by taking $\mathbf{j} \sim \sigma \mathbf{u} \times \mathbf{B}$ from Ohm's law.

These numbers can be estimated for Earth's core as follows. 
If we accept that the geomagnetic field is produced by dynamo effect in the core, then the magnetic Reynolds number must be larger that $\mathcal{O}(10)$ (the critical magnetic Reynolds number for dynamo action in a spherical shell is typically $\sim 50$).
This implies that $Re$ must be at least of $\mathcal{O}(10^{7})$ and Rossby above $\mathcal{O}(10^{-8})$.
The order of magnitude of the velocity corresponding to $Rm \sim 10$ is $10 \eta/H \sim 5 \times 10^{-6}$ m.s$^{-1}$.
If instead we take a velocity scale of $5 \times 10^{-4}$ m.s$^{-1}$ as obtained from core flow inversions, we obtain $R_{m}\sim 10^{3}$, $Re \sim 10^{9}$, and $Ro \sim 10^{-6}$.
With a magnetic field of $\sim 4$ mT \citep{Gillet2010}, the Elsasser number is $\Lambda \sim 10$.

The values of these numbers suggest that in the Navier-Stokes equation the dominant forces would be the Coriolis force, the Lorentz force, and presumably the buoyancy force  (which is difficult to estimate from simple arguments, but which is likely non-negligible since it is the source of motion).
This corresponds to the so-called MAC balance (Magnetic, Archimedes, Coriolis).

Solving numerically these equations in regimes which are relevant to the geodynamo is difficult;
it has in fact not yet been possible to solve them with parameter values approaching that of the Earth.
They are two main reasons for this:
\begin{enumerate}
\item
The low viscosity of molten iron means that the velocity field likely develops small scale turbulent fluctuations, which would require a high spatial resolution and fine time-stepping to be fully resolved.
We have seen in section \ref{Section:RotatingConvection} that thermal convection would initiate at a length scale on the order of $H E^{1/3}$, or about $20$ m for $E\sim 10^{-15}$. This is $10^{5}$ smaller than the outer core thickness.
Resolving this scale in 3D numerical simulations would necessitate at least $10^{5}$ grid points in each direction, or $10^{15}$ grid points in total.
Today's most resolved  numerical simulations of  the geodynamo have a spatial resolution of about 2 km, which allows to reach $E = 10^{-7}$ \citep{Schaeffer2017}. 

\item
The magnetic Prandtl number is quite small in liquid metals (probably $\sim 10^{-6}$ in Earth's core), but much larger values are being used in dynamo simulations.
Solving the induction equation with a low $P_{m}$ present no intrinsic difficulty: a high magnetic diffusivity usually means smooth variations of the magnetic field and small magnetic fields gradients.
What is difficult is to obtain dynamo action at a low $P_{m}$ in numerical simulations.
The reason for this is easily understood by noting that the magnetic Reynolds number can be written as the product of the Reynolds number and magnetic Prandtl number: $Rm=Re \times P_{m}$.
Since dynamo operation requires reaching  $Rm > \mathcal{O}(10)$, doing this with a low $P_{m}$ requires a high value of $Re$.
This implies the development of turbulent velocity fluctuations down to small length scales, which are difficult to resolve numerically.
\end{enumerate}

\subsection{Successes and challenges}

Though numerical dynamos are still quite far from Earth's conditions in terms of non-dimensional parameters, this does not mean that relevant numerical simulations cannot be done.
In the past few decades, the geodynamo modelling community has been quite successful in strengthening the case for a convectively powered geodynamo.

As already discussed in section \ref{Section:GeodynamoHypothesis}, a major achievement  has been to demonstrate that dynamos can indeed be sustained by rotating convection \citep{Zhang1988,Glatzmaier1995a,Kageyama1995}.
Since the first few numerical models of the dynamo (which were done at relatively high $E$ and $P_{m}$), the non-dimensional parameter space has been explored toward lower $E$ and $P_{m}$, and higher $Ra$.
By performing geodynamo simulations at various values of the governing non-dimensional parameters, scaling laws for typical velocity and magnetic field strength have been obtained.
When extrapolated to the core condition, the predictions of these scaling laws are in reasonable agreement with estimates of the core flow (from inversion of secular variation of the CMB magnetic field) and observed field strength \citep[\textit{e.g.}][]{starchenko2002,olson2006,Christensen2006,christensen2010b}. 
In addition, it has been shown that the magnetic field can be ``Earth like'' in a well-defined region of the parameter space, which includes the estimated state of Earth's core \citep{christensen2010}. 
``Earth-like'' geodynamo simulations also often exhibit polarity reversals.

In spite of the important successes described above, the question of whether numerical simulations do reach a dynamical regime similar to planetary cores has been a constant preoccupation of the geodynamo community.
In other words, are geodynamo simulations ``Earth-like'' for good reasons, or is it just coincidental?
Numerical simulations are still far from the Earth in terms of non-dimensional numbers: the most massive numerical simulations of the geodynamo have reached $E=10^{-7}$, $P_{m}=0.1$, with a Rayleigh number $6\times 10^{3}$ times supercritical  \citep{Schaeffer2017}. 
Though quite impressive, this is still far from the Earth (seven or eight orders of magnitude in terms of $E$, five orders of magnitude in $P_{m}$). 
And most of available geodynamo simulations are significantly further away from Earth's core parameters values.
The main question is whether viscous effects in state of the art simulations play a significant role or not.
\cite{Christensen2006} have argued, from a large set of numerical dynamos ($E\in 10^{-6}-3\times 10^{-4}$, $P_{m}\in 0.06-10$), that the magnetic field strength and mean velocity follow scaling laws which are independent of diffusivities. 
This may suggest that an asymptotic regime has been reached, and that further decreasing the viscosity down to Earth's value would not change the dominant force balance, allowing to use the obtained scaling laws to extrapolate outputs of numerical simulations to planetary cores conditions.
However, several authors have questioned whether viscosity effects in current geodynamo simulations are indeed small enough, and whether diffusivity-free scaling laws correctly describe the available set of numerical geodynamos \citep{stelzer2013,king2013,cheng2016}.
The ratio of magnetic to kinetic energies, which is estimated to be  $\sim 10^{3}-10^{4}$ in Earth's core, is much smaller in numerical models at low ($\lesssim 1$) $P_{m}$: $\mathcal{O}(10)$ at best, and often below 1 \citep{Schaeffer2017}.

\chapter{Energetics of the geodynamo}
\label{Section:Energetics}

\begin{figure}[t]
\centering
\includegraphics[width=0.75\linewidth]{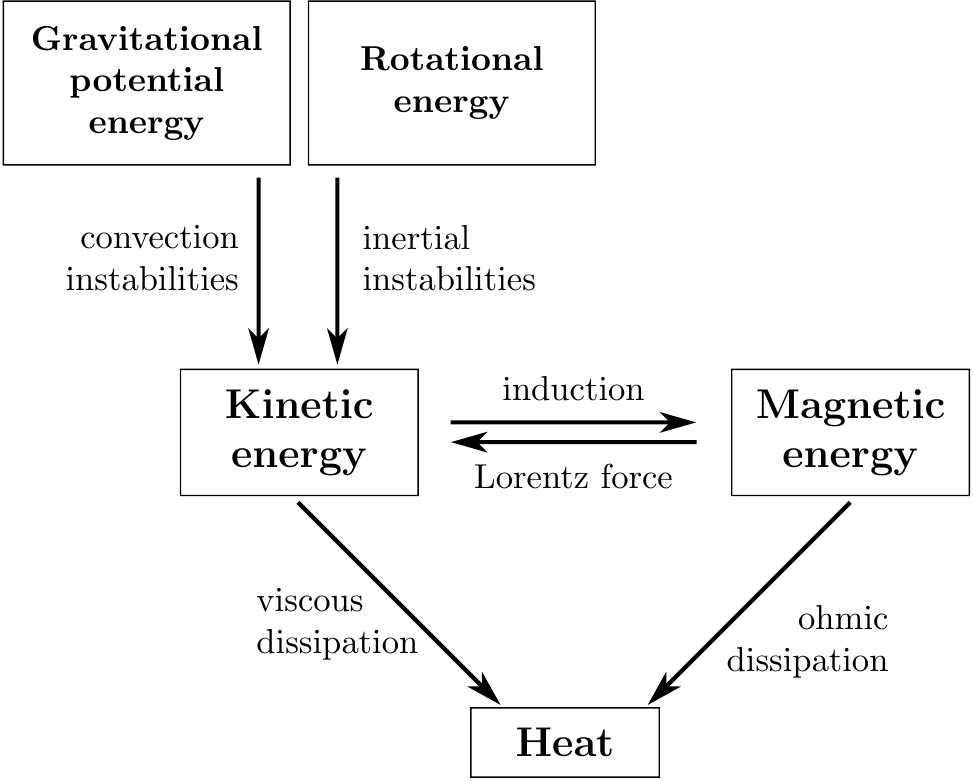}
\caption{Energy flow of the geodynamo.}
\label{Fig:EnergyFlow}
\end{figure}

Figure \ref{Fig:EnergyFlow} shows schematically the energy flow of a dynamo powered by natural convection (thermal or compositional) or astronomical forcing (tides, precession, nutation, or libration).
Natural convection  converts the gravitational potential energy of unstable density gradients into kinetic energy; inertial instabilities excited by astronomical forcing converts rotational energy into kinetic energy.
If the magnetic Reynolds number is large enough for dynamo action, a fraction of this kinetic energy is converted into magnetic energy through induction.
Magnetic energy can be converted back into kinetic energy (through the work of the Lorentz force).
Kinetic and magnetic energies are dissipated into heat through viscous and ohmic dissipation.

Dynamo action requires the rate of release of gravitational or rotational energies to be high enough to maintain the magnetic field against viscous and ohmic dissipation.
The two energy reservoirs are huge: the gravitational potential energy of the Earth is $\simeq 2 \times 10^{32}$ J, and the rotational energy is $\simeq 2\times 10^{29}$ J. 
Only a fraction of these would be enough to maintain Earth's magnetic field for the last 4 Gy if the amount of dissipation is in the range 0.1-3 TW as discussed later in this section.
However, the fact that enough energy is available in Earth's system is by no means a sufficient condition for sustaining the geomagnetic field.
Only a small fraction of this energy can actually be made available to sustain the geomagnetic field; the goal of the energetics approach of the geodynamo is to estimate this fraction.

An equation for the evolution of the magnetic energy $E_{m}=\int_{\mathcal{V}_{\infty}} \frac{B^{2}}{2 \mu_{0}}d{\mathcal{V}}$ can be obtained from Ohm's law and Maxwell's equations, or equivalently from the dot product of the induction equation with $\mathbf{B}$.
Integration over the whole space $\mathcal{V}_{\infty}$ gives 
\begin{equation}
\underbrace{\frac{d}{dt} \int_{\mathcal{V}_{\infty}} \frac{B^{2}}{2 \mu_{0}}d{\mathcal{V}}}_{\substack{\text{Rate of change}\\ \text{of magnetic energy}}} = 
- \underbrace{\int_{\mathcal{V}} \mathbf{u} \cdot \mathbf{f}_{L} d{\mathcal{V}} }_{\substack{\text{power of}\\ \text{Lorentz force}}}
-  \underbrace{\int_{\mathcal{V}} \frac{j^{2}}{\sigma} d{\mathcal{V}}}_{\substack{\text{Ohmic dissipation}\\ \text{ (=Joule heating)}}} ,
\end{equation}
where $\mathbf{f}_{L} $ is the Lorentz force, and $\mathcal{V}$ the volume of the core. 
In this equation, the magnetic energy includes the contribution of the magnetic field outside the core.
On the other hand, the integrals of the power of the Lorentz force and of ohmic dissipation can be restricted to the core volume $\mathcal{V}$ under the assumption that the electric conductivity is equal to 0 outside the core. 
This equation cannot be used alone to predict whether the geodynamo can be dynamically sustained, because the (unknown) velocity field $\mathbf{u}$ appears in the equation.
We therefore need a second conservation equation involving $\mathbf{u}$, which is obtained by taking the dot product of Navier-Stokes equation with $\mathbf{u}$, and again integrating over the volume of the core.
This gives an equation for the time derivative of the total kinetic energy of the core:
\begin{equation}
\frac{d E_{k}}{dt} 
 =
\int_\mathcal{V} \left( \mathbf{f}_{L}+  \rho \mathbf{g} - \boldsymbol{\nabla} P \right) \cdot \mathbf{u}\, d\mathcal{V} 
+ \oint_\mathcal{S} \mathbf (\underline{\tau} \cdot \mathbf{u}) \cdot  \mathbf{n}\, d \mathcal{S} 
- \Phi_{\nu},
\end{equation}
where $\mathcal{S}$ is the surface area of the core-mantle boundary, $\underline{\tau}$ is the deviatoric stress tensor and $\Phi_{\nu}=\int_{\mathcal{V}} \underline{\tau}:\nabla \mathbf{u}\, d\mathcal{V}$ is the viscous dissipation.
There is no power associated with the Coriolis force since $(\rho \Omega \times \mathbf{u}) \cdot \mathbf u = 0$.

Combining the kinetic and magnetic energy equations, we can get rid of the power of the Lorentz force and obtain:
\begin{equation}
\frac{d}{dt}\left( E_{m} + E_{k} \right) = \underbrace{\int_\mathcal{V} \left( \rho \mathbf{g} - \boldsymbol{\nabla} P \right) \cdot \mathbf{u}\, d\mathcal{V} 
+ \oint_\mathcal{S}  (\underline{\tau} \cdot \mathbf{u}) \cdot  \mathbf{n}\, d {\mathcal{S}} }_{\text{\normalsize available power $P_{a}$}}
- \underbrace{( \Phi_{\nu} + \Phi_J)}_{\text{\normalsize $\Phi$}}
\label{Eq:TotalEnergyEquation}
\end{equation}
In this equation:
\begin{enumerate}
\item
The term $\int_{\mathcal{V}} \left(\rho \mathbf{g} - \boldsymbol{\nabla} P \right) \cdot \mathbf{u}\, d{\mathcal{V}} $ is the rate of work of non-hydrostatic processes.
It is equal to 0 if the core is in a hydrostatic state.
In planetary cores, deviations from a hydrostatic state can be due to  density, pressure, or gravity field fluctuations, which can arise from either natural convection or astronomical forcing.
The integral is positive in the case of natural convection; it can be either positive or negative in the case of astronomical forcing.
\item
The term $\oint_{\mathcal{S}} \mathbf (\underline{\tau} \cdot \mathbf{u}) \cdot  \mathbf{n}\, d {\mathcal{S}}$  is the rate of work of deviatoric stresses (viscous stresses) at the CMB.
It is different from zero if the mantle rotates at a different rate than the core (in particular due to precession/nutation/libration) or if the CMB is deformed (tides).
\item
$\Phi$ is the total dissipation due to Joule heating (Ohmic dissipation) and viscous heating (viscous dissipation). It is always positive.
\end{enumerate}

Starting from a state at rest with $E_{k}=E_{m}=0$, the sum of the kinetic and magnetic energy would increase if the sum of the two first terms on the RHS of equation \eqref{Eq:TotalEnergyEquation}, which we denote by $P_{a}$, is positive. 
Increasing $\mathbf{u}$ and $\mathbf{B}$ will increase the dissipation  $\Phi$ up to a point of saturation at which the rate of energy production is on average balanced by dissipation, \textit{i.e.}  $P_{a} \simeq \Phi$.
In a statistically steady state, time averaging equation \eqref{Eq:TotalEnergyEquation} gives
\begin{equation}
\left\langle  \Phi \right\rangle = P_{a} = \left\langle \int_{\mathcal{V}} \left( \rho \mathbf{g} - \boldsymbol{\nabla} P \right) \cdot \mathbf{u}\, d{\mathcal{V}} 
+ \oint_{\mathcal{S}} \mathbf (\underline{\tau} \cdot \mathbf{u}) \cdot  \mathbf{n}\, d {\mathcal{S}} \right\rangle.
\end{equation}
From an energetic point of view, the geodynamo problem amounts to estimate whether the rate  of energy energy release $ P_{a}$ %(the RHS term) 
is large enough to sustain Earth's magnetic field against dissipation.
To answer this question, we need to estimate: (i) how dissipative is the geodynamo, and (ii)  what is the available power $ P_{a}$  in the core.

\subsection{How dissipative is the geodynamo?}

The amount of ohmic dissipation in the core is given by
\begin{equation*}
\Phi_{J} = \int_{\mathcal{V}} \frac{j^{2}}{\sigma} d{\mathcal{V}}= \int_{\mathcal{V}} \frac{|\boldsymbol{\nabla} \times \mathbf{B}|^{2}}{\sigma\mu_{0}^{2}} d{\mathcal{V}}.
\end{equation*}
To estimate how dissipative is the geodynamo, we thus need to know how $\mathbf B$ varies spatially in the core.
This is difficult. 
As discussed in section \ref{Section:GeomagneticFieldGeometry}, direct observations of the core magnetic field are restricted to the poloidal field at the core-mantle boundary, up to spherical harmonic degree $l_\mathrm{max}=13$, which corresponds to a spatial wavelength of about 1600 km.
We have no direct observation of the harmonic components of the poloidal field at $l > 13$, and no constraint on the spatial variations of the toroidal part of the magnetic field. 
Numerical simulations of the geodynamo suggest that ohmic dissipation is dominated by the contributions of smaller scale components of the magnetic field \citep{Roberts2003}, so estimating the geodynamo dissipation from direct observations seems hopeless. 
Published estimates of $\Phi_{J}$ have been obtained  from extrapolations of numerical or experimental dynamos results \citep{Buffett2002,Roberts2003,Christensen2004,christensen2010,stelzer2013}, and range from $\sim 0.1$ TW to  several TW.

\subsection{Estimating the available power}

The energetics approach of the geodynamo problem has so far been restricted to the case of a convectively driven dynamo; the case of mechanical forcing has not been treated in a rigorous and usable way, in part because of the difficulty of estimating the $\int_{\mathcal{V}} \left(\rho \mathbf{g} - \boldsymbol{\nabla} P \right) \cdot \mathbf{u}\, d{\mathcal{V}} $ term.
We will thus restrict ourselves to the case of a convectively driven dynamo. 
In this situation the surface integral in equation \eqref{Eq:TotalEnergyEquation} is equal to 0.

Two different methods have been classically used: either work on the entropy budget rather than on equation \eqref{Eq:TotalEnergyEquation} \citep[\textit{e.g.}][]{Gubbins2004,Labrosse2015}, or estimate directly the contribution of convective motions on $P_{a}$. 
These two approaches are equivalent \citep{Lister2003} and in both cases require assumptions on the state of the core, which is usually assumed to be continuously well mixed by convective motions.
Here we use the later approach, which is perhaps a bit more physically intuitive, and follow the approach of \cite{Buffett1996}. 
We will only sketch the derivation of the model; the details can be found in \cite{Buffett1996}. 

Denoting by $\psi$ the gravitational potential, with $ \mathbf{g}=-\boldsymbol{\nabla} \psi$, one can show by using mass conservation that 
\begin{align}
P_{a} &=  \int_{\mathcal{V}} \left(- \rho \boldsymbol{\nabla} \psi - \boldsymbol{\nabla} P \right) \cdot \mathbf{u}\, d\mathcal{V}  \\
&= - \int_{\mathcal{V}} \psi(r) \frac{\partial \rho}{\partial t} d\mathcal{V}
\end{align}
where $\partial \rho/\partial t$ is the rate of change of the density due solely to convective re-arrangement.
The next step is to calculate the change of density in the core associated to the convective re-distribution of heat and chemical elements, under the assumption that convection keeps the core well mixed and isentropic.
By doing this, one can write $P_{a}$ as
\begin{equation}	
P_{a} = 
\underbrace{  \left( \beta \rho {F}_\mathrm{icb}  +  \frac{\alpha}{c_{p}} Q_\mathrm{icb}^\mathrm{>ad}   \right)   \left( \bar \psi - \psi_\mathrm{icb}\right) }_{\text{Contribution of IC solidification}} 
+\underbrace{ \frac{\alpha}{c_{p}} Q_\mathrm{cmb}^\mathrm{>ad} \left( \psi_\mathrm{cmb} - \bar \psi\right) }_{\text{Contribution of CMB flux}}
\label{Eq:ConvectivePower1}
\end{equation}
where $Q_\mathrm{icb}^\mathrm{>ad}=Q_\mathrm{icb} - Q_\mathrm{icb}^\mathrm{ad}$ and $Q_\mathrm{cmb}^\mathrm{>ad}=Q_\mathrm{cmb} - Q_\mathrm{cmb}^\mathrm{ad}$ are the super-adiabatic heat flux at the inner core boundary (ICB) and core-mantle boundary (CMB), ${F}_\mathrm{icb} =4\pi r_\mathrm{icb} ^{2} \dot  r_\mathrm{icb}  c$ is the flux of light elements concentration (in wt.\%.m$^{3}$.s$^{-1}$) at the ICB ($c$ being the concentration in light elements of the outer core),  
$\bar \psi$ is the mass-averaged value of the gravitational potential in the outer core, and $\psi_\mathrm{cmb}$ and $\psi_\mathrm{icb}$ are the values of the gravitational potential at the CMB and ICB.
The factors  $\bar \psi - \psi_\mathrm{cmb}$ and $\psi_\mathrm{icb} - \bar \psi $ come from the fact that density perturbations originating from either the CMB or ICB are redistributed over the whole core. 
The superadiabatic heat flux at the ICB happens to be well approximated by the release of latent heat due to inner core solidification, $Q_\mathrm{icb}^\mathrm{>ad} \simeq 4\pi r_\mathrm{icb} ^{2} \rho L  \dot  r_\mathrm{icb} $, where $\dot  r_\mathrm{icb} $ is the time-derivative of the inner core radius $r_\mathrm{icb} $, and $L$ is the latent heat of iron at core conditions.
Since ${F}_\mathrm{icb} $ is also proportional to $\dot r_\mathrm{icb} $, the whole ICB contribution is proportional to the rate of growth of the inner core, which is itself controlled by the rate at which heat is extracted from the core.

\begin{figure}
\centering
\includegraphics[width=0.49\linewidth]{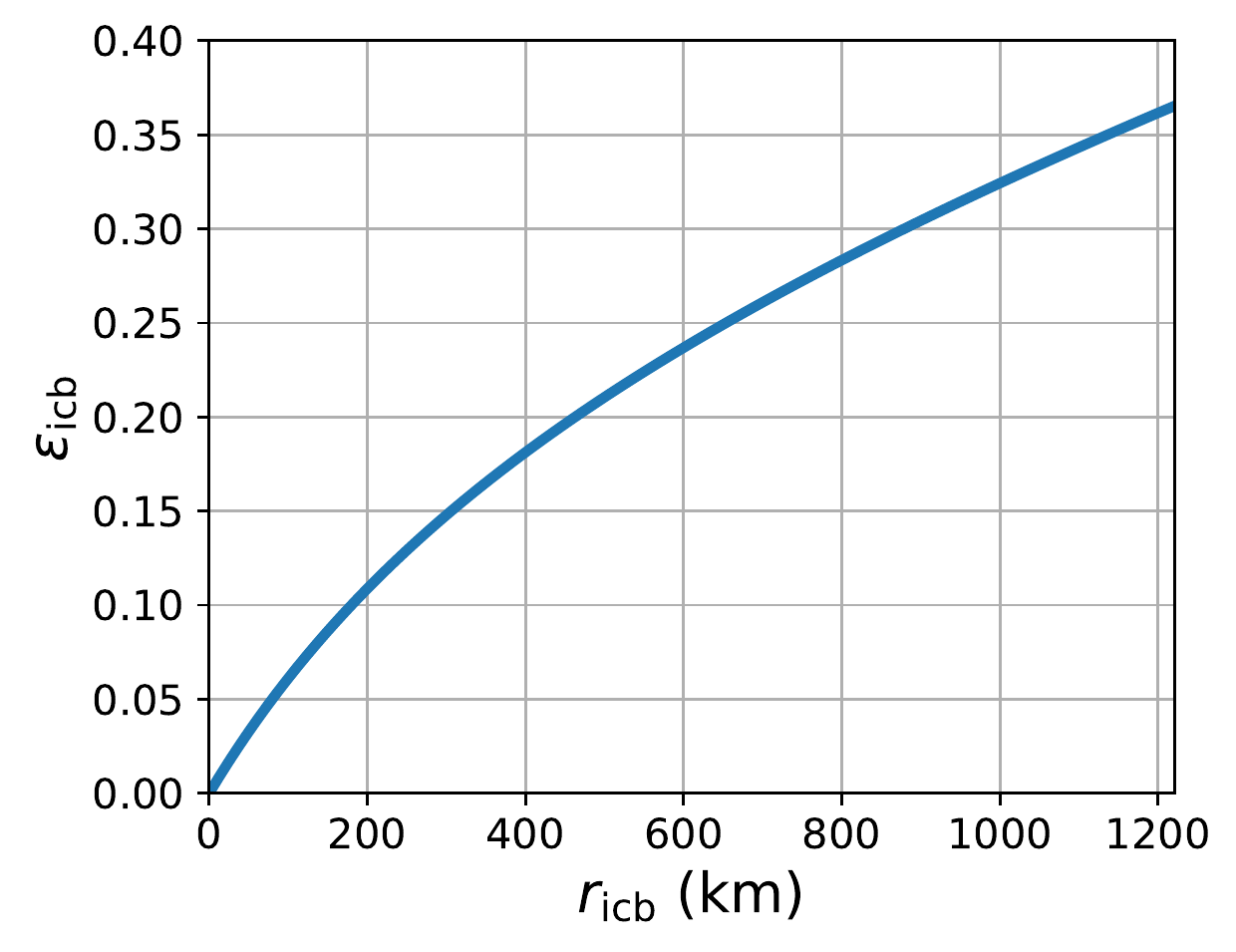} \\
\caption{\textit{Dynamo efficiency} $\epsilon_\mathrm{icb}$ associated with compositional and thermal buoyancy fluxes from the inner core boundary, as a function of inner core size (calculated from \cite{Buffett1996}).}
\label{Fig:DynamoEfficiencies}

\end{figure}

The terms in equation \eqref{Eq:ConvectivePower1} are all tied to the core thermal evolution and controlled by the heat flux at the CMB.
To use equation \eqref{Eq:ConvectivePower1} in a predictive way, one needs to relate the rate of inner core growth to the heat flux at the CMB, which can be done by considering the energy balance of the core (coming from the first law of thermodynamics), which simply states that the heat flux extracted from the core at the CMB is equal to the sum of sensible heat released by the cooling of the core, latent heat of inner core crystallisation, and compositional energy (associated with the mixing in the core of light elements released by inner core solidification).
Doing this allows to write equation \eqref{Eq:ConvectivePower1} as
\begin{equation}
P_{a} =  \epsilon_{\mathrm{icb}} Q_\mathrm{cmb} + \epsilon_\mathrm{cmb}   Q_\mathrm{cmb} ^{>\mathrm{ad}},
\label{Eq:ConvectivePower2}
\end{equation}
where $\epsilon_{\mathrm{icb}}$ and $\epsilon_\mathrm{cmb}$ are the so-called \textit{dynamo efficiencies}.\footnote{The CMB efficiency is usually defined in another way, such that $P_{a} = \left( \epsilon_{\mathrm{icb}} + \epsilon_\mathrm{cmb}  \right) Q_\mathrm{cmb}$. This has the advantage of linking $P_{a}$ directly to $Q_\mathrm{cmb}$, but then $\epsilon_\mathrm{cmb}$ is itself a function of $Q_\mathrm{cmb}$ (it includes a factor $Q_\mathrm{cmb} ^{>\mathrm{ad}}/Q_\mathrm{cmb}$).}
Expressions for these efficiencies can be found in \cite{Buffett1996} or \cite{Lister2003}.
One important point is that $\epsilon_{\mathrm{icb}}$ is an increasing function of inner core size (figure \ref{Fig:DynamoEfficiencies}), which means that the contribution of compositional convection increases with inner core size.
An approximate expression of $\epsilon_\mathrm{cmb}$ is 
\begin{equation}
\epsilon_\mathrm{cmb} \simeq \frac{1}{5}\frac{\alpha g_\mathrm{cmb} r_\mathrm{cmb}}{c_{p}} \simeq 0.1.
\end{equation}

\begin{figure}
\includegraphics[width=\linewidth]{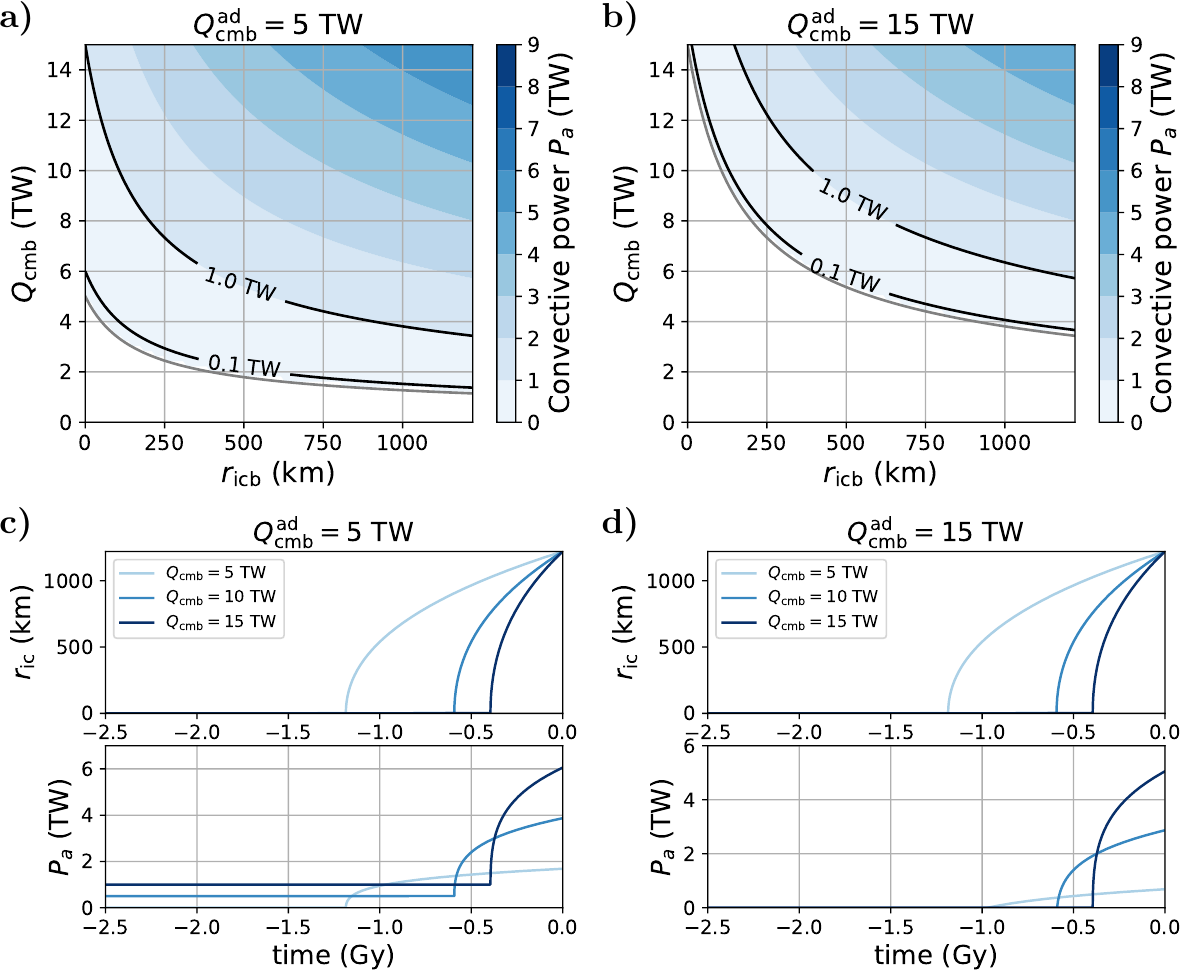}
\caption{\textbf{a)} convective power $P_{a}$ as a function of $r_\mathrm{icb}$ and $Q_\mathrm{cmb}$, with $Q_\mathrm{cmb}^\mathrm{ad}=5$ TW; \textbf{b)} same as \textbf{a} with $Q_\mathrm{cmb}^\mathrm{ad}=15$ TW; \textbf{c)} evolution with time of inner core size $r_\mathrm{icb}$ and convective power $P_{a}$, for $Q_\mathrm{cmb}^{\mathrm{ad}}=5$ TW and $Q_\mathrm{cmb}$ equal to 5, 10, or 15 TW; \textbf{d)} same as \textbf{c} with  $Q_\mathrm{cmb}^{\mathrm{ad}}=15$ TW.
}
\label{Fig:ConvectivePower}

\end{figure}

Having obtained estimates of $\epsilon_{\mathrm{icb}}$ and $\epsilon_\mathrm{cmb}$, it is now possible to estimate the convective power $P_{a}$. 
One key parameter is the heat flux conducted along the adiabat, $Q_\mathrm{cmb}^{\mathrm{ad}}$, which depends mostly on the thermal conductivity of the outer core (see equation \eqref{Eq:AdiabaticFlux}).
Figures \ref{Fig:ConvectivePower}a and b show $P_{a}$ as a function of the CMB heat flux and inner core size, for $Q_\mathrm{cmb}^{\mathrm{ad}}$ equal to either 5 TW or 15 TW.
The 5 TW and 15 TW values are representative of the low ($\sim 30$ W.m$^{-1}$.K$^{-1}$) and high ($\sim 100$ W.m$^{-1}$.K$^{-1}$) thermal conductivity estimates. 
Figures \ref{Fig:ConvectivePower}c and d show the evolution with time of inner core size $r_\mathrm{icb}$ and convective power $P_{a}$, for $Q_\mathrm{cmb}^{\mathrm{ad}}$ equal to either 5 TW or 15 TW, and imposed CMB heat flux $Q_\mathrm{cmb}$ equal to 5, 10, or 15 TW, the current CMB heat flux being thought to be in the range 5-15 TW \citep{Lay2006}.
On figures \ref{Fig:ConvectivePower}a and b, the grey line corresponds to the limit above which $P_{a}> 0$. 

One can see that a positive $P_{a}$ requires the CMB heat flux to be above a threshold value, which is equal to $Q_\mathrm{cmb}^{\mathrm{ad}}$ at $r_\mathrm{icb}=0$, and decreases with increasing $r_\mathrm{icb}$ due to the increasing importance of the ICB buoyancy flux.
From equation \eqref{Eq:ConvectivePower2}, a dynamo dissipating at a given rate $\Phi_{0}$ requires a CMB heat flux
\begin{equation}
Q_\mathrm{cmb} (P_{a}=\Phi_{0}) = \frac{\epsilon_\mathrm{cmb} Q_\mathrm{cmb} ^{\mathrm{ad}} + \Phi_{0}}{\epsilon_\mathrm{cmb}+\epsilon_\mathrm{icb}} .
\label{eq:CriticalQcmb}
\end{equation}
The CMB heat flux above which $P_{a}>0$ is given by setting $\Phi_{0}=0$ in equation \eqref{eq:CriticalQcmb}.
On figures \ref{Fig:ConvectivePower}a and b are also shown the $Q_\mathrm{cmb}$ values corresponding to $\Phi_{0}=0.1$ TW and 1 TW.

One can see that in the current state of the core, the convective power can easily be of a few terawatts if the CMB heat flux is in the range 5-15 TW as currently believed \citep{Lay2006}. 
There is therefore no difficulty in powering the geodynamo 
with the inner core at its current size. 
However, powering the geodynamo happens to be more problematic when the inner core is smaller, and before its nucleation.
With a low core thermal conductivity and $Q_\mathrm{cmb} ^{\mathrm{ad}}=5$ TW, the convective power $P_{a}$ would be of a few tens of TW. 
Driving the geodynamo with thermal convection seems therefore possible, though the convective power would be much lower than at present. 
Whether this would have a significant impact on the large scale part of the geomagnetic field remains an open question.  
If the high estimates of the core thermal conductivity are correct, the CMB heat flux may well be sub-adiabatic. 
In this situation thermal convection is not possible, and driving the dynamo requires another source of motion and energy.
Possible additional sources of energy include exsolution of light elements from the core \citep{ORourkeStevenson2016,Badro2016,ORourke2017} and astronomical forcing \citep{andrault2016}.
In addition, adding some radioactive heating in the core (possible due to $^{40}$K) would help by decreasing the rate of cooling of the core and increasing the age of the inner core \citep{Labrosse2015}.

%--------------------------------------------------------------------------------------------------------------------------------------------%
%--------------------------------------------------------------------------------------------------------------------------------------------%
\chapter{Inner core dynamics}
\label{Section:InnerCore}
%--------------------------------------------------------------------------------------------------------------------------------------------%
%--------------------------------------------------------------------------------------------------------------------------------------------%

The Earth's inner core is the deepest layer of our planet: a 1221-km-radius sphere of solid iron-alloy surrounded by molten metal. Its existence was unknown until the first observations of seismic reflexions at the inner core boundary by Inge Lehmann, in one of the shortest-title paper ever: P' \citep{Lehmann1936}. The arrivals of P-waves in the core shadow zone, where P-waves are refracted away by the presence of the 3600-km-radius core, have been explained by the existence of a new discontinuity inside the core, the inner core boundary. 

\begin{figure}[t]
\centering
\includegraphics[width=0.65\linewidth]{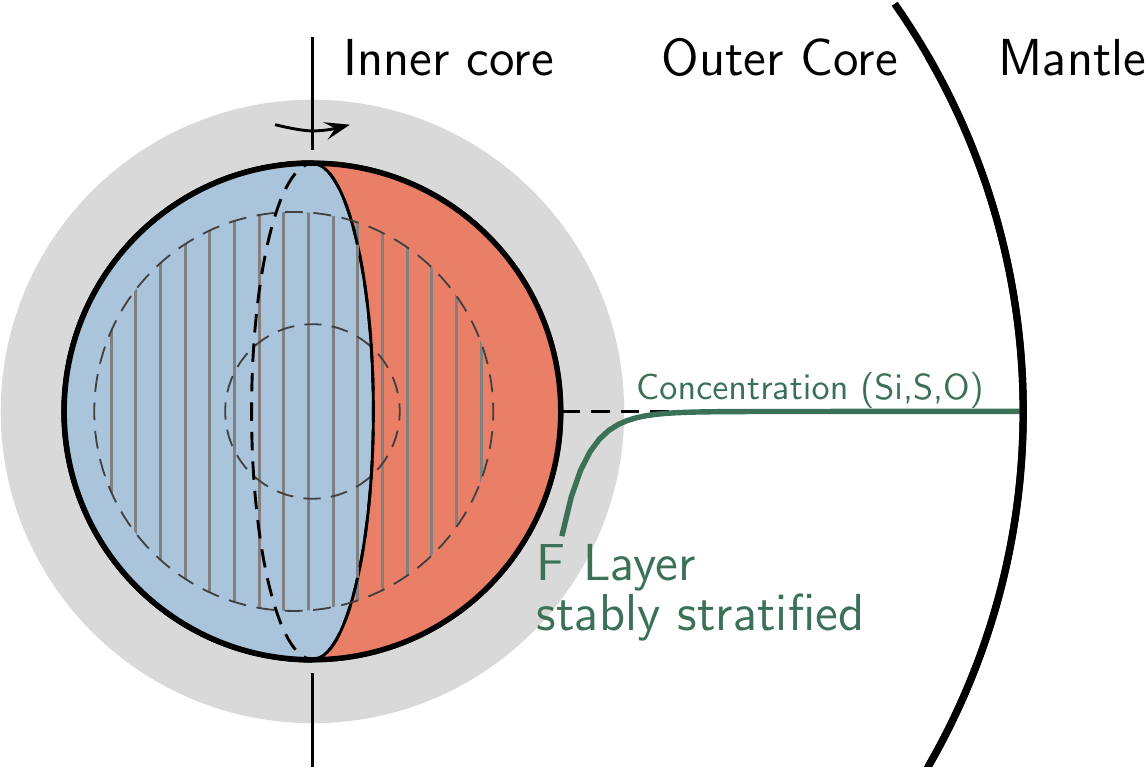}
\caption{A schematic view of the structure of the inner core.
In addition to a global elastic anisotropy oriented parallel to the axis of rotation
of the Earth (grey lines), the inner core has radial and horizontal variations of its seismic properties. 
There is a strong asymmetry between its western and eastern hemispheres
(approximately defined by the Greenwich meridian), which have different seismic waves propagation velocity, attenuation, and degree of anisotropy. 
Anisotropy is weak or non-existent near the surface of the inner core, and increases in depth. Finally, the inner core is
surrounded by a layer that appears to be stably stratified, the F-layer.
}
\label{Fig:ICStructure}
\end{figure}

Since these first observations, the study of seismic waves travelling through the inner core and normal modes sampling the deepest layers have provided a blurry image of the inner core structure (figure \ref{Fig:ICStructure}). \citet{Birch1940} and \cite{Jacobs1953} have proposed that the inner core is a solid sphere of the same metal constituting the outer core, while the actual solidity has been demonstrated only several years later by \citet{Dziewonski1971}. \citet{Poupinet1983} were the first to note the different propagation velocities of waves travelling parallel to the rotation axis and perpendicular to it. This anisotropy of seismic properties has since been extensively studied, demonstrating the existence of a complex structure of the inner core. Among the most robust and surprising features of the inner core structure, we can cite two: a strong anisotropy for the bulk of the inner core, and an uppermost layer of the inner core with a strong hemispherical dichotomy in P-waves velocity, but no detectable anisotropy. 

The existence of anisotropy in the inner core is an evidence for crystal orientation within the bulk of the inner core (iron crystals being elastically anisotropic). Such crystal orientation is the main motivation for studying flows in the inner core, as lattice-preferred orientation (LPO) may be deformation-induced \citep{karato2012deformation}. Thus, it is thought that the observed structure may be an evidence for flows within the inner core, likely to be combined with initial crystallisation-induced LPO. 

We will discuss here the constitutive equations and some of the aspects of inner core dynamics, focusing on the flow forced in the inner core by the magnetic field diffused from the outer core. 

\section{Constitutive equations}

We consider an incompressible fluid in a spherical domain of radius $r_\text{icb}$,  with a newtonian rheology and uniform viscosity $\eta$.  Neglecting inertia and rotation, the equation of continuity and the conservation of momentum are written as 
\begin{align}
    \nabla \cdot \mathbf{u} &= 0, \label{eq:continuity}\\
  \mathbf{0}&=-\boldsymbol{\nabla} p+\frac{\eta}{\rho_s}\boldsymbol{\nabla} ^2 \mathbf{u} +\mathbf{F},  \label{eq:momentumequation_1} 
\end{align}
where  $\mathbf{F}$ denotes volume forces,  $p$ is the dynamic pressure (including gravity potential) and $\mathbf{u}$ the velocity field. 

Among the volume forces, the buoyancy force is written as $\mathbf{F}_\text{buoyancy} = \Delta\rho \mathbf{g}$, where $\Delta \rho$ is the density difference compared to a neutral density profile and $\mathbf{g} = g_\text{icb}r/r_\text{icb}\mathbf{e}_{r}$ the acceleration of gravity. $g_\text{icb}$ is the acceleration of gravity at the surface of the inner core.

In this model, the density variations only play a role in the buoyancy term, and they can be related to variations in temperature or concentration in light elements (Si, S, O, ...) compared to a reference state. We take as the reference temperature profile the adiabatic temperature profile $T_\text{ad}(r)$ anchored at the melting temperature at the radius of the inner core $r_\text{icb}$. The deviation of the temperature field compared to this reference profile is the potential temperature $\Theta = T-T_\text{ad}$. The potential composition field is defined as $C=c^{s}-c^{s}_\text{icb}$, where $c^{s}$ is the concentration of light elements in the inner core and $c^{s}_\text{icb}$ its value at the inner core boundary. The density variations are thus $\rho \alpha_T \Theta$ or $\rho\alpha_C C$ for respectively thermal or compositional stratification, where $\alpha_T$ and $\alpha_C$ are the thermal and compositional expansion coefficients. As both potential temperature and potential composition are solutions of an advection-diffusion equation and are both set to zero by construction at the inner core boundary, we will consider a general equation for a quantity $\chi$, representing either of these quantities. $\chi$ is solution of
\begin{equation}
\frac{\partial \chi}{\partial t}+\mathbf{u} \cdot \boldsymbol{\nabla} \chi = \kappa_\chi \nabla ^  2\chi + S_\chi(t), 
\label{Eq:Transport_chi}
\end{equation}
where $\kappa_\chi$ is the diffusivity and $S\chi(t)$ a source term built from the evolution of the reference profile as 
\begin{align}
S_T=& \kappa_T \nabla ^ 2 T_\text{ad}-\frac{\partial T_\text{ad}}{\partial t }, \label{Eq:ST} \\
S_C = &-\frac{d c^{s}_\text{icb}}{d t}. \label{Eq:SC}
\end{align}

The continuity and momentum equations can be solved using a poloidal-toroidal decomposition of the velocity field $\mathbf{u}=\nabla \times (T\mathbf{r}) + \nabla \times \nabla \times (P\mathbf{r})$, where $\mathbf{r}=r \mathbf{e}_{r}$ is the position vector and $T$ and $P$ respectively the toroidal and poloidal components. In the following, we will only consider boundary conditions with a zero vertical vorticity and volume forces without toroidal components. The flow is thus expected to have only non-zero poloidal component, and applying $\mathbf{r}\cdot (\nabla \times \nabla \times )$ to equation \eqref{eq:momentumequation_1}, we obtain
\begin{equation}
    0 = -(\nabla ^ 2)L^2 P + \mathbf{r}\cdot (\nabla \times \nabla \times \mathbf{F}),
\end{equation}where $L^2$ is the Laplace horizontal operator defined as
\begin{equation}
L^2 =  -  \frac{1}{\sin \theta} \frac{\partial}{\partial \theta} \left ( \sin \theta \frac{\partial}{\partial 
\theta} \right ) -\frac{1}{\sin ^2\theta} \frac{ \partial ^2 }{\partial \phi ^2 }.
\end{equation}
One could note that for a volume force of the form $\mathbf{F}=F_r r\mathbf{e}_{r}$ such as the buoyancy forces, the second term of the right-hand side of the equation simplifies as $\mathbf{r}\cdot (\nabla \times \nabla \times \mathbf{F})=L^2F_r$.  Splitting the volume force term as one term for the buoyancy forces and one term for the other forces, we have $\mathbf{F} = \rho \alpha_\chi \chi r g_\text{icb}\mathbf{e}_{r}/r_\text{icb} + \mathbf{F}_\text{volume}$. 
Expanding the scalar fields $P$ and $\chi$ with horizontal spherical harmonics $Y_l ^ m$ satisfying $L^ 2 Y_l^ m = - l(l+1) Y_l^m$ of degree $l$ and order $m$, as $P=\sum_{l,m}P_l^ mY_l ^m$ and $\chi =\sum_{l,m}\chi_l^ mY_l ^m$, the equation of interest is eventually written for each $(l, m)$ as
\begin{equation}
\text{D}_l^ 2 P_l^ m  -\rho \alpha_\chi g_\text{icb}\frac{r}{r_\text{icb}}  \chi_l^m  - \frac{f_l^ m}{l(l+1)}=0, 
\end{equation}where $f_l^ m$ is the spherical harmonics decomposition of the poloidal component of the volume force $\mathbf{F}$, and $\text{D}_l$ a second-order differential operator defined as 
\begin{equation}
   \text{D}_l =  \frac{\text{d}}{\text{d} r^2}+\frac{2}{r}\frac{\text{d}}{\text{d}r} -\frac{l(l+1)}{r^2}.
\end{equation}

\subsection{Boundary conditions}
\label{BC}

The inner core boundary is a crystallisation front, where the iron-alloy of the outer core freezes due to the slow secular cooling of our planet. Its exact position is determined by the intersection of the melting temperature profile of the iron-alloy and the temperature profile in the core. Any solid material pushed dynamically further from this intersection would melt, while any liquid pushed inward would freeze. The timescale $\tau_\phi$  involved in the freezing or melting of a small topography at the inner core boundary can be estimated from the timescale needed by  outer core convection to extract the latent heat released by crystallisation. We note $h$ the topography at the inner core boundary.

From continuity of stress at the ICB, the mechanical boundary conditions are written at $r=r_\text{icb}(t)$. We consider that the dynamical topography $h$ is small (compared to the horizontal wavelength) and that the vector normal to the boundary is close to the radial unit vector. The tangential and normal components of the stress tensor are written as
\begin{align}
\tau_{r \theta} = & \eta \left [ r\frac{\partial }{\partial r } \left ( \frac{u_\theta}{r}\right ) +\frac{1}{r}\frac{\partial u_r}{\partial \theta}\right ],     \\
\tau_{r \phi} = & \eta \left [ r\frac{\partial }{\partial r } \left ( \frac{u_\phi}{r}\right ) +\frac{1}{r\sin \theta}\frac{\partial u_r}{\partial \phi}\right ],     \\
\tau_{rr} = &2 \eta \frac{\partial u_r}{\partial r} -p, 
\end{align}
where $p$ here is the total pressure. 
The viscosity of the outer core being much smaller than the viscosity of the inner core, we can assume tangential stress free conditions.
The boundary conditions are then written as $\tau_{r\theta}(r=r_\text{icb}) = \tau_{r\phi}(r=r_\text{icb}) = 0$ and continuity of $\tau_{rr}$ across the ICB, which for a small topography amounts to state that the normal stress on the inner core side at $r=r_\text{icb}$ is equilibrated by the weight of the topography:
\begin{equation}
    \underbrace{2 \eta \frac{\partial u_r}{\partial r} - p'}_{\text{normal stress}} =  \underbrace{\Delta \rho g h}_{\substack{\text{topography}\\ \text{weight}}}, 
\end{equation}
$p'$ being the dynamical pressure on the inner core side of the ICB. 

To close the system of equations, we consider the time evolution of the topography.  The topography can be formed by deformation of the inner core boundary by the underlying flow $u_r-\dot{r}_\text{icb}$ and is eroded by phase change, such that we can write that $\text{D}h/\text{D}t=u_r-\dot{r}_\text{icb} + V_r$, where $V_r$ is the velocity of phase change in the radial direction. $V_r$ at first order is $-h/\tau_\phi$, where $\tau_\phi$ is a typical timescale for the phase change. Considering a dynamical equilibrium for the topography, we obtain $u_r-\dot{r}_\text{icb} = h/\tau_\phi$ and the continuity of normal stress is written as 
\begin{equation}
-\Delta \rho g_\text{icb}\tau_\phi (u_r-\dot{r}_\text{icb}) -2\eta \frac{\partial u_r}{\partial r} + p'=0. 
\end{equation}

\subsection{Non-dimensionalisation and final set of equations}

The governing equations are made dimensionless using characteristic scales for time, length, velocity, pressure and $\chi$ (potential temperature or composition) as, respectively, the diffusion time $r_\text{icb}^2/(6\kappa_\chi)$, its radius $r_\text{icb}$, $\kappa_\chi/r_\text{icb}$, $\eta\kappa_\chi / r^2_\text{icb}$ and $S_\chi r_\text{icb}^2/(6\kappa)$.
Using the same symbols for dimensionless quantities, the non-dimensional set of equations is 
\begin{align}
\boldsymbol{\nabla} \cdot \mathbf{u} = 0, \label{eq:divu_adim}\\
-\boldsymbol{\nabla} p'+ Ra\, \chi\, \mathbf{r} + \boldsymbol{\nabla} ^2 \mathbf{u} + \mathbf{F}_\text{volume}=0, \label{eq:ic_adim_mom}\\
\frac{\partial \chi}{\partial t} = \nabla ^2 \chi - \mathbf{u} \cdot \boldsymbol{\nabla} \chi +6. \label{eq:temperature_adim}
\end{align}
The dimensionless number $Ra$ is a Rayleigh number expressed as 
\begin{equation}
Ra = \frac{\alpha_\chi \rho g_\text{icb}S_\chi r_\text{icb}^5}{6\kappa_\chi^2\eta}.\label{Ra_ic}
\end{equation}
The momentum equation \eqref{eq:ic_adim_mom} can also be written for the poloidal decomposition in spherical harmonics as 
\begin{equation}
    \text{D}_l^ 2 P_l^ m  - Ra\, \chi_l^m  - \frac{f_l^ m}{l(l+1)}=0. 
\end{equation}

The last step is to express the boundary conditions in term of the poloidal decomposition and in non-dimensional form. Noting that $u_r = L^2P/r$ and that the horizontal integration of the momentum equation taken at $r=1$ gives $-p' +\partial (r\nabla ^2 P)/\partial r = \text{cste}$, the stress free condition takes the form
\begin{equation}
    \frac{\text{d}^2 P_l^m}{\text{d}r^2}+[l(l+1)-2]\frac{P_l^m}{r^2}=0,
\end{equation}
and the normal stress balance is
\begin{equation}
    r^2 \frac{\text{d}^3 P_l^m}{\text{d}r^3}-3l(l+1)\frac{\text{d}P_l^m}{\text{d}r} = \left [ l(l+1) \mathcal{P} -\frac{6}{r}\right ]P_l^m,  
\end{equation}
where $\mathcal{P}$ is a dimensionless number comparing the timescale of viscous relaxation of the boundary $\eta / \Delta \rho g_\text{icb} r_\text{icb}$ and the time scale of phase change $\tau_\phi$, \textit{de facto} quantifying the permeability of the inner core boundary. It is defined as
\begin{equation}
    \mathcal{P} = \frac{\Delta \rho g_\text{icb} r_\text{icb} \tau_\phi }{\eta}. 
\end{equation}

\section{Unstable or stable stratification in the inner core?}

The core crystallises from the center outward because the solidification temperature of the core mixture increases with depth
faster than the (adiabatic) core geotherm \citep{Jacobs1953}. 
One consequence of this solidification mode is that the inner core is cooled from above, a configuration which is potentially prone to thermal convection. 
Thermal convection further requires  the inner core temperature profile to be superadiabatic, which depends on a competition between extraction of the inner core internal heat by diffusion and advection, and cooling at the ICB.
Equation \eqref{Eq:Transport_chi} shows that superadiabaticity in the inner core (\textit{i.e.} a potential temperature increasing with depth) requires $S_{T}$ to be positive.
Fast cooling and a low inner core thermal diffusivity ($S_{T}>0$) promotes superadiabaticity; slow cooling and high thermal diffusivity ($S_{T}<0$) results in a stable thermal stratification.

The expression of $S_T$ (equation \ref{Eq:ST}) can be rewritten by writing the time derivative of the temperature at the ICB as a function of the rate of inner core growth \citep{Deguen2011a}.
This gives 
\begin{equation}
    S_T = - \frac{1}{r_\text{icb}} \left.\frac{\text{d}T_\mathrm{ad}}{\text{d}r}\right|_\text{icb} \left ( \left [ \frac{\text{d}T_s}{\text{d}T_\text{ad}}-1 \right ]r_\text{icb}\dot{r}_\text{icb} - 3\kappa_{T} \right ) , 
\end{equation}
where $\text{d}T_s / \text{d}T_\text{ad}$ is the ratio of the Clapeyron slope to the adiabat. It is then straightforward that the term $S_{T}$ is positive only if
\begin{equation}
    \frac{d r_\text{icb}^{2}}{dt} > \frac{6\kappa_{T}}{\dfrac{\text{d}T_s}{\text{d}T_\text{ad}}-1}. 
\end{equation}
Unfortunately, the uncertainties on the thermal conductivity and inner core growth rate are such that the sign of $S_{T}$ is not known with much certainty.
The high value of the thermal conductivity currently favoured \citep{de-Koker2012,pozzo2012,Gomi2013,gomi2016} results in a stable thermal stratification.

The inner core may also have developed a compositional stratification. 
The concentration in light elements of newly crystallised solid, $c^{s}_\text{icb}$, is linked through the partition coefficient $D$ to the concentration in the liquid from which it crystallises, $c^{l}_\text{icb}$, as
\begin{equation}
c^{s}_\text{icb} = D c^{l}_\text{icb},
\end{equation}
while its derivative with respect to inner core size is 
\begin{align}
\frac{d c^{s}_\text{icb}}{d r_\mathrm{icb}} &= D \frac{d c^{l}_\text{icb}}{d r_\mathrm{icb}} + c^{l}_\text{icb} \frac{d D}{dr_\mathrm{icb}}, \\
&=  D c^{l}_\text{icb} \left[ \frac{d \ln c^{l}_\text{icb}}{d r_\mathrm{icb}} + \frac{d \ln D}{d r_\mathrm{icb}}  \right].  \label{Eq:dcicb}
\end{align}
A stable compositional stratification would develop if $c^{s}_\text{icb}$ increases with increasing inner core size (more light elements in the upper part of the inner core); conversely, an unstable stratification would develop if $c^{s}_\text{icb}$ decreases with increasing inner core size. 
The first term on the right-hand-side term of equation \eqref{Eq:dcicb} is very likely positive, due to the  gradual enrichment of the outer core in light elements expelled during crystallisation. 
The second term depends on how $D$ varies with pressure and temperature along the $(P,T)$ path defined by the evolution of the position of the inner core boundary.
\textit{Ab initio} calculations \citep{Gubbins2013} suggest that it is negative, and of the same order of magnitude as the first term on the r.h.s. of equation \eqref{Eq:dcicb}.
The relative importance of the two terms depend on the exact composition of the inner core \citep{Gubbins2013,Labrosse2014}, which is not very well constrained.
Again, it is difficult to be definitive: given our current knowledge of the composition of the core and of the partitioning behaviour of its light elements, stable and unstable compositional stratifications seem equally plausible.

Natural (thermal or compositional) convection in the inner core has been studied in details \citep[\textit{e.g.}][]{Weber1992,Wenk2000b,Alboussiere2010, Monnereau2010,Deguen2011a,Cottaar2012,Deguen2013a, Mizzon2013,deguen2018}.
In the limit $\mathcal{P}\rightarrow 0$, the convection instability takes the form of a translation of the inner core, with melting on one hemisphere and solidification on the other  \citep{Alboussiere2010, Monnereau2010, Deguen2013a, Mizzon2013,deguen2018}.
We will focus here on the case of neutral or stable stratification, and consider the flow forced by the Lorentz force associated with the magnetic field diffused in the inner core from the outer core. 

\section{Deformation induced by the Lorentz force}
\label{Section:ICLorentz}

As discussed in section \ref{Section:GeodynamoHypothesis}, the flow in the outer core sustains a magnetic field extending upward to the surface of the Earth but also inward inside the inner core. The magnetic Reynolds number of the inner core is likely small: assuming for example a velocity $10^{-10}$~m.s$^{-1}$ gives a magnetic Reynolds number on the order of $10^{-5}$.
This shows that the magnetic field is only diffused inside the inner core, with no net advection or generation of the field. A diffused magnetic field in the inner core will add two terms in the set of equations: the Lorentz force  in the momentum equation, and Joule heating in the energy equation. We are interested here in the flow driven by the Lorentz force in the inner core \citep{Lasbleis2015}.

Geodynamo simulations often exhibits a strong toroidal magnetic field close to the inner core boundary.
As we are interested in the largest effect on the inner core dynamics, we consider here only  low-order toroidal components of the magnetic field at the ICB, which have the largest  penetration length scale.  

We thus add in the momentum equation the Lorentz force due to a purely toroidal and axisymmetric magnetic field of degree two at the ICB $\mathbf{B}|_\text{icb} = B_0 \sin \theta \cos \theta \mathbf{e}_{\phi} $ \citep{Buffett2000}. 
Imposing this field at the ICB and solving for its diffusion inside the inner core  assuming it does not vary with time ($\boldsymbol{\nabla} ^2 \mathbf{B}=0$), we obtain $\mathbf{B} = B_0 r^2/r_\text{icb}^2\cos \theta \sin \theta \mathbf{e}_{\phi}$. This field is associated to an electric current density $\mathbf{j} = \frac{1}{\mu_0} \boldsymbol{\nabla} \times \mathbf{B}$, with $\mu_0$ the magnetic permeability. The associated Lorentz force is $\mathbf{f}_L = \mathbf{j} \times \mathbf{B} $  which non-potential part (magnetic tension) can be written as 
\begin{equation}
\mathbf{\tilde f}_L = \frac{B_0^2 }{\mu_0 r_\text{icb}}\frac{r^3 }{r^3_\text{icb}}\left [ f_r \mathbf{e}_{r} + f_\theta \mathbf{e}_{\theta} \right ], \label{Lorentz}
\end{equation}with $f_r$ and $f_\theta$ two functions of $\theta$ expressed as 
\begin{align}
f_r(\theta) = 3\cos ^4 \theta -\frac{15}{7}\cos ^2 \theta +\frac{4}{35},\\
f_\theta(\theta) = \cos \theta \sin \theta \left ( \frac{4}{7} -3\cos ^2 \theta  \right ) .
\end{align}
Injecting this force into the dimensionless Stokes equation, we obtain 
\begin{equation}
\mathbf{0} = -\boldsymbol{\nabla} p + Ra\, \chi\, \mathbf{r} + \boldsymbol{\nabla} ^2 \mathbf{u} + M\mathbf{\tilde f}_L, \label{momentum_mag}
\end{equation}where
\begin{equation}
M = \frac{B_0^2 r^2_\text{icb} }{\mu_0 \eta \kappa}
\end{equation}
is similar to a Hartmann number, quantifying the ratio of the Lorentz force to the viscous force, and $\mathbf{\tilde f}_L$ is defined as in \eqref{Lorentz} without the prefactor $\frac{B_0^2 }{\mu_0 r_\text{icb}}$.

Equation \eqref{momentum_mag} is solved using a poloidal decomposition and horizontal spherical harmonics decomposition. The term corresponding to the Lorentz force gives
\begin{equation}
\mathbf{r}\cdot (\boldsymbol{\nabla} \times \boldsymbol{\nabla} \times \mathbf{\tilde f}_L) = 8r^2 (1-3\cos ^2 \theta) = -\frac{16}{\sqrt{5}} r^2 Y_2^0, 
\end{equation}
where $Y_2^0= \sqrt{5}(3\cos ^2 \theta -1)/2$. 
The momentum equation can thus be written as an equation for the spherical harmonics components $P_l^m$ and $t_l^m$of respectively the poloidal component of  the velocity and the temperature as 
\begin{equation}
D_l^2 P_l^m - Ra\, t_l^m + \frac{16}{\sqrt{5}l(l+1)}M r^2 \delta_{2l}\delta_{0m} =0, \; \; l\geq 1, \label{mag_poloidal}
\end{equation}
where $\delta$ is the Kronecker symbol. 

\subsection{Neutral stratification}

We first consider the neutral stratification end-member where $Ra=0$. In that case, the only force driving flows in the system is the Lorentz force, and we do not need to solve for the temperature or composition fields. The flow results from a balance between the Lorentz and viscous forces.
Since the characteristic length scale of velocity variations must be the size of the inner core (1 in dimensionless form), we have
\begin{equation}
   \underbrace{\nabla ^2 \mathbf{u}}_{\sim u} \sim  \underbrace{M\mathbf{\tilde f}_L }_{\sim M}, 
\end{equation}
which implies that the magnitude of the velocity field should be proportional to $M$.

We can now solve analytically the flow field for a neutral stratification. With $Ra=0$, equation \eqref{mag_poloidal} reduces  for $l=2$ and $m=0$ to 
\begin{equation}
D_2^2 P_l^m + \frac{8}{3\sqrt{5}}M r^2 = 0, \label{mag_poloidal_neutral}
\end{equation}
which we solve with the boundary conditions at $r=1$ described in section \ref{BC}:
\begin{align}
        \frac{\text{d}^2 P_2^0 }{\text{d} r}+4\frac{P_2^0}{r^2}=0, \\
        r\frac{\text{d}r^3P_2 ^0 }{\text{d}r^3}-18\frac{1}{r}\frac{\text{d}P_2^0 }{\text{d}r} = \left ( \mathcal{P}-\frac{1}{r^2} \right)6 P_2 ^0.  
\end{align}

Equation \eqref{mag_poloidal_neutral} is a fourth order non-homogeneous differential equation, which solution can be obtained by solving the homogeneous equation and noticing that 
\begin{equation}
P_2^0 =     -\frac{1}{3^3 7 \sqrt{5} }M r^6
\end{equation}
is one solution of the complete equation. 
Searching for a polynomial solution, we find that $r^\alpha$ is solution of the homogeneous equation $D_2^2 P_2^0 = 0$ if $\alpha$ is a zero of the polynomial expression $[\alpha (\alpha +1) -6][(\alpha-2)(\alpha-1)-6]$. We then obtain the general solution of equation \eqref{mag_poloidal_neutral} as 
\begin{equation}
    P_2 ^0 (r) =  -\frac{M}{3^3 7 \sqrt{5} } r^6 +Ar^{-3}+ Br^{-1} + Cr^2 + D r^4.  
\end{equation}
$A$ and $B$ must be equal to 0 for the velocity to remain finite at $r=0$.
$C$ and $D$ are obtained from the boundary conditions at $r=1$, and we finally obtain
\begin{equation}
    P_2^0 (r) =  \frac{1}{3^3 7 \sqrt{5} }M \left ( -r^6+ \frac{14}{5}r^4 - \frac{9}{5}r^2 + \frac{1}{19+5\mathcal{P}} \left [ \frac{204}{5}r^4-\frac{544}{5}r^2 \right ]  \right ). 
\end{equation}

From the expression of $P_2^0$, we can now obtain the expressions for the velocity field from 
\begin{align}
  \label{eq:ur}
  u_r= 3\frac{P_2^0}{r}Y_2^0,\\
  \label{eq:utheta}
  u_{\theta}=\frac{1}{r}\frac{d}{dr}(r\, P_2^0)\frac{\partial Y_2^0}{\partial \theta}. 
\end{align}
Defining the r.m.s. velocity as 
\begin{equation}
    V_\text{rms}^2 = \frac{3}{4\pi} \int_0^{2\pi}\int_0^{\pi}\int_0^1 (u_r^2+u_{\theta}^2)\sin \theta r^2\,dr\,d\theta\,d\phi, 
\end{equation}
we obtain 
\begin{equation}
  \label{eq:rmsV_anal}
V_\text{rms}=M\frac{4}{189}\sqrt{\frac{34}{715}}\frac{\sqrt{74029-1576\, P+76\, P^2}}{19+5\,P}.
\end{equation}

This expression for the RMS velocity gives insight on the effect of $\mathcal{P}$ on the global dynamics. As predicted at the beginning of the subsection, the velocity is indeed a linear function of  $M$, with the boundary conditions modifying the prefactor. For $\mathcal{P}\rightarrow 0$, $V_\text{RMS}\sim 0.066 M$, and for $\mathcal{P}\rightarrow \infty$, $V_\text{RMS}\sim 0.008 M$. Permeable boundary conditions ($\mathcal{P}\rightarrow 0$) give velocities about one order of magnitude higher than impermeable boundary conditions ($\mathcal{P}\rightarrow \infty$).

\begin{figure}
    \centering
    \includegraphics[width=0.7\linewidth]{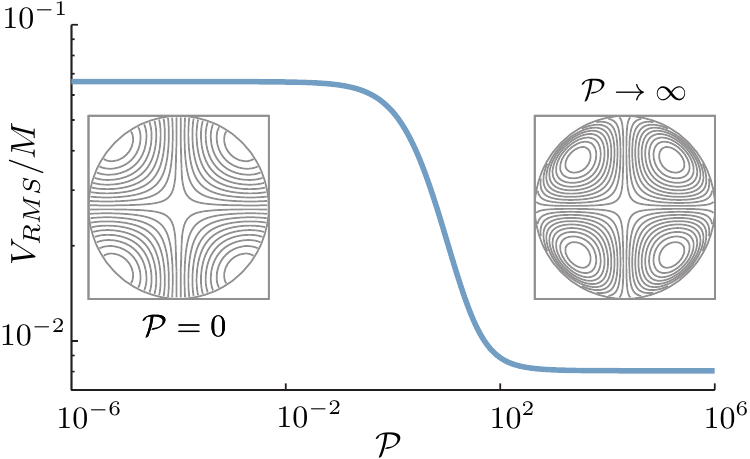}
    \caption{Neutral stratification: r.m.s. velocity as a function of the parameter $\mathcal{P}$, and meridional cross-section of the streamlines of the two end-members $\mathcal{P}\rightarrow 0$ and $\mathcal{P}\rightarrow \infty$. (Modified from \citet{Lasbleis2015}.)}
    \label{fig:neutral_strat}
\end{figure}

\subsection{Stable stratification}

If the inner core is stably stratified ($Ra<0$), the buoyancy forces resulting from the deformation of constant density surfaces would tend to oppose further deformation, and inhibit vertical motions.
We can thus anticipate that the flow obtained in the limit of $Ra = 0$ can be significantly altered by a strong stratification. %, which will tend to inhibit vertical motions.

\begin{figure}
    \centering
    \includegraphics{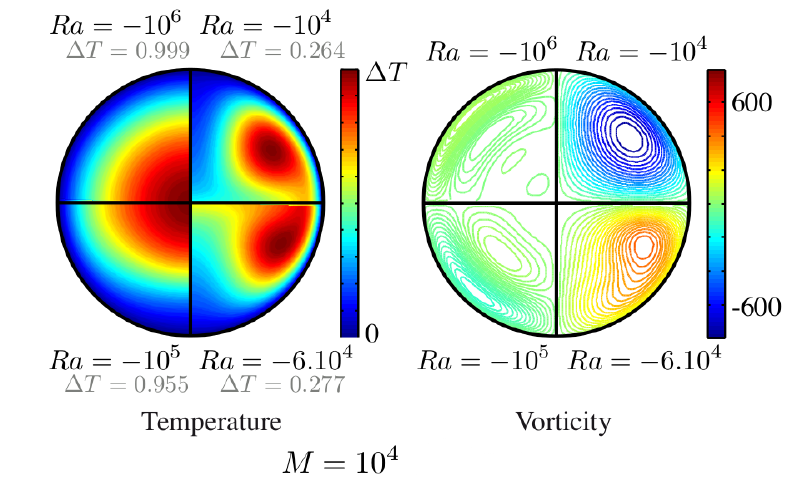}
    \caption{Meridional cross-sections of the temperature and the vorticity fields for $M=10^4$ and four different values of the Rayleigh number. At low $Ra$, the flow is similar to the neutral stratification case. If $-Ra\gtrsim M$ (strong stratication) the flow is confined in a layer at the top of the inner core. (Modified from \citet{Lasbleis2015}.)}
    \label{fig:T_vorticity}
\end{figure}

To estimate to what extent the neutral stratification solution can be altered by the presence of a stable density stratification, we consider the vorticity equation (obtained by taking the curl of equation \eqref{momentum_mag}), which is
\begin{equation}
    \mathbf{0} = -Ra \frac{\partial \chi}{\partial \theta} \mathbf{e}_{\phi}+ M \boldsymbol{\nabla} \times \mathbf{\tilde f}_L + \boldsymbol{\nabla} ^2 \boldsymbol{\zeta},
\end{equation}
where $\boldsymbol{\zeta}=\boldsymbol{\nabla} \times \mathbf{u}$ is the vorticity. 
Because the form of the magnetic field considered here forces a degree 2 flow, and because in non-dimensional form $\chi$ varies between 0 and 1, the $\theta$ derivative of $\chi$ is $\lesssim 1$.
The magnitude of the baroclinic vorticity production  $-Ra \frac{\partial \chi}{\partial \theta}$ is therefore $\lesssim -Ra$. 
Since the vorticity production associated with the Lorentz force is on the order of $M$, a stable stratification can affect significantly the flow only if $-Ra \gtrsim M$.
As one of the main unknown for inner core dynamics is the viscosity, it is interesting to note that the ratio $M/Ra$,
\begin{equation}
\frac{M}{Ra} = \frac{B_0^2}{\mu_0 \Delta \rho\, g_\text{icb}\, r_\text{icb}} ,
\end{equation}
does not depend on the viscosity, so that the boundary between a strongly stratified regime and a neutral stratification does not depends on the viscosity. 

Figure \ref{fig:T_vorticity} shows the temperature and vorticity fields obtained by solving numerically equations \eqref{eq:temperature_adim} and \eqref{mag_poloidal} at $M=10^{4}$ and $Ra=-10^{4}$, $-6\times 10^{4}$, $-10^{5}$, and $-10^{6}$.
The stratification has a negligible effect on the flow at $Ra=-10^{4}$, but already has a significant effect at $Ra=-6\times 10^{4}$, which is consistent with the criterion we just derived ($-Ra\gtrsim M$ for a strong effect of the stratification).
At $Ra=-10^{5}$ and $-10^{6}$, the flow induced by the Lorentz force is essentially confined to a thin layer below the inner core boundary, in which the flow direction is essentially horizontal.
\citet{Lasbleis2015} found that the thickness of this layer is $\propto (-Ra) ^{-1/6}$ and the strain rate in the layer is $\propto M (-Ra)^{-1/3}$. 
For application to Earth's inner core, we take $\Delta \rho \sim -1$ kg.m$^{-3}$ and $B_{0}\sim 4$ mT \citep{Gillet2010}, and find $M/(-Ra) \sim 10^{6}$, which would imply that stratification strongly affects the flow forced by the Lorentz force.

%--------------------------------------------------------------------------------------------------------------------------------------------%
%--------------------------------------------------------------------------------------------------------------------------------------------%
\chapter{Core formation}
\label{Section:CoreFormation}
%--------------------------------------------------------------------------------------------------------------------------------------------%
%--------------------------------------------------------------------------------------------------------------------------------------------%

The final section of this chapter concerns the formation of Earth's core. 
The Earth formed about 4.56 Gy ago through the accretion of solar nebula material, a process which is estimated to have taken a few tens of millions of years.
Accretion in the solar system went through different dynamical phases which  involved increasingly energetic and catastrophic impacts and collisions \citep{Lunine2011,walsh2011low}.
The last phase of accretion, in which most of the Earth mass was accreted, involved extremely energetic collisions between already differentiated planetary embryos ($\sim 100 - 1000$ km size), which resulted in  widespread melting and the formation of magma oceans. 

The basic ingredients of a terrestrial planet -- an iron-rich metal and silicates -- are immiscible, and can separate under the action of gravity to form an iron-rich core surrounded by a rocky mantle.
The timescale of phase separation is  much smaller than the accretion timescale if at least one of the two phases is liquid, and it is thus believed that core formation has been concomitant with Earth's accretion.  
Importantly, the metal added to the Earth by each impact had no reason to be in thermodynamic equilibrium with the silicate mantle of the growing Earth,
which implies that heat  and chemical elements would have been transferred from one phase to the other  when the two phases were in contact.
%Heat and chemical elements would therefore have been transferred from one phase to the other.
Chemical elements and heat released during accretion have thus been partitioned between the core and mantle, in a way which depends on the exact physical mechanism by which metal and silicate have separated.

This has important implications for the  state of the planet at the end of its accretion, and its subsequent evolution.
The partitioning between core and mantle of the heat released during accretion has set the initial temperature contrast between the mantle and core, a key parameter for the early dynamics of the planet, with implications for the possibility of forming a basal magma ocean \citep{Labrosse2007}, the existence of an early dynamo \citep{Williams2004,monteux2011might}, and the subsequent thermal and magnetic evolution of the planet. % 
The partitioning of chemical elements between the core and mantle has been used to constrain the timing of differentiation and the physical conditions under which it occurred \citep[\textit{e.g.}][]{Yin2002,Halliday2004,Wood2006,Rudge2010}.
It has also important implications for a number of geodynamical issues.
One example is the identity and abundance of light elements and radioactive elements in the core \citep{Corgne2007,badro2015}, which depend on the conditions (pressure $P$, temperature $T$, oxygen fugacity $f{O_2}$) at which metal and silicate have interacted for the last time. 

The large impacts which dominated the last stages of Earth formation injected enormous amounts of kinetic energy into the magma oceans, creating highly turbulent environments in which it has been conjectured that the cores of the bodies impacting the Earth  would  fragment down to  centimetre scale, at which metal and silicates can efficiently exchange chemical elements and heat \citep{Stevenson1990,Karato1997,Rubie2003}. 
The metal  is envisioned to disperse in the magma ocean and equilibrate with the silicates, before raining out.
It would then collect at the base of the magma ocean, and finally migrate toward the core as large diapirs \citep{Stevenson1990,Karato1997,Monteux2009,Samuel2010} or by the propagation of iron dykes \citep{stevenson2003}, at which point further chemical equilibration is unlikely to be significant. 

Most geochemical models of core formation are based on this so-called \textit{iron rain} scenario, but  the fluid dynamics involved is actually poorly understood, even at a qualitative level.
Efficient chemical exchange requires a high metal-silicates interfacial area-to-volume ratio, which requires fragmentation or stretching of the metal down to $\sim$ cm size. 
Differentiation of terrestrial bodies started early, and it is  now recognized that most of the mass of the Earth  was accreted from already differentiated planetary bodies, with cores of their own.
Whether or not these large volumes of iron ($\sim 100$ to 1000 km) would indeed fragment down to cm scale at which chemical equilibration can occurs  therefore remains an open question, and a matter of much speculation  \citep{Dahl2010,Deguen2011c,Samuel2012,Deguen2014,Wacheul2014}.

We have yet no well-tested and self-consistent theory of fragmentation and thermal or compositional equilibration in the context of metal-silicate separation in a magma ocean. 
For this reason we have here made the choice of focusing on the basic concepts and mechanisms which we believe are important to know and understand to tackle this problem, rather than trying to give a definitive answer to this question.

\section{Problem set-up and non-dimensional numbers}
\label{Section:SetUp}

To keep things reasonably simple, we will ignore here the dynamics of the impact itself and consider the fate of an initially spherical mass of molten iron falling into silicates, which can be either solid or molten.
This molten iron mass may be either the core of a planetary body impacting the Earth, or a fragment of the core if it has been significantly dispersed by the impact. 
As an additional simplifying hypothesis, we will even assume that the metal mass has no initial velocity. 

The volume of molten iron is assumed to be close to spherical, and has a radius $R_{0}$.
We denote by $\rho_{m}$ and $\rho_{s}=\rho_{m}-\Delta \rho$ the densities of metal and silicates, and by $\eta_{m}$ and $\eta_{s}$ their viscosities.
The metal and silicates phases are immiscible and we denote by $\gamma$ the interfacial tension of the metal-silicates interface. 
We denote by $g$ the acceleration of gravity.

The evolution of the metal mass depends on 7 dimensional parameters ($R_{0}$, $\rho_{m}$, $\rho_{s}$, $\eta_{m}$, $\eta_{s}$, $\gamma$) involving 3 fundamental units (length, weight, and time).
According to Vashy-Buckingham's theorem, the number of independent non-dimensional numbers to be used to describe the problem is equal to $7-3=4$.
One possible set is the following:
\begin{align}
&\text{Bond number: } Bo = \frac{\Delta\rho\, g\, R_{0}^{2}}{\gamma} , \\
&\text{Grashof number: } Gr = \frac{\Delta \rho}{\rho_{s}}   \frac{g R_{0}^{3}}{\nu_{s}^{2}} ,\\
&\text{density ratio: }  \frac{\rho_{m}}{\rho_{s}}, \\
&\text{viscosity ratio: } \frac{\eta_{m}}{\eta_{s}}.
\end{align}
The Bond number is a measure of the relative importance of buoyancy and interfacial tension.
The Grashof number is basically a Reynolds number obtained by taking Stokes' velocity as a velocity scale.
Additionnal useful numbers include Reynolds and Weber numbers, which can be defined as
\begin{align}
&\text{Reynolds number: } Re = \frac{\rho_{s }U R_{0}}{\eta_{s}}, \\
&\text{Weber number: } We = \frac{\rho_s U^2 R_{0}}{\gamma},
\end{align}
where $U$ is a velocity scale to be defined. 
The Reynolds and Weber numbers compare inertia to viscous forces and interfacial tension, respectively. 
The density ratio is close to 2.

If in addition we consider heat or mass transfer between metal and silicates, then two additional dimensional parameters enter the problem: the diffusivities (thermal or compositional) $\kappa_{m}$ and $\kappa_{s}$ in the metal and silicates. 
Since there is no additional fundamental units, Vashy-Buckingham's theorem implies that two additional non-dimensional numbers must be used.
One possible choice is to use the ratio of the diffusivities and a P\'eclet number:
\begin{align}
&\text{diffusivity ratio: } \frac{\kappa_{m}}{\kappa_{s}}, \\
&\text{P\'eclet number: } Pe = \frac{U R_{0}}{\kappa_{s}}.
\end{align}
In what follows, we will assume that $\kappa_{s}=\kappa_{m}$ for the sake of simplicity. % (they are in general not widely different).

\section{Preliminary considerations}

\subsection{Terminal velocity}

Since we have chosen to focus on the case of metal mass falling with no initial velocity, a relevant velocity scale is its terminal velocity, reached when 
the buoyancy force ($\sim \Delta \rho g R_{0}^{3}$) is balanced by the drag on the surface of the metal mass. 
Two different scalings can be obtained depending on whether the drag is dominated by viscous stresses ($\sim \eta_{s} U/R_{0}$) or dynamic pressure ($\sim \rho_{s} U^{2}$).
The ratio of the dynamic pressure and viscous stresses contributions to the total drag is
\begin{equation}
\frac{\text{dynamic pressure}}{\text{viscous stress}} \sim \frac{\rho_{s} U^{2}}{\eta_{s} U/R_{0}} = Re.
\end{equation}
The drag on the metal mass is obtained by multiplying the dominant stress by the surface area of the metal mass: it is on the order of $\eta_{s} U R_{0}$ if the drag is dominated by the contribution of viscous stresses (low $Re$), and on the order of $\sim \rho_{s} U^{2} R_{0}^{2}$ if it is dominated by the contribution of dynamic pressure (high $Re$). 
The force balance on the metal mass can thus be written as
\begin{equation}
 \underbrace{\Delta \rho g R_{0}^{3}}_{\text{\normalsize buoyancy}} \sim \underbrace{ \max\left(\rho_{s} U^{2} R_{0}^{2}, \eta_{s} U R_{0} \right)}_{\text{\normalsize drag}} .
\end{equation}
The terminal velocity obtained from this balance is
\begin{equation}
U \sim \min \left( \frac{\Delta \rho g R_{0}^{2}}{\eta_{s}}, \left( \frac{\Delta \rho}{\rho_{s}} g R_{0} \right)^{1/2}  \right).
\end{equation}
The first scaling gives Stokes' settling velocity, and corresponds to the low $Re$ limit; the second scaling is the so-called newtonian scaling, and corresponds to the high $Re$ limit.
The two velocities are on the same order of magnitude when
\begin{equation}
Gr   \sim 1,
\end{equation}
which defines the boundary between the two scalings (here $\nu_{s}=\eta_{s}/\rho_{s}$).
The terminal velocity is thus
\begin{align}
U \sim \frac{\Delta \rho g R_{0}^{2}}{\eta_{s}} \quad & \text{ if } Gr \ll 1, \label{Eq:StokesVelocity} \\
U \sim \left( \frac{\Delta \rho}{\rho_{s}} g R_{0} \right)^{1/2}\quad & \text{ if } Gr \gg 1. \label{Eq:NewtonianVelocity}
\end{align}
With these scalings, the Reynolds and Weber numbers based on the terminal velocity are given by
\begin{align}
\begin{cases}
Re&\sim Gr \\
We&\sim Gr  Bo
\end{cases}
  \quad \text{ if } Gr \ll 1, \\
\begin{cases}
Re &\sim Gr^{1/2}  \\
We &\sim Bo
\end{cases}
  \quad \text{ if } Gr \gg 1.	
\end{align}

\subsection{Maximal stable size of a falling drop}

Interfacial tension (unit J.m$^{-2}$) can be interpreted as an energy per unit area. 
Deforming an interface in a way which results in an increase of the interfacial area costs energy, and interfacial tension effects will tend to minimise the surface area of the interface.
If no other force acts on a drop, interfacial tension would keep it spherical, hence minimising its surface area.

Interfacial tension can also be seen as a force per unit of length (it can be verified that J.m$^{-2}=$N.m$^{-1}$): if a piece of an interface is divided into two parts, the force imparted by one part of the surface on the other is parallel to the interface and has a magnitude given by the product of the interfacial tension with the length of the curve separating the two parts of the surface.
If integrated over a curved surface, 
one can also show that interfacial tension induces a pressure jump across the interface equal to
\begin{equation}
\Delta P = \gamma \left( \frac{1}{R_{1}} + \frac{1}{R_{2}} \right),
\end{equation}
where $R_{1}$ and $R_{2}$ are the principal radii of curvature. This pressure jump is called Laplace's pressure. 
Across a spherical interface (the surface of a drop or bubble of radius $R_{0}$), Laplace's pressure is equal to $2\gamma/R_{0}$.

A falling drop can be deformed by the stresses imparted by the surrounding fluid onto the drop, or in other words by the fluid drag.
If the total drag on the drop is $F_\mathrm{drag}$, the mean stress on the surface of the drop is  $\sim F_\mathrm{drag}/R_{0}^{2}$.
One can expect significant deformation of the drop if the hydrodynamic stress variations due to the drag exceed Laplace's pressure:
\begin{equation}
\frac{F_\mathrm{drag}}{R_{0}^{2}} \gtrsim \frac{\gamma}{R_{0}}.
\label{Eq:DragSurfaceTensionBalance}
\end{equation}
If the drop reached its terminal velocity, then the drag must be equal to the total buoyancy of the drop, $F_\mathrm{drag} \sim \Delta \rho g R_{0}^{3}$.
Combining this with equation \eqref{Eq:DragSurfaceTensionBalance}, we find that strong deformation of the drop will happen if its radius is larger than a critical radius $R_{c}$ given by
\begin{equation}
R_{c} \sim \left( \frac{\gamma}{\Delta \rho g} \right)^{1/2}.
\label{Eq:maxDropSizeTerminalVelocity}
\end{equation}
This length is also known as a \textit{capillary length}: it is the length scale over which buoyancy effects dominate over surface tension effects.
Interfacial tension will keep the drop close to spherical if its radius is smaller than $R_{c}$.
Equation \eqref{Eq:maxDropSizeTerminalVelocity} is equivalent to writing that deformation is significant if the Bond number of the drop is large compared to 1.
The interfacial tension between metal and silicates is on the order of 1 J.m$^{-2}$ and $\Delta \rho \sim 4000$ kg.m$^{-3}$.
With $g\sim 10$ m.s$^{-2}$, we thus have $R_{c} \sim 5$ mm. 

Strong drop deformation may happen before reaching terminal velocity, and in this case the above scaling will not be the most relevant.
If drag is dominated by viscous effects ($F_\mathrm{drag} \sim \eta_{s} U R_{0}$), then we find that deformation of the drop may happen if its velocity is larger than a critical velocity 
\begin{equation}
U_{c} \sim \frac{\gamma}{\eta}.
\label{Eq:ViscousDragSurfaceTensionBalance}
\end{equation}
This criterion is of limited use since if the drag is dominated by viscous effect (which means that $Re \ll 1$), then the drop velocity will very quickly reach a terminal velocity equal to Stokes' settling velocity. Using Stokes' velocity for $U$ in equation \eqref{Eq:ViscousDragSurfaceTensionBalance} gives equation \eqref{Eq:DragSurfaceTensionBalance}.

If instead drag is dominated by the contribution of dynamic pressure ($F_\mathrm{drag} \sim \rho_{s} U^{2} R_{0}^{2}$), then we find that strong deformation requires the drop Weber number is large:
\begin{equation}
We = \frac{\rho_{s} U^{2} R_{0}}{\gamma} \gtrsim 1.
\end{equation}
This criterion reduces to equation \eqref{Eq:maxDropSizeTerminalVelocity} when the drop reaches its terminal velocity ($U \sim \sqrt{ ({\Delta \rho}/{\rho_{s}}) g R_{0} }$ since in this limit $Re \gg 1$).

\section{The low Reynolds limit: diapirism}
\label{Section:Diapirism}

A first relevant limit of the problem described in section \ref{Section:SetUp}  corresponds to molten metal diapirs travelling through a solid, or partially molten, silicate layer.
The radius of these diapirs may be similar to the size of the core of the impactors, say somewhere between 1 km and 1000 km.
The acceleration of gravity is smaller or equal to its current value in Earth's mantle, $g \sim 10$ m.s$^{-2}$.
The viscosity $\eta_{s}$ of the silicates is a strong function of temperature, and can also be significantly decreased if the silicate layer is partially molten. 
A reasonable range is $10^{15}$ to $10^{21}$ Pa.s.
The viscosity of molten iron is $\sim 10^{-2}$ Pa.s. 

With this parameter values,  $Bo \gtrsim 10^{10}$, $Gr \lesssim 10^{-4}$, $\rho_{m}/\rho_{s}\sim 2$, $\eta_{m}/\eta_{s} \lesssim 10^{-18}$. 
The Grashof number being small, we are well into the low Reynolds regime: viscous forces  dominate over inertia in the silicates.
The limit of low $Re$ and high $Bo$ has been studied numerically in the context of core formation \citep{Samuel2008,Monteux2009,Samuel2010}, and experimentally in other contexts \citep[\textit{e.g.}][]{Ribe1983,Bercovici1997}.
In this limit surface tension is unimportant, but the volume of metal is kept roughly spherical because $Re\ll 1$. Since viscous forces are so important in the silicates, the flow around the metal mass is limited to spatial scales on the order of $R_{0}$, which limits the deformation of the metal mass (in other words, small scale perturbations of the metal-silicate interface shape are damped viscously). 

The falling velocity is thus simply given by Stokes' velocity (equation \eqref{Eq:StokesVelocity}).
The law of heat or mass transfer between the diapir and its surrounding is also well known \citep[\textit{e.g.}][]{Clift1978,Ribe2007,Ulvrova2011}: the heat flux is given by
\begin{equation}
\text{heat flux}=a 4\pi R_{0}^{2}k_{s} \frac{\bar T - T_{s}(z)}{R_{0}} Pe^{1/2}
\end{equation}
where $k_{s}$ is the thermal conductivity in the silicates, $\bar T$ the mean temperature in the diapir, $T_{s}(z)$ the temperature of the surrounding silicate layer at depth $z$, and $a$ is a constant on the order of 1.
A similar expression can be written for chemical elements transfer. 

The heat balance of the diapir writes
\begin{equation}
\rho_{m} c_{p,m} \frac{4\pi}{3} R_{0}^{3} \frac{d \bar T}{dt} = - 4\pi R_{0}^{2} a k_{s} \frac{\bar T - T_{s}(z)}{R_{0}} Pe^{1/2},
\end{equation}
where $c_{p,m} $ is the heat capacity of the metal phase. 
Transforming the time derivative into a derivative with respect to the distance $z$ travelled by the diapir (using $d(...)/dt = U d(...)/dz$) and re-arranging gives
\begin{equation}
\frac{d \bar T}{dz} + \frac{\bar T}{\ell} = \frac{T_{s}(z)}{\ell}, \label{Eq:DiapirTEvolution}
\end{equation}
where the characteristic length $\ell$ is given by
\begin{equation}
\ell = \frac{1}{3\, a} \frac{\rho_{m} c_{p,m}}{\rho_{s} c_{p,s}} R_{0} Pe^{1/2}.
\end{equation}
$\ell$ is the characteristic distance over which the temperature of the diapir responds to changes of the surrounding temperature, the \textit{thermal equilibration distance}.
The general solution of equation \eqref{Eq:DiapirTEvolution} is 
\begin{equation}
\bar T = T_{0} \mathrm{e}^{-z/\ell} + \int_{0}^{z} \frac{\mathrm{e}^{(z'-z)/\ell} }{\ell} T_{s}(z') dz',
\end{equation}
where $T_{0}$ is the initial temperature of the diapir.
In practice the amount of heat transfer is small because $\ell$ is typically larger than the mantle thickness. 
The compositional diffusivity in the solid silicates being perhaps four orders of magnitude smaller that the thermal diffusivity, exchange of chemical elements would be even smaller (the equilibration distance being $\propto\, Pe^{1/2}\propto\, \kappa^{-1/2}$). 
Diapirs migrating through a solid part of the mantle would therefore exchange a negligible amount of heat and chemical elements with the surrounding mantle.

\section{The high Reynolds limit: metal-silicates separation in a magma ocean}

Let us now consider the case of a volume of molten metal falling into molten rocks - a \textit{magma ocean}.
The parameter values are similar to what we have considered when discussing the case of diapirism, except that the viscosity of the silicates is much smaller, on the order of $10^{-1}$ Pa.s.
With these parameter values,  $Bo \gtrsim 10^{10}$, $Gr \gtrsim 10^{22}$, $\rho_{m}/\rho_{s}\sim 2$, $\eta_{m}/\eta_{s} \lesssim 10^{-18}$. 
Since $Gr\gg 1$, the relevant velocity scale is the newtonian scaling given by equation \eqref{Eq:NewtonianVelocity}. 
This gives $Re \sim Gr^{1/2} \gtrsim 10^{11}$ and $We\sim Bo \gtrsim 10^{10}$. 
The very large values of $Re$, $Bo$, and $We$ imply that neither viscous forces nor interfacial tension can keep the metal volume spherical: a molten mass of metal falling into a magma ocean should suffer significant deformation, possibly resulting in its fragmentation into drops. 

\begin{figure}[t!]
\includegraphics[width=\linewidth]{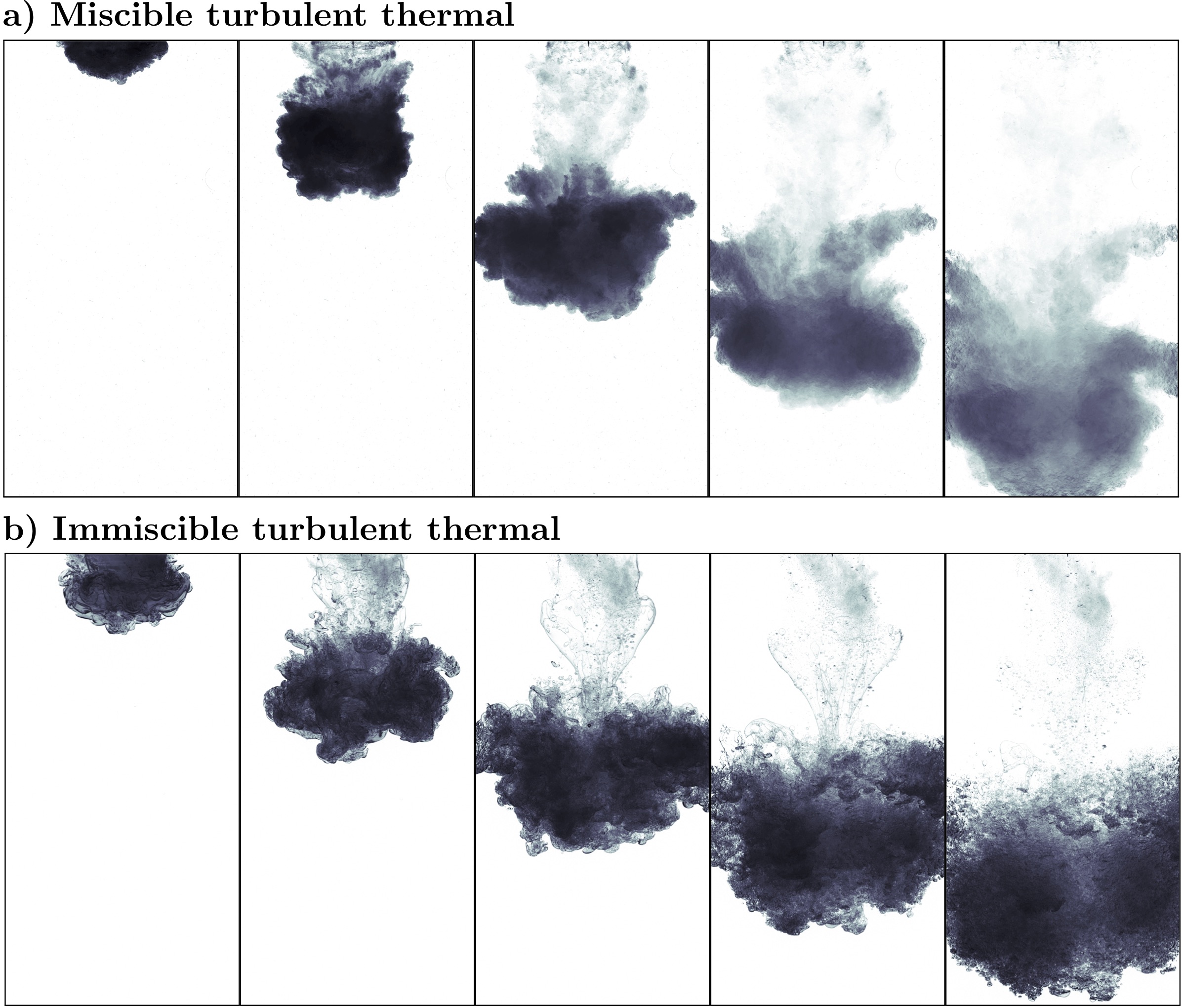}
\caption{
\textbf{a)} 
A 169 mL volume of an aqueous solution of NaI ($\rho=1502$ kg.m$^{-3}$)  falling into fresh water, at $\mathrm{Re}=4\times 10^{4}$, $\rho_{m}/\rho_{s}=1.5$, $\eta_{m}/\eta_{s}=1$.
The time interval between each image is 0.3 s. 
\textbf{b)} 
A 169 mL volume of an aqueous solution of NaI ($\rho=1601$ kg.m$^{-3}$) falling into a low viscosity silicon oil ($\rho=821$ kg.m$^{-3}$, $\eta = 8.2\times 10^{-4}$ Pa.s), 
at $Bo=3.4\times10^{4}$, $\mathrm{Re}=5.5\times 10^{4}$, $\rho_{m}/\rho_{s}=1.95$, $\eta_{m}/\eta_{s}=1.2$ \citep{DeguenRisso2019prep}. 
The time interval between each image is 0.2 s. 
}
 \label{Fig:TurbulentThermal1}
\end{figure}

\begin{figure}[t]
\centering
\includegraphics[width=0.6\linewidth]{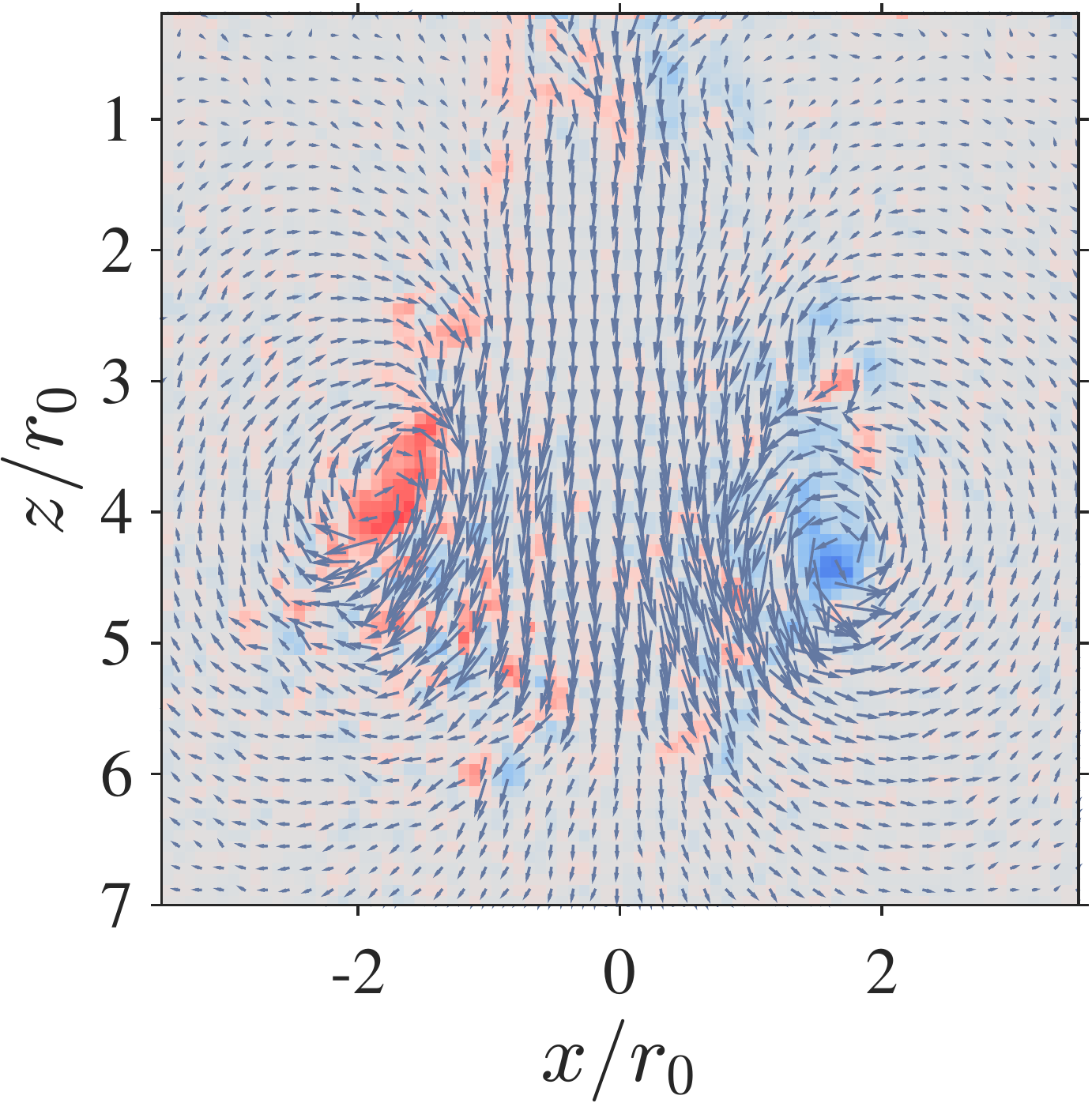}

\caption{Velocity field obtained from PIV measurements in an experiment in which a 169 mL of aqueous solution of NaI ($\rho=1280$ kg.m$^{-3}$) is released in a 1cst viscosity silicon oil.
The colorscale gives the vorticity field (red is clockwise, blue counterclockwise).
The concentration of the NaI solution has been chosen so that its optical index matches that of the silicon oil, in order to avoid optical distorsion.
The non-dimensional parameter values are $\mathrm{Bo}=2\times10^{4}$, $\mathrm{Re}=4.2\times 10^{4}$, $\rho_{m}/\rho_{s}=1.56$, $\eta_{m}/\eta_{s}=1.2$. 
}
\label{Fig:TurbulentThermalPIV}
\end{figure}

\subsection{Observations from laboratory experiments}

The very large values of the Bond and Weber numbers imply that interfacial tension should be unimportant for the large scale dynamics.
This suggests to first look at the case of infinite Bond and Weber numbers, which corresponds to the limit of miscible fluids.

Figure \ref{Fig:TurbulentThermal1}a shows snapshots from an experiment in which a negatively buoyant volume of an aqueous solution of sodium iodide is released into fresh water.
The volume of the (dyed) negatively buoyant fluid is seen to increase as it falls, which indicates that it entrains and incorporates ambient fluid, resulting in its gradual dilution.
Measurements show that the mean radius of the dyed mixture increases linearly with the distance from the point of release. 
This is what is known as a \textit{turbulent thermal} in the fluid mechanics and atmospheric science communities \citep[\textit{e.g.}][]{Batchelor1954,Scorer1957,Woodward1959}.
The name \textit{thermal} is inherited from the usage of glider pilots, for whom a thermal is an isolated mass of warm air rising through the lower atmosphere. % 
Though the buoyancy in atmospheric thermals is due to temperature differences, the nature of the source of buoyancy (thermal or compositional) happens to be of secondary importance and introduces no qualitative difference.
The term thermal has since been used to denote an isolated buoyant mass of a fluid rising or falling (depending on the sign of the buoyancy), irrespectively of the nature of the source of buoyancy. 
Here we will also often use the term \textit{buoyant cloud} instead of thermal.

Figure \ref{Fig:TurbulentThermal1}b shows snapshots from a similar experiment in which a negatively buoyant volume of sodium iodide  is now released into a low-viscosity silicon oil.
The NaI solution and the silicon oil are \textit{immiscible}, so we are one important step closer to the core-mantle differentiation configuration.
The experimental fluids and configuration have been chosen so as to maximize the values of the Bond and Reynolds numbers, which are $Bo=3.4\times10^{4}$ and $\mathrm{Re}=5.5\times 10^{4}$.
The density ratio is $\rho_{m}/\rho_{s}=1.95$ (close to metal-silicate), and the viscosity ratio is $\eta_{m}/\eta_{s}=1.2$. 
The large-scale evolution of the negatively buoyant volume is strikingly similar to what has been observed in the \textit{miscible} experiment: the volume of the negatively buoyant fluid increases linearly with distance, which indicates that it entrains and incorporates silicon oil.
PIV measurements on a similar experiment (figure \ref{Fig:TurbulentThermalPIV}) show that the velocity field has a vortex ring structure, with most of the entrainment of silicon oil probably occurring from the rear of the cloud.

In contrast, the small scale structure in the {immiscible} experiment is qualitatively different from what we can observe in the {miscible} experiment.
In the miscible experiment, the negatively buoyant solutions \textit{mixes} with the entrained water, diffusion of the NaI salt allowing homogenisation at the molecular scale. 
In the immiscible experiment, the NaI solution of course does not mix with the entrained silicon oil since the two liquids are immiscible. 
A close inspection of the last snapshot of the immiscible experiment reveals that the dense phase has been fragmented into droplets. %, while no discontinuity are apparent in the miscible case.

\subsection{Large-scale dynamics: turbulent entrainment model}

\begin{figure}[t]
\centering
\includegraphics[width=0.5\linewidth]{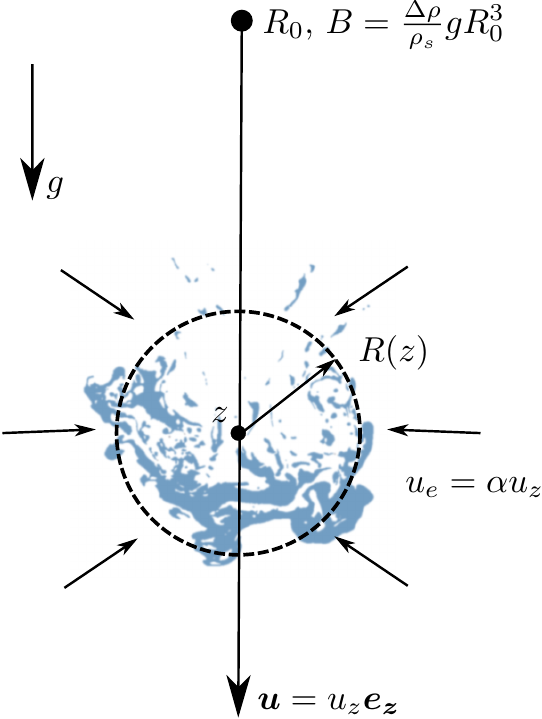}
\caption{The \textit{turbulent thermal} model: a volume of fluid with initial radius $R_{0}$ and density $\rho_{a}+\Delta\rho$ is released at $z=0$ in a fluid of density $\rho_{a}$.
The thermal has a mean vertical velocity $u_{z}$.  
Its mean radius $R$ increases with the distance $z$ from the source due to entrainment of ambient fluid at a rate $u_{e}=\alpha u_{z}$.
 }
\label{Fig:TurbulentEntrainment}
\end{figure}

We consider the evolution of a mass of negatively buoyant fluid falling into another one (figure \ref{Fig:TurbulentEntrainment}). 
The mass of negatively buoyant fluid has an initially spherical shape and an initial radius $R_{0}$, and has a density $\rho_{a}+\Delta\rho$, where $\rho_{a}$ is the density of the surrounding fluid. 
We denote by  $u_{z}$  the vertical velocity of the center of mass of the negatively-buoyant cloud, and by  $R(z)$ its spatial extension.
Following \cite{Batchelor1954}, we assume that far from the source $u_{z}$ and $R$  only depend on either distance $z$ or time $t$, and on its total amount of buoyancy defined as
\begin{equation*}
B = g \frac{\Delta \rho}{\rho_{a}} V_{0},
\end{equation*}
where $V_{0}$ is the initial volume of the released buoyant mass.
We thus assume that surface tension has no effect of the evolution of $u_{z}$ and $R$.
Dimensional analysis then shows that 
\begin{equation*}
\left\{
\begin{array}{c}
u_{z} \sim B^{1/2}z^{-1}  \\
R \sim z 
\end{array}
\right.
\quad \text{or, equivalently} \quad
\left\{
\begin{array}{c}
u_{z} \sim B^{1/4}\, t^{-1/2} \\
R \sim B^{1/4}\, t^{1/2}
\end{array}
\right.
,
\end{equation*}
which predicts that the spatial extension of the cloud increases linearly with $z$: the cloud must therefore entrain ambient fluid.
The prediction that the mean velocity decreases as $z^{-1}$, or, equivalently, as $R^{-1}$, is consistent with the fact that the total buoyancy of the cloud is conserved, but not its volume.
The buoyancy is ``diluted'' by the incorporation of neutrally buoyant ambient fluid to the cloud.

A more physical (and more general) way to obtain the evolution of a turbulent thermal has been given by \cite{Morton1956}, who based their analysis on the assumption that the rate of entrainment of ambient fluid within the buoyant cloud is simply proportional to the mean vertical velocity $u_{z}$ and to the surface area of the cloud (figure \ref{Fig:TurbulentEntrainment}).
This is the basic assumption of the \textit{turbulent entrainment} models used to describe the dynamics of turbulent clouds, plumes, and jets.

This assumption implies that the time derivative of the cloud volume is given by
\begin{equation}
\frac{d}{dt}\left( \frac{4\pi}{3} R^{3} \right) = 4\pi R^{2} \alpha\, u_{z},
\label{Eq:EntrainmentAssumptionMacroscopic}
\end{equation}
where $\alpha$ is the \textit{entrainment coefficient}. 
Noting that $d(...)/dt=u_{z} d(...)/dz$, where $z$ is the vertical position of the center of mass of the cloud, integration of equation \eqref{Eq:EntrainmentAssumptionMacroscopic} gives
\begin{equation}
R = R_{0} + \alpha z,
\end{equation}
which is consistent with the prediction of dimensional analysis.

Conservation of momentum allows to obtain a predictive law for the vertical velocity $u_{z}$ of the center of mass of the cloud.
Ignoring fluid drag on the cloud and a possible loss of buoyancy in the wake of the cloud, conservation of momentum simply states that
\begin{equation}
\frac{d}{dt} \left(  \frac{4\pi}{3} R^{3} \bar\rho u_{z}  \right) = \frac{4\pi}{3} R_{0}^{3} \Delta\rho g
\label{Eq:ThermalMomentumConservation1}
\end{equation}
where $\bar\rho$  is the mean density of the cloud, given by
\begin{equation}
\bar\rho =  \rho_{a} + \Delta \rho \frac{R_{0}^{3}}{R^{3}} = \rho_{a} + (1+\alpha z)^{-3} \Delta \rho.
\end{equation}
Conservation of mass implies that
\begin{equation}
\frac{d}{dt}\left( \frac{4\pi}{3} R^{3} \bar\rho \right) = 4\pi \rho_{a} R^{2} \alpha u_{z}.
\end{equation}
Using this relation in equation \eqref{Eq:ThermalMomentumConservation1}, using the transformation $d(...)/dt=u_{z} d(...)/dz=\alpha u_{z} d(...)/dR$, and re-arranging gives
\begin{equation}
   \frac{d u_{z}^{2}}{dR}  + \frac{6}{R } \frac{\rho_{a}}{\bar\rho}   u_{z}^{2} =  \frac{2 g}{\alpha}  \frac{\Delta \rho}{\bar\rho} \frac{R_{0}^{3}}{R^{3} },
\end{equation}
a linear first order ordinary differential equation with varying coefficients. 
The solution, written for $u_{z}$ as a function of $z$, is 
\begin{equation}
u_{z} = \left(\frac{g}{2 \alpha^{3}}\frac{\Delta\rho}{\rho_{a}}R_{0}^{3}\right)^{1/2}
\mathcal{F}\left(\dfrac{R_{0}}{\alpha z} , \frac{\Delta\rho}{\rho_{a}}\right)
\frac{1}{z}
,
\end{equation}
where
\begin{equation}
\mathcal{F}\left(\dfrac{R_{0}}{\alpha z} , \frac{\Delta\rho}{\rho_{a}}\right) = 
\frac{
\left[
1 + 4 \dfrac{R_{0}}{\alpha z} + 6 \left(\dfrac{R_{0}}{\alpha z}\right)^{2} + 4 \left( 1 + \dfrac{\Delta\rho}{\rho_{a}} \right) \left(\dfrac{R_{0}}{\alpha z}\right)^{3}
\right]^{1/2}
}{
1 + 3 \dfrac{R_{0}}{\alpha z} + 3 \left(\dfrac{R_{0}}{\alpha z}\right)^{2} + \left( 1 + \dfrac{\Delta\rho}{\rho_{a}} \right) \left(\dfrac{R_{0}}{\alpha z}\right)^{3}
} .
\end{equation}
The function $\mathcal{F}$ tends toward 1 at large  $\alpha z / R_{0}$.
The full solution is thus consistent with the dimensional analysis prediction when far from the source.
Close to the source, the velocity is given (at first order in $\alpha z / R_{0}$), by
\begin{equation}
u_{z} = \left( 2 \frac{\Delta \rho}{\rho_{a}} g z\right)^{1/2} .
\end{equation}

Though the turbulent thermal model just described has been developed to model \textit{miscible} flows, experiments suggest that it can also be applied to immiscible fluids in situations where the Bond and Weber numbers are large \citep{Deguen2014,Landeau2014,Wacheul2014,Wacheul2017}.
A qualitative comparison of figures \ref{Fig:TurbulentThermal1}a and \ref{Fig:TurbulentThermal1}b suggests that it is indeed the case.
The similarity between the miscible and immiscible experiments also holds on a quantitative level: from a series of experiments similar to that presented on figure \ref{Fig:TurbulentThermal1}, we have been able to show that the evolution of both $R$ and $u_{z}$ are very well described by the turbulent entrainment model with $\alpha=0.25 \pm 0.05$, similar to miscible flows \citep{Deguen2014,Landeau2014}. 
This demonstrates that there is indeed no effect of surface tension on the large scale part of the flow.

\subsection{Fragmentation}

\subsubsection{Qualitative observations from experiments}

\begin{figure}[t!]   %[th]
\includegraphics[width=\linewidth]{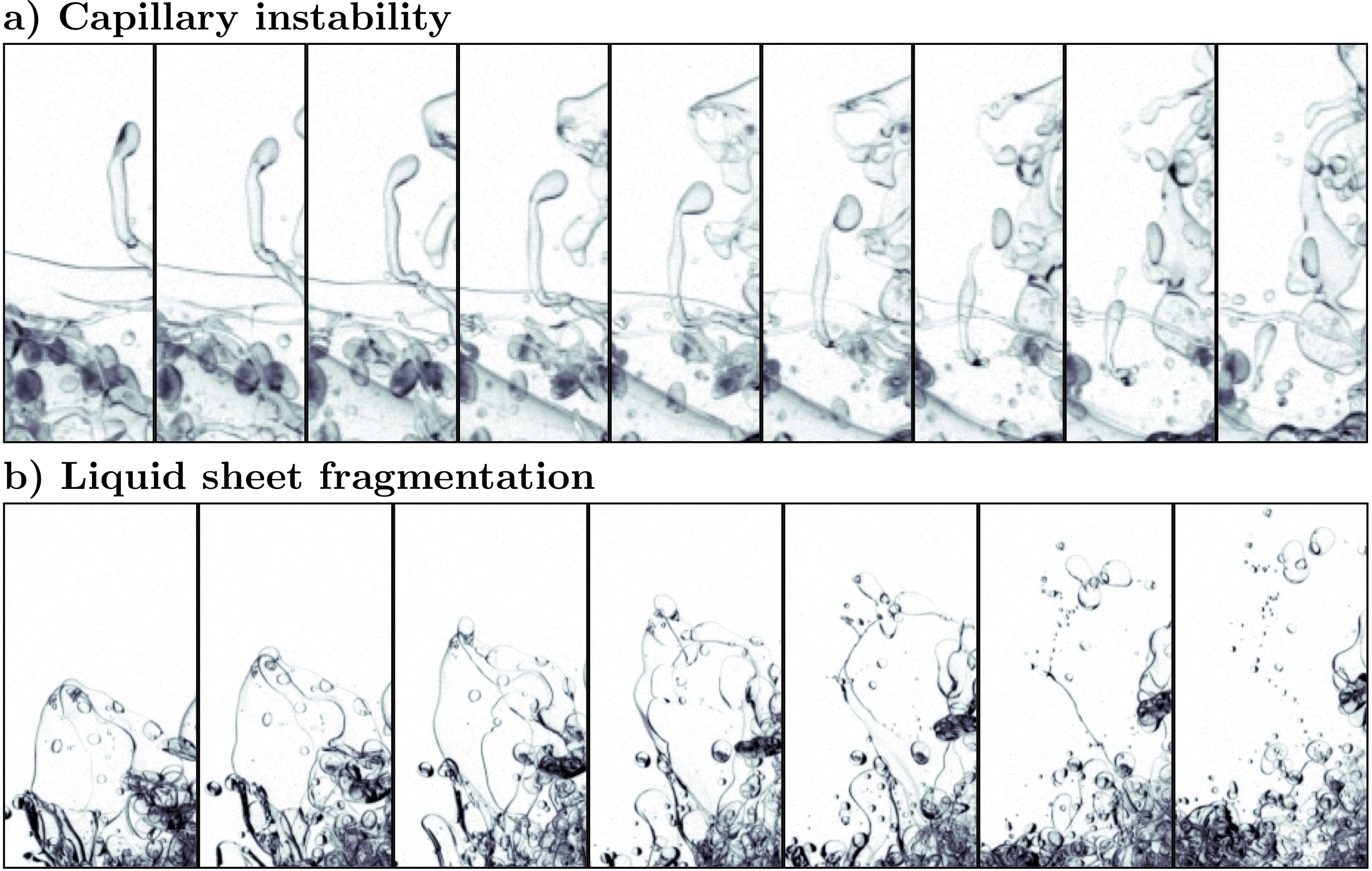}
\caption{
\textbf{Two fragmentations mechanisms} observed in the experiment shown in figure \ref{Fig:TurbulentThermal1}b. 
\textbf{a)}
Fragmentation of a liquid (NaI solution) ligament.
The time interval between two images is $\Delta t=10$ ms, and the width of each image is 1cm.
\textbf{b)}
Fragmentation of liquid film.
The time interval between two images is $\Delta t=20$ ms, and the width of each image is 1.8cm.
\label{Fig:Fragmentation}
}
\end{figure}

In the experiment shown on figure \ref{Fig:TurbulentThermal1}b and in similar experiments, most of the fragmentation of the dense liquid occurs during a relatively short time span.
In figure \ref{Fig:TurbulentThermal1}b, the dense phase is essentially continuous until the third snapshot, and almost entirely fragmented into drops at the fourth snapshot. 
The analysis of images obtained with a high-speed camera (1kHz) shows that drops formation results from two mechanisms:
\begin{enumerate}
\item
the fragmentation of stretched cylindrical ligaments of aqueous solution through the Rayleigh-Plateau capillary instability, as shown on figure \ref{Fig:Fragmentation}a (the mechanism of the Rayleigh-Plateau instability is explained below), 
\item
the fragmentation of thin liquid films, as shown in figure \ref{Fig:Fragmentation}b.
In this regime, thin films of aqueous solutions are stretched by the flow before eventually being punctured. 
The film then quickly retracts, the liquid forming the film gathering into ligaments which then fragment into drops due to the Rayleigh-Plateau instability.
\end{enumerate}
These two modes of fragmentation are classically observed in fragmentation problems in a variety of contexts. 
In fact, liquid fragmentation necessitates a capillary instability, irrespectively of the nature of the flow \citep{villermaux2007fragmentation}. 
What varies from one problem to another is the sequence of mechanisms resulting in the formation of ligaments which can fragment as a result of the Rayleigh-Plateau capillary instability.
In experiments such as shown on figure \ref{Fig:TurbulentThermal1}b, the observed sequence is the following: 
(i) the interface is destabilised and deformed by the combined effect of shear and Rayleigh-Taylor instabilities; 
(ii) three-dimensional structures generated by the destabilisation of the interface are stretched and stirred by the mean flow and velocity field fluctuations; 
(iii) stirring produces ligaments and films, which will then break up and produce a population of drops.

\subsubsection{Rayleigh-Plateau capillary instability}

In many situations, the deformation of an interface results in an increase of its surface area, and hence of its interfacial energy.
In this case deformation is not energetically favoured, and mechanical work therefore has to be provided to deform the interface.
This is for example the case of initially planar interface: any perturbation of the interface results in an increase of its surface area and energy.

In contrast, the deformation of a cylindrical interface can, under certain conditions, result in a decrease of its surface area, and hence of its interfacial energy.
Take a cylinder of one liquid into another, of length $L$ and radius $R_{0}$.
Its surface area is $2\pi R_{0} L$ and its interfacial energy is $2\pi R_{0} L \gamma$.
It is easy to see that the cylindrical shape is not very favourable from an energetic point of view: if the liquid of the cylinder is re-arranged to form a sphere of the same volume ($\pi R_{0}^{2} L$), the sphere will have a radius equal to $\left[(3/4) R_{0}^{2} L \right]^{1/3}$, and a surface area equal to $4 (3/4)^{2/3} \pi R_{0}^{4/3} L^{2/3}$, which is smaller than the cylinder surface area if the length of the cylinder is larger than $(9/2) R_{0}$.
This shows that the fragmentation into drops of liquid cylinder is energetically favoured if the ratio of its length and radius is larger than $9/2$.

\begin{figure}
\centering
\includegraphics{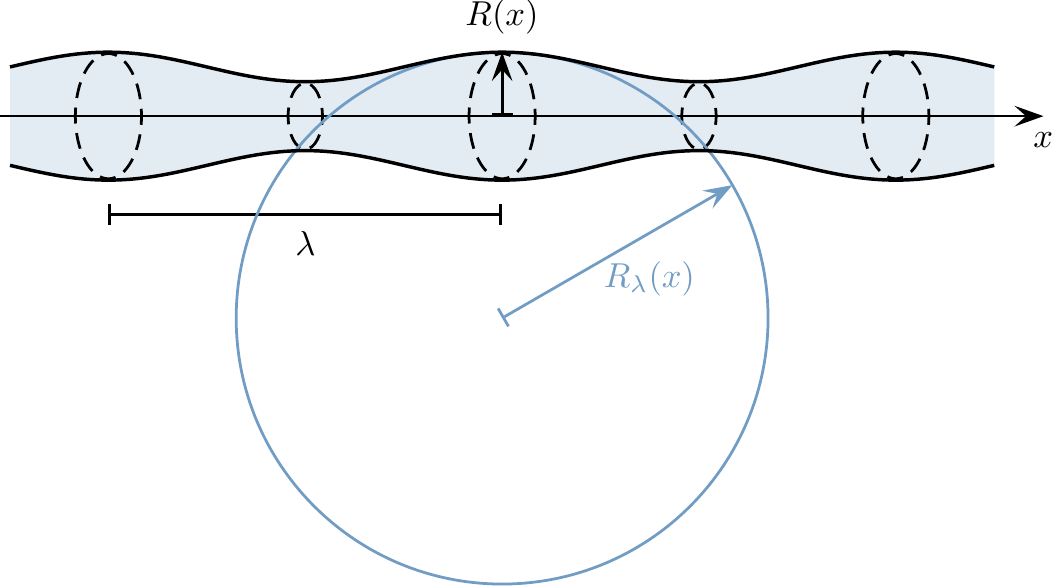}
\caption{A liquid cylinder (blue) with its surface perturbed by a axisymmetric sinusoidal perturbation.}
\label{Fig:CapillaryInstability}
\end{figure}

Is fragmentation dynamically possible? 
To see if it is, let us consider again a liquid cylinder of radius $R_{0}$, and assume now that its surface is perturbed from its initial shape as
\begin{equation}
R(x) = \bar R + \epsilon \sin \left( 2\pi \frac{x}{\lambda}\right),
\end{equation}
where $x$ is the coordinate along the axis of symmetry of the cylinder, and $\epsilon$ and $\lambda$ the amplitude and wavelength of the perturbation (figure \ref{Fig:CapillaryInstability}).
Note that conservation of mass implies that $\bar R < R_{0}$: comparing the volume of a section of length $\lambda$ of the unperturbed and perturbed states indeed shows that 
\begin{equation}
\bar R = R_{0} \left( 1 - \frac{\epsilon^{2}}{4 R_{0}^{2}}\right)^{1/2}.
\end{equation}
Calculating the surface area of the perturbed cylinder shows that the perturbation induces a decrease of the surface area (and hence of energy) if $\lambda > 2\pi R_{0}$, which suggests that the cylindrical shape may be unstable to perturbations with wavelengths larger than $2\pi R_{0}$.

To see how the instability works, let us consider the two principal curvatures of the interface (figure \ref{Fig:CapillaryInstability}).
One is the curvature associated with the radius of the cylinder, $1/R(x)$, and the other is the curvature associated with the longitudinal perturbation, $1/R_{\lambda}(x)$, which is equal to the divergence in the $x-$direction of the normal $\mathbf{n}$  of the interface. 
The Laplace pressure jump across the interface is equal to
\begin{equation}
\Delta P = \gamma \left( \frac{1}{R(x)} + \frac{1}{R_{\lambda}(x)} \right).
\end{equation}

The contribution of $1/R_{\lambda}$ is positive where $R>R_{0}$, and negative where $R<R_{0}$. It thus produces a pressure gradient from regions of large $R$ to regions of small $R$, which can drive a flow from large to small $R$ that would decrease the amplitude of the radius perturbation. It thus has a stabilizing effect. 
In contrast, the pressure jump associated with $1/R$ is larger in regions of small $R$. It thus produces a pressure gradient from small to large $R$, which may drive a flow that would increase the amplitude of the perturbation.
The amplitude of the radius perturbation can therefore grow if the pressure gradient associated with the curvature $R(x)$ is larger in magnitude that the pressure gradient associated with the curvature $R_{\lambda}(x)$.

At first order in $\epsilon$, one find  
\begin{equation}
\frac{1}{R} = \frac{1}{R_{0}} \left[ 1 -  \frac{\epsilon}{R_{0}} \sin \left( 2\pi \frac{x}{\lambda}\right) \right] + \mathcal{O}(\epsilon^{2} ),
\end{equation}
and
\begin{equation}
\frac{1}{R_{\lambda}} = \frac{-1}{\sqrt{1+(dR/dx)^{2}}} \frac{d^{2} R}{dx^{2}} = \frac{4\pi^{2}}{\lambda^{2}} \epsilon \sin \left( 2\pi \frac{x}{\lambda}\right) + \mathcal{O}(\epsilon^{2} ),
\end{equation}
which gives a pressure jump across the interface given by
\begin{equation}
\Delta P = \gamma \left[ \frac{1}{R_{0}}    + \epsilon \left(  \frac{4\pi^{2}}{\lambda^{2}} -  \frac{1}{R_{0}^{2}}  \right) \sin \left( 2\pi \frac{x}{\lambda}\right) \right] + \mathcal{O}(\epsilon^{2} ).
\end{equation}
Its gradient along $x$ is given by  
\begin{equation}
\frac{\partial \Delta P}{\partial x} = \epsilon \frac{2\pi}{\lambda} \gamma \left(  \frac{4\pi^{2}}{\lambda^{2}} -  \frac{1}{R_{0}^{2}}  \right) \cos \left( 2\pi \frac{x}{\lambda}\right) + \mathcal{O}(\epsilon^{2} ).
\end{equation}
This shows that the pressure gradient inside the liquid cylinder is from small to large $R$ if $\frac{4\pi^{2}}{\lambda^{2}} < \frac{1}{R_{0}^{2}}$, and from large to small $R$ instead.
In other words, the initial perturbation will grow if
\begin{equation}
\lambda > \lambda_{c} = 2 \pi R_{0},
\end{equation}
which will eventually lead to the fragmentation of the cylinder into drops.
This dynamical criterion is slightly more restrictive that the energy criterion, since $2\pi \simeq 6.26 > 9/2=4.5$.  

Things get more complicated in situations where liquid ligaments are deformed and stretched by the ambient flow. 
Fragmentation can be significantly protracted by stretching effects \citep{taylor_1934,tomotika_1936,mikami_1975,eggers2008physics}, which can be understood as follows.
Let us consider a stretched ligament, the surface of which is modulated by a longitudinal perturbation of wavelength $\lambda $.
The stretching will affect the disturbance, which will see its wavelength increase in proportion to the amount of stretching.
If the perturbation wavelength initially corresponds to the optimal wavelength for the growth of the capillary instability, increasing the wavelength will decrease the rate of growth of the disturbance. 

\subsection{Chemical and heat transfer at the drop scale}

If the metal phase ends up being fragmented into drops of size equal or smaller than the maximal stable size $R_{c}\sim \sqrt{ \gamma/(\Delta \rho g) }$ (equation \eqref{Eq:maxDropSizeTerminalVelocity}), then thermal and chemical equilibration of the metal phase with the surrounding silicates is not an issue \citep[\textit{e.g.}][]{Stevenson1990,Karato1997,Rubie2003,Ulvrova2011}, as shown below.

One can show \citep{lherm2018} that the timescale of chemical equilibration  of a falling metal drop with its surrounding is given by
\begin{equation}
\tau_\text{eq} \sim \frac{R^2}{6 \kappa_s} D_\text{m/s} \mathrm{Pe_s}^{-1/2}\left[1+\frac{1}{D_\text{m/s}}\left(\frac{\kappa_s}{\kappa_m}\right)^{1/2}\right] ,
\end{equation}
where $\kappa_s$ and $\kappa_m$ are the compositional diffusivities in the silicates and metal, $D_\text{m/s}$ the metal/silicates partitioning coefficient of a given chemical element, and $Pe_{s} = U R/\kappa_{s}$.
The distance $\ell_\text{eq}$ fallen by the drop during a time $\tau_\text{eq}$ is given by
\begin{equation}
\ell_\text{eq} = \tau_\text{eq} U \sim \frac{R}{6} D_\text{m/s} \mathrm{Pe_s}^{1/2}\left[1+\frac{1}{D_\text{m/s}}\left(\frac{\kappa_s}{\kappa_m}\right)^{1/2}\right] 
\end{equation}
In the case of siderophile elements ($D_\text{m/s}\gg 1$), this is approximated by
\begin{equation}
\ell_\text{eq} \sim \frac{R}{6} D_\text{m/s} \mathrm{Pe_s}^{1/2}.
\end{equation}

For a metal drop falling into a magma ocean, $R_{c}$ is about 5 mm.
The corresponding Grashof number is 
\begin{equation}
Gr = \frac{\Delta \rho}{\rho_{s}}   \frac{g R_{c}^{3}}{\nu_{s}^{2}} \sim \left( {\frac{g}{10\, \mathrm{m.s}^{-2}}} \right) \times \left( \frac{10^{-2}\, \text{Pa.s}}{\eta_{s}} \right)^{2} \times 10^{5}.
\end{equation}
Magma ocean viscosity is estimated to be in the range $10^{-3}$-$10^{-2}$ Pa.s, which implies that the drop is in the newtonian regime and has a terminal velocity given by $\sim \sqrt{(\Delta\rho/\rho_{s} g R_{c}}$.
With a compositional diffusivity $\kappa_{s}\sim 10^{-9}$ m.s$^{-2}$, this gives a P\'eclet number around $10^{6}$ and we thus have $(R/6)  \mathrm{Pe_s}^{1/2} \simeq 1$ m.
Siderophile elements like Nickel or Cobalt have partitioning coefficients $D_\text{m/s}$ around $10^{3}$ at low pressure, and as low as $\sim 10$ when approaching the pressure of the core-mantle boundary.
This would give $\ell_\text{eq} \sim 1$ km at low pressure, and $\ell_\text{eq} \sim 10$ m at high pressure.
This is in both cases much smaller than typical magma ocean depth, which implies that drops of metal a few mm in size will readily equilibrate with the surrounding molten silicates.

However, the above conclusion rests on the assumption that the metal phase  fragments into drops of a few mm in radius. 
We have no well tested fragmentation model that can be used in the context of core formation, so whether the metal phase would fragment or not is still an open question.
It is also possible that efficient equilibration does not require fragmentation of the metal phase. 
The metal phase would necessarily be intensely stirred and stretched before fragmentation, and this may allow for efficient chemical transfer between the metal and silicates phase \citep{lherm2018}.

\subsection{Acknowledgments}

This project has received funding from the European Research Council (ERC) under the European Unions Horizon 2020 research and innovation programme (grant agreement No 716429).

\end{document}